\documentclass[a4paper,10pt]{article}
\usepackage[cm]{fullpage}
\usepackage{xcolor}
\usepackage{amsmath}
\usepackage{amsfonts}
\usepackage{amssymb}
\usepackage{mathtools} 
\usepackage{afterpage}
\usepackage{transparent}
\usepackage{tabularx}
\usepackage{dsfont}
\usepackage{graphicx}
\usepackage{epstopdf}
\usepackage[bookmarks=false]{hyperref}
\usepackage{appendix}
\usepackage[authoryear]{natbib}
\usepackage{cleveref} 
\usepackage{bm}
\usepackage{subfig} 
\crefname{equation}{}{}

\newcommand{\vecZeroes}{\boldsymbol{0}}
\newcommand{\vecOnes}{\boldsymbol{1}}

\newcommand{\vecA}{\boldsymbol{a}}

\newcommand{\vecB}{\boldsymbol{b}}
\newcommand{\vecbigB}{\boldsymbol{B}}

\newcommand{\vecbigI}{\boldsymbol{I}}

\newcommand{\vecbigR}{\boldsymbol{R}}
\newcommand{\vecU}{\boldsymbol{u}}

\newcommand{\vecX}{\boldsymbol{x}}
\newcommand{\vecbigX}{\boldsymbol{X}}
\newcommand{\vecXi}{\boldsymbol{\xi}}
\newcommand{\vecY}{\boldsymbol{y}}

\newcommand{\vecZ}{\boldsymbol{z}}
\newcommand{\vecbigZ}{\boldsymbol{Z}}
\newcommand{\E}{{\rm I\kern-.18em E}}
\newcommand{\Cov}{\mathrm{Cov}}
\newcommand{\Var}{\mathrm{Var}}
\newcommand{\ACF}{\mathrm{ACF}}
\newcommand{\CCF}{\mathrm{CCF}}
\definecolor{darkgreen}{rgb}{0,0.6,0}
\def\f{summary}
\def\s{a}

\DeclarePairedDelimiter\abs{\lvert}{\rvert}%
\DeclarePairedDelimiter\norm{\lVert}{\rVert}%

\makeatletter
\let\oldabs\abs
\def\abs{\@ifstar{\oldabs}{\oldabs*}}
\let\oldnorm\norm
\def\norm{\@ifstar{\oldnorm}{\oldnorm*}}
\makeatother

\newenvironment{algo}[1]
{
   \framebox[\textwidth][l]{
       \begin{minipage}{\textwidth}
           \begin{tabbing}
           \settabs
            #1
           \end{tabbing}
        \end{minipage}
    }
}{}
\newcommand{\settabs}{mmm\=mmm\=mmm\=mmm\=mmm\=mmm\=\kill}

\numberwithin{equation}{section} 

\hypersetup{
	pdfstartview = {XYZ null null 1.0},
    unicode      = false,            
    pdftoolbar   = true,             
    pdfmenubar   = true,             
    colorlinks   = false
}

\title{Stochastic parameterization with VARX processes}
\author{N. Verheul$^{1}$\footnote{Corresponding author. E-mail address: nick.verheul@cwi.nl, Postal address: 1090 GB Amsterdam, P.O. Box 94079, Netherlands} , \ D.T. Crommelin$^{1,2}$\footnote{E-mail address: daan.crommelin@cwi.nl, Postal address: 1090 GB Amsterdam, P.O. Box 94079, Netherlands} \\ \mbox{}\\ \textit{\mbox{}$^1$ Centrum Wiskunde \& Informatica (CWI), Amsterdam, Netherlands}
\\ \textit{\mbox{}$^2$ Korteweg-de Vries Institute for Mathematics, University of Amsterdam}
}
\date{\today}

\begin{document}

\maketitle
\abstract{In this study we investigate a data-driven stochastic methodology to parameterize small-scale features in a prototype multiscale dynamical system, the Lorenz '96 (L96) model. We propose to model the small-scale features using a vector autoregressive process with exogenous variable (VARX), estimated from given sample data. To reduce the number of parameters of the VARX we impose a diagonal structure on its coefficient matrices. We apply the VARX to two different configurations of the 2-layer L96 model, one with common parameter choices giving unimodal invariant probability distributions for the L96 model variables, and one with non-standard parameters giving trimodal distributions. We show through various statistical criteria that the proposed VARX performs very well for the unimodal configuration, while keeping the number of parameters linear in the number of model variables. We also show that the parameterization performs accurately for the very challenging trimodal L96 configuration by allowing for a dense (non-diagonal) VARX covariance matrix.}

\vspace{.1in}

 
\vspace{.1in}


\section{Introduction}\label{sec:introduction}
\subsection{Background}
\ifx\s\f \noindent {\color{darkgreen} Summary: insufficient scales in equations of motion, consider reduced systems, need additional dynamical term: parameterization} \fi

For many spatially extended dynamical systems the equations of motion cannot be solved on sufficiently fine scales because of unfeasible computational costs. The typical approach for dealing with this problem is to formulate a reduced system that describes the variables of interest, usually the large-scale degrees of freedom. To compensate for the missing dynamical effects (feedback) that arise from the small scales, some dynamical term that represents or approximates these missing effects needs to enter the reduced system. Following common terminology in ocean-atmosphere science where this is an important problem, we call such terms parameterizations. In a previous study, we considered discrete resampling-based methods \citep{verheul2016datadriven,verheul2017covariatebased}. These methods were successful in reproducing various important statistical and physical aspects in the reduced models. These promising results notwithstanding, their capability to model spatial correlations in the dynamical feedback from the small scales is limited. Here, instead, we investigate parameterizations that are better able to reproduce the spatio-temporal correlations explicitly, without significant computational cost.

\ifx\s\f \noindent {\color{darkgreen}Summary: specific parameterization we use with traits that set it apart} \fi

Specifically, we propose to use a vector autoregressive process with exogenous parameters (VARX) for parameterization. We include endogenous and exogenous variables in the VARX process with coefficient matrices that have sparse structure, e.g. (tri)-diagonal. The main aim is that the reduced model with the parameterization accurately reproduces the statistical properties of the reference (fully resolving, non-reduced) model, including its spatial correlations. Moreover, our stochastic parameterization assumes no knowledge of the underlying physical structure of the system. We use available sample data from the fully resolving reference model to infer the VARX model, similar in spirit to the data-driven approaches in \citet{crommelin2008subgrid,porta2014toward,verheul2016datadriven,verheul2017covariatebased}. In the context of ocean-atmosphere modeling, various other forms of stochastic parameterizations have been considered, e.g. stochastic cellular automata \citep{shutts2005kinetic,bengtsson2013stochastic,crommelin2018cellular}, and Markov chain approaches \citep{majda2002stochastic,crommelin2008subgrid,khouider2010stochastic,dorrestijn2016stochastic}, see also \citep{berner2017stochastic} for a recent overview.

\ifx\s\f  \noindent {\color{darkgreen} Summary: the form of multiscale models we consider, including reduction} \fi

We consider multiscale models wherein the state vector $\vecZ := (\vecX, \vecY_1, \dots, \vecY_J)$  evolves over time according to a set of coupled ordinary differential equations (ODEs) that include a constant forcing $\mathcal{F}$, a linear operator $\mathcal{L} \vecZ$, and some nonlinear operator $\mathcal{B} (\vecZ)$. This set of ODEs can result from the spatial discretization of a partial differential equation; in this study we focus on the ODE formulation in which the elements of the state vector are associated with, e.g, values on a spatial grid. We consider nonlinear ODEs of the following form (occurring in e.g. ocean models \citep{hua1986numerical,berloff2005dynamically}):
\begin{align}
\frac{d\vecX}{dt} &= \mathcal{F} + \mathcal{L}_{x} \vecX + \mathcal{B}_{xx}(\vecX) + \mathcal{B}_{xy}(\vecX, \vecY_1, \dots, \vecY_J) \label{eq:generalresolvedprocess}\\
\frac{d\vecY_j}{dt} &= \mathcal{L'}_{x} \vecX + \mathcal{L}_{y} \vecY_j + \mathcal{B}_{yy}(\vecY_1, \dots, \vecY_J) + \mathcal{B}_{yx}(\vecY_1, \dots, \vecY_J, \vecX) \label{eq:generalunresolvedprocess},
\end{align}

\noindent where the vector $\vecX := (x_1, \dots, x_K)$ represents the large-scale processes, the vectors $\vecY_j := (y_{j,1}, \dots, y_{j,k})$ represent the small-scale processes, where $1 \le j \le J$, $1 \le k \le K$ are spatial grid indices, and each $x_k$ is coupled to $J$ small-scale $y_{j,k}$. Thus, $K$ is the total number of gridpoints on which the large-scale processes are defined. This number can be very large for spatially extended systems (e.g. $K=10^6$ for a system with 2 spatial dimensions, specified on a $1000 \times 1000$ grid). The $y_{j,k}$ can be thought of as being defined on a micro-grid (with $J$ gridpoints) associated with each macro-gridpoint $k$.

The operator $\mathcal{B}_{yy}$ denotes the nonlinear self-interaction of the $\vecY_j$ variables, and $\mathcal{B}_{xy}$ denotes the nonlinear feedback of the $\vecY_j$ variables on the $\vecX$ variables. The operators $\mathcal{B}_{xx}$ and $\mathcal{B}_{yx}$ have analogous interpretations. We assume that an analytic solution to (\ref{eq:generalresolvedprocess})-(\ref{eq:generalunresolvedprocess}) is not available, so that we have to resort to numerical integration.
The computational bottleneck for numerical integration of \Crefrange{eq:generalresolvedprocess}{eq:generalunresolvedprocess} is evolving all $\vecY_j$ variables for each $x_j$. Therefore, we construct a reduced model involving only the variables of interest $\vecX$. This reduced model consists of (\ref{eq:generalresolvedprocess}) with $\mathcal{B}_{xy}$ replaced by a stochastic (VARX) parameterization $\widetilde{\vecB}:= \widetilde{\vecB}(\widetilde{\vecX})$ that is meant to emulate $\vecB := \mathcal{B}_{xy}(\vecX, \vecY_1, \dots, \vecY_J)$. To distinguish between variables in the original deterministic model (e.g. $\vecX$), and their analogues in the reduced stochastic model (e.g. $\widetilde{\vecX}$) we use the tilde-notation for all variables in the stochastic model. Thus, the reduced model is 
\begin{equation}
\frac{d\widetilde{\vecX}}{dt} = \mathcal{F} + \mathcal{L}_{x} \widetilde{\vecX} + \mathcal{B}_{xx}(\widetilde{\vecX}, \widetilde{\vecX}) + \widetilde{\vecB}(\widetilde{\vecX}) \, . \label{eq:generalstochasticmodel}
\end{equation}

\ifx\s\f \noindent {\color{darkgreen}Summary: properties of parameterization: state-dependency, emulate the missing dynamical effects, arbitrarily many covariates, Lorenz 96, overview} \fi

The state-dependence of $\widetilde{\vecB}(\widetilde{\vecX})$ allows the properties of the stochastic process to evolve together with the resolved variables ($\widetilde{\vecX}$). The parameters of the process $\widetilde{\vecB}(\widetilde{\vecX})$ are inferred from reference simulation data $(\vecbigX, \vecbigB)$, obtained by numerical integration of \Crefrange{eq:generalresolvedprocess}{eq:generalunresolvedprocess}. Here $\vecbigX := (\vecX^1, \dots, \vecX^N)$, $\vecbigB := ( \vecB^1, \dots, \vecB^N)$
in which $\vecX^n := \vecX(t_n) := \vecX(n \Delta t)$ denotes the $n$-th time-instance of $\vecX$ and $\vecB^n := \mathcal{B}_{xy}(\vecX^n, \vecY_1^n, \dots, \vecY_J^n)$ denotes the $n$-th time-instance of $\vecB$. Finally, $N$ denotes the number of sample points (or time steps).


In section \ref{sec:varx} we present a simple and straightforward VARX framework that uses sparse coefficient matrices. Then in Section \ref{sec:l96} we apply our parameterization to the Lorenz '96 (L96) model \citep{lorenz1996predictability}, a frequently used test bed for developing parameterization methods \citep{palmer2001nonlinear,wilks2005effects,crommelin2008subgrid,chorin2015discrete}. Next, we discuss technical details of our parameterization in Section \ref{sec:practical} and present numerical results in Section \ref{sec:results}.

\section{VARX representation}\label{sec:varx}
\ifx\s\f \noindent {\color{darkgreen}Summary: VARX model defined by its matrices } \fi

We model the stochastic term $\widetilde{\vecB}$ in (\ref{eq:generalstochasticmodel}) as a VARX process (see, e.g, \citet{lutkepohl2005new}). Numerical implementation of such a process is straightforward (e.g. \citet{pavliotis2016stochastic}).  We make no assumptions about the underlying physics of $\vecB$, instead we infer the VARX process from the second-order statistics of $\vecB$ estimated from the available sample data $(\vecbigX, \vecbigB)$. 

\subsection{Mean: linear combination of covariates}\label{sec:mean}

\ifx\s\f \noindent {\color{darkgreen} Summary: VARX definition, state-dependency introduction} \fi

A complete characterization of a VARX$(p)$, i.e. VARX of order $p$, is given by its drift matrices $A_i$, $i=1,\dots,p$ and $D$ and covariance matrix $\Sigma \Sigma^T$:
\begin{equation}\label{eq:varx}
\widetilde{\vecB}^{n} = \vecA_0 + A_1 \widetilde{\vecB}^{n-1} + \dots + A_p \widetilde{\vecB}^{n-p} + D \vecX^n + \Sigma \vecXi^n
\end{equation}

\noindent where $\vecA_0$ is the linear offset, $A_1, \dots, A_p$ represent the \emph{endogenous} drift matrices, $D$ is the \emph{exogenous} drift matrix, $\Sigma \Sigma^T$ is the covariance matrix, and $\vecXi^n$ is a vector of independent normally distributed random variables, $\vecXi^n \sim \mathcal{N} (0,I)$. The matrices $A_i$, $D$, and $\Sigma$ all have size $K \times K$. 

Borrowing some terminology from statistics, the variable $\widetilde{\vecB}^n$ is known as the \emph{regressand} and the variables $\widetilde{\vecB}^{n-1}, \dots, \widetilde{\vecB}^{n-p}, \vecX^n$ are known as the \emph{regressors}. By choosing regression coefficient matrices $A_i$ or $D$ to be nonzero, the variable $\widetilde{\vecB}^n$ becomes dependent on those regressors. By imposing certain sparsity patterns on the drift matrices in (\ref{eq:varx}) we can choose to make $\widetilde{\vecB}^{n}$ conditional on $\widetilde{\vecB}$ or $\vecX$ at specific space or (past) time points. For example, $\widetilde{b}^n_k$ can be made conditionally dependent on its previous state ($n-1$) at neighboring gridpoints ($k \pm 1$) by letting the matrix elements $(A_1)_{l,m}$ be nonzero if $(l, m) = (k\pm 1, k)$ or if $(l, m)=(k, k\pm 1)$. Similarly, if $D$ is diagonal, $\tilde b^n_k$ is conditionally dependent on $x^n_k$ (i.e., at the same spatial grid point with index $k$) but not on $x^n_{k'}$ at grid points $k' \neq k$. 

\ifx\s\f \noindent {\color{darkgreen} Summary: Cost-efficiency of training phase, iterations} \fi

Typically, the matrices $A_i$ and $D$ in (\ref{eq:varx}) are obtained through maximum likelihood estimation. We apply the weighted least squares procedure \citep{strutz2010data} to obtain accurate estimators. The training phase of our proposed algorithm consists primarily of calculating the regression coefficients (i.e., the elements of the matrices $A_i$, $D$ and $\Sigma \Sigma^T$).
Since the weighted least squares procedure is highly optimized, this training phase is very cost-efficient. Generalizations of this approach are possible by modeling $\widetilde{\vecB}$ as realizations of a Gaussian process where the drift components are represented as generalized linear models (GLIMs) (see e.g. \citet{mccullagh1989generalized} for a detailed description on GLIMs). 

\ifx\s\f \noindent {\color{darkgreen} Summary: no curse of dimensionality, linear relation} \fi

With data-driven approaches, the number of parameters that must be estimated and/or the amount of sample data needed, can grow exponentially in the number of conditioning variables, see, e.g, the binning approach in \citet{verheul2016datadriven,verheul2017covariatebased} or the conditional Markov chain setups in \citet{crommelin2008subgrid,dorrestijn2013data,gottwald2016data}. Here we mitigate this problem by imposing the structure of VARX, wherein, even if all matrices in (\ref{eq:varx}) are fully dense (and thus $\tilde b_k^n$ is conditional on the entire vector $\vecX^n$ as well as on all vectors $\widetilde{\vecB}^{n'}$ with $n-p\leq n' \leq n-1$), we still only have $K+K^2(p+2)$ parameters. By restricting the drift matrices in (\ref{eq:varx}) to be sparse, the number of parameters reduces further. For example, if we choose all $A_i$, $D$ and $\Sigma \Sigma^T$ to be diagonal matrices, then the number of parameters grows linearly in $p$ and $K$.

With this approach, estimation and order selection are nontrivial issues. For example, order selection is difficult because $\vecB$ will be very strongly correlated with itself at short lag times and  effectively decorrelated at long lag times. For a more detailed discussion of the order selection problem, see Section \ref{sec:results_lagchoice}. Another difficulty for estimation is that the model needs to satisfy the stationarity constraints, otherwise the trajectory of the model variables can diverge to infinity. In order for the VAR$(p)$ to be stationary, the matrix elements of the $A_i$ must satisfy the VAR$(p)$ stability constraint \citep{lutkepohl2006forecasting}: 
\begin{equation}\label{eq:varxstationarity}
\forall \lambda : \hspace{.1in} \abs{I^n \lambda^p - A_1 \lambda^{p-1} - A_2 \lambda^{p-2} \dots - A_p} = 0 \hspace{.1in} \Rightarrow \hspace{.1in} \abs{\lambda} < 1.
\end{equation}

Equivalently, the VAR$(p)$ is stationary if the eigenvalues of the \emph{companion matrix} $F$ have modulus less than one, where the companion matrix of (\ref{eq:varx}) is defined as:
\begin{equation}\label{eq:companionmatrix}
\begin{bmatrix} 
A_1 & A_2 & \dots & A_n \\
\vecOnes_n & \vecZeroes & \dots & \vecZeroes \\
\vecZeroes & \ddots & \vecZeroes & \vdots \\
\vecZeroes & \dots & \vecOnes_n & \vecZeroes \\
\end{bmatrix}.
\end{equation}

The development of regression methods that explicitly enforce this stability constraint is beyond the scope of this study. We only verify that our models satisfy the stability constraint \emph{a posteriori}. 

\subsection{Covariance and resulting VARX}\label{sec:covariance}
\ifx\s\f \noindent {\color{darkgreen} Summary: covariance definition} \fi


We consider two different forms of the covariance $\Sigma \Sigma^T$ of the VARX process  (\ref{eq:varx}). In one, the covariance matrix is a multiple of the identity matrix, i.e. all cross-covariances are ignored and auto-covariances do not depend on $k$. In the other, the covariance matrix is fully dense, allowing for nonzero cross-covariances and $k$-dependent auto-covariances.

Given the constant offset $\vecA_0$ and the matrices $\{ A_i \}_{1 \le i \le p}$ and $D$, we calculate the residuals $\vecB^{n} - \vecA_0 - A_1 \vecB^{n-1} - \dots - A_p \vecB^{n-p} - D \vecX^n$ of the regression fit from the  sample time series $(\vecbigX, \vecbigB)$. For the first form of the covariance we set $\Sigma_D := \sigma I$, where $\sigma$ is the averaged standard deviation over the residuals over all $k$. Although this form is extremely simple, it has only a single parameter $(\sigma)$ so that it can easily be used even when $K$ is very large.
For a dense covariance matrix we compute all the pairwise sample covariances from residuals.  $\Sigma_L$ is then obtained from the Cholesky decomposition of the sample covariance matrix. This is straightforward and general but becomes unfeasible for large $K$ (we recall that the covariance matrix and hence also $\Sigma_L$ is of size $K \times K$). However, we include this covariance structure as an ``optimal'' reference for the current study.  

\ifx\s\f \noindent {\color{darkgreen} Summary: stochastic setup, properties, state-dependency} \fi

Applying this VARX$(p)$ model as forcing $\widetilde{\vecB}$ to the reduced model (\ref{eq:generalstochasticmodel}) results in the following stochastic model:
\begin{equation}\label{eq:completestochasticmodel}
\widetilde{\vecX}^{n+1} = T \left(\Delta t, \, \mathcal{F} + \mathcal{L}_{x} \widetilde{\vecX}^n + \mathcal{B}_{xx}(\widetilde{\vecX}^n, \widetilde{\vecX}^n) + \widetilde{\vecB}^n \right), \qquad \widetilde{\vecB}^{n} = \vecA_0 + A_1 \widetilde{\vecB}^{n-1} + \dots + A_p \widetilde{\vecB}^{n-p} + D \widetilde{\vecX}^n + \Sigma \vecXi^n,
\end{equation}

\noindent where $T$ represents a numerical integration scheme of choice (see Section \ref{sec:practical}), and $\Sigma$ can be either $\Sigma_D$ or $\Sigma_L$.
We emphasize the coupling between $\widetilde{\vecX}$ and $\widetilde{\vecB}$ goes in both directions: $\widetilde{\vecB}$ enters as a forcing term in the time integration of $\widetilde{\vecX}$, whereas the time evolution of $\widetilde{\vecB}$ depends on $\widetilde{\vecX}$ through the dependence of the VARX process on $\widetilde{\vecX}$. Such a state-dependence allows for the modeling of different dynamical regimes of $\widetilde{\vecB}$. If $\vecX$ and the chosen lagged $\vecB$ are adequate predictors, such regimes can occur in a  similar fashion as in the sample data $(\vecbigX, \vecbigB)$.

We note that while the VARX process allows for a spatially varying (i.e., $k$-dependent) mean and covariance, only the mean is able to vary temporally. Therefore, we expect our parameterization to be less suitable for cases where the small-scale processes have multiple variance regimes under the same large-scale state $\vecX$.

\subsection{Computational complexity}\label{sec:modelingchoices}

\ifx\s\f \noindent {\color{darkgreen} Summary: low online computational costs} \fi

The methodology we propose here requires very little computational cost in the training stage. First, the regression matrices $A_i$ and $D$ in (\ref{eq:completestochasticmodel}) are calculated with a single least squares call. The least squares algorithm is very efficient with computational complexity $O(K^2 N)$, and a well-optimized routine on many computational platforms. Second, the covariance $\Sigma \Sigma^T$ is estimated straightforwardly with the sample (co)variances calculated from the residuals, also with complexity $O(K^2 N)$. In the case of VARX models with diagonal covariance, the matrix root $\Sigma_D$ of $\Sigma_D \Sigma_D^T$ is computed directly with sample standard deviations. In the alternate case of fully dense covariance, the matrix root $\Sigma_L$ is computed with a Cholesky decomposition. For $N > K$, the Cholesky decomposition is a less costly operation with $O(K^3)$ complexity that only needs to be calculated once in the initialization phase because our covariance is constant over time.

The motivation for restricting the regression matrices $A_i, D$ and $\Sigma$ (by imposing sparsity, e.g. a diagonal form) has two origins: first, the memory usage. Many ocean-atmosphere studies consider models with very large grids, e.g. $K = 512^2$ gridpoints. Full covariance matrices for such grids would contain upward of $512^4$ nonzeroes. Such matrices typically are too large to fit in the computing platform's work memory, making efficient online computations unfeasible. Second, the cost of numerically integrating $\widetilde{\vecX}$ over time, i.e. the online costs of the stochastic methodology.  The online cost of our stochastic methodology is dominated by the matrix vector products (MVPs) required to simulate $\widetilde{\vecB}$ (\ref{eq:completestochasticmodel}). If we restrict the number of nonzero conditioning variables, the drift matrices $A_i$ and $D$ become sparse, e.g. $(K \times K)$-band matrices. This reduces the complexity of the drift MVPs in (\ref{eq:completestochasticmodel}) to linear in $K$. The structure of the covariance has a different impact on the computational complexity of (\ref{eq:completestochasticmodel}). The diagonal $(K \times K)$-matrix $\Sigma_D$ gives linear (in $K$)  complexity of the MVPs in (\ref{eq:completestochasticmodel}). By contrast, the lower-triangular $\Sigma_L$ gives $O(K^2)$ complexity of the MVPs in (\ref{eq:completestochasticmodel}), causing a computational bottleneck for large $K$. Imposing sparsity (other than diagonality) on $\Sigma$ in a statistically and dynamically consistent way is nontrivial yet important for systems with large $K$; we leave this topic for future study.

\subsection{Comparison to other stochastic parameterizations}\label{sec:varx_comparisons}
In this study we compare different stochastic parameterizations in terms of their effect on the long-term statistical behavior of the resolved model variables (see Section \ref{sec:results}). Besides the VARX model proposed here, this comparison includes parameterizations based on AR(1) and on NARMAX processes that have been proposed before in the literature. For clarity, we label the different parameterizations with short descriptive names (e.g. (\ref{eq:b_double_tri_dense})) instead of referring to equation numbers. We compare the following parameterizations: 
\begin{itemize}
\item WN: white noise process. This is an ``unconditioned'' parameterization (no conditioning on  $\widetilde{\vecX}$ or on past values of $\widetilde{\vecB}$). It is included as it represents the simplest stochastic model, and enables us to assess the merit of more complicated stochastic models.

\item AR$(1)$: autoregressive process, independently applied to each of the grid points $k$. Discussions on AR processes can be found in standard text books on time series analysis. In \citet{arnold2013stochastic}, parameterization with AR$(1)$ is proposed and discussed in more detail. They consider a parameterization consisting of both a deterministic and stochastic part: a regressed polynomial dependent on $x$ (deterministic) and a one-step autoregression (stochastic) with varying options for noise models (we compare to their "additive" noise model). They show both that the stochastic parameterizations improve significantly over deterministic parameterizations and that the autoregression models  are a major improvement over WN. We include this parameterization as it is a special case of the VARX models proposed here. 

\item VARX$(p)$ $\Sigma_D$: vector autoregressive process with exogenous variable. We choose all matrices $A_i = 0$ for $i\ne p$ (see (\ref{eq:varx})) and we enforce sparsity by requiring the drift matrices $A_p, D$ and the noise matrix $\Sigma_D$ all to be diagonal. We choose a single nonzero drift matrix $A_i$ to circumvent parameter estimation issues, as resolving these would require a study of itself, see Section \ref{sec:results_lagchoice} for a detailed discussion. As discussed above, imposing sparsity on the regression coefficient matrices is intended to limit the number of parameters and to make this parameterization approach more tractable for high-dimensional ocean and atmosphere models. 

\item VARX$(p)$ $\Sigma_L$: similar as VARX$(p)$ $\Sigma_D$, however with  a lower triangular (non-diagonal) root covariance matrix $\Sigma_L$ instead of a diagonal one ($\Sigma_D$). This allows us to explicitly model the cross-correlations between spatial points. $\Sigma_L$ is not sparse; we leave the case of a non-diagonal but sparse covariance matrix for a follow-up study (nearing completion).

\item $\text{NARMAX}_{1,2,0,1}$ and $\text{NARMAX}_{1,1,1,0}$: nonlinear autoregression moving average with exogenous input models, proposed for parameterization by \citet{chorin2015discrete}. The subscripts denote the values of parameters $(p,r,s,q)$ that define the specific NARMAX structure (e.g. the number of endogenous variables, or the number of moving average terms). The NARMAX parameterization in \citet{chorin2015discrete} is applied independently to each grid point. Thus, NARMAX is scalar-valued, whereas VARX is vector-valued. When the matrices $A_p$, $D$ and $\Sigma$ are all multiples of the identity matrix, VARX can be seen as a specific case of NARMAX: in addition to the VARX description, NARMAX includes moving average noise and nonlinearities in the regressed terms. 

While model selection for NARMAX (selecting the structure of nonzero model variables in its general form) is a nontrivial problem, we compare to the specific two NARMAX models proposed in \citet{chorin2015discrete}. These models were selected for the exact same test configuration as the unimodal configuration in this study (see Table \ref{tab:parameters}) and the configuation in \citet{crommelin2008subgrid}. Here we test how these NARMAX models perform in case of the trimodal configuration. We refer to \citet{chorin2015discrete} for the extensive algorithmic details of the NARMAX parameterizations and model choices.



\end{itemize} 

\section{Lorenz '96 model}\label{sec:l96}
\ifx\s\f \noindent {\color{darkgreen} Summary: definition deterministic L96, periodicity} \fi

The 2-layer Lorenz '96 (L96) model \citep{lorenz1996predictability} is frequently used to test and develop stochastic parameterizations. It was formulated as an idealized representation of atmospheric flow, but has similarities to various multiscale models. The L96 model equations from \citet{lorenz1996predictability} were reformulated in \citet{fatkullin2004computational} to explicitly express the time scale gap $\epsilon$ between the variables $x_k$ and variables $y_{j,k}$:
\begin{align}
\frac{d\, x_k}{dt} &= x_{k-1} (x_{k+1} - x_{k-2}) - x_k + F + b_k \label{eq:l96slow} \\
\frac{d\, y_{j,k}}{dt} &= \frac{1}{\epsilon} \left[ y_{j+1,k} (y_{j-1,k} - y_{j+2,k}) - y_{j,k} + h_y x_k \right] \label{eq:l96fast} \\
&\text{with } b_k := \frac{h_x}{J} \sum_{j=1}^J y_{j,k}, \label{eq:l96defresidual}
\end{align} 

\noindent where $k=1, \dots, K$ and $j=1, \dots J$ can be interpreted as spatial indices for the variables $x_k$ and $y_{j,k}$ on a circle with constant latitude. Because of the circle's periodicity, the following periodic boundary conditions hold:
\begin{equation}
x_k = x_{k+K}, \quad y_{j,k} = y_{j,k+K}, \quad y_{j+J,k} = y_{j,k+1}.
\end{equation}

\ifx\s\f  \noindent {\color{darkgreen} Summary: interpretation variables, scale-gap, and choice of parameters} \fi

\subsection{Model parameter configurations}
Generally, the variables $x_k$ and $y_{j,k}$ are referred to as the ``large-scale'' and ``small-scale'' variables. When setting $\epsilon \ll 1$ there is clear time scale separation, with $x_k$ and $y_{j,k}$ the fast and slow variables, respectively. Instead, we choose $\epsilon = 0.5$, so that no clear temporal scale gap exists, as is more realistic for oceanic and atmospheric flows (see also \citet{crommelin2008subgrid}). This choice also provides a more challenging setup for parameterizations because it does not allow for parameterization by averaging of the fast variables. We test two L96 model configurations, with different parameters, as detailed below. The parameter choices for these two configurations are also listed in Table \ref{tab:parameters} for clarity.

For the first configuration we follow the setup from \citet{crommelin2008subgrid} and \citet{chorin2015discrete}, with parameters $(\epsilon, K, J, F, h_x, h_y) = (0.5, 18, 20, 10 ,-1 ,1 )$. This configuration of the L96 model results in a reference distribution for $x_k$ that is unimodal and not too far from Gaussian (see, for example, Figure \ref{fig:distr_controlG}). We refer to this as the unimodal configuration. 

To put our suggested parameterization approach further to the test we also use a nonstandard configuration of the L96 model that we call the trimodal configuration. By increasing the forcing $F$, the number of spatial points $K$, and the feedback parameter from the fast to the slow scales $h_x$, the stationary distribution and dynamics of the L96 model become significantly more difficult to reproduce with the reduced model with stochastic parameterization (see Figure \ref{fig:distr_controlN} for the stationary distribution of $x_k$). The model parameters that define the unimodal and trimodal configurations are listed in Table \ref{tab:parameters}.

\ifx\s\f \noindent {\color{darkgreen} Summary: L96 properties, covariate $x_k$} \fi
\subsection{Stochastic model}

The L96 system is ergodic \citep{fatkullin2004computational} and invariant under spatial translations. The statistical properties of each $x_k$ are identical. As a direct consequence, the cross-correlations are the same for each spatial point $k$, this satisfies the assumptions of the simple covariance form discussed in Section \ref{sec:covariance}. 

\begin{figure}[htb]
\centering
\captionsetup[subfloat]{justification=centering}
\subfloat[\label{fig:scatterplot}]{\includegraphics[width=0.49\linewidth]{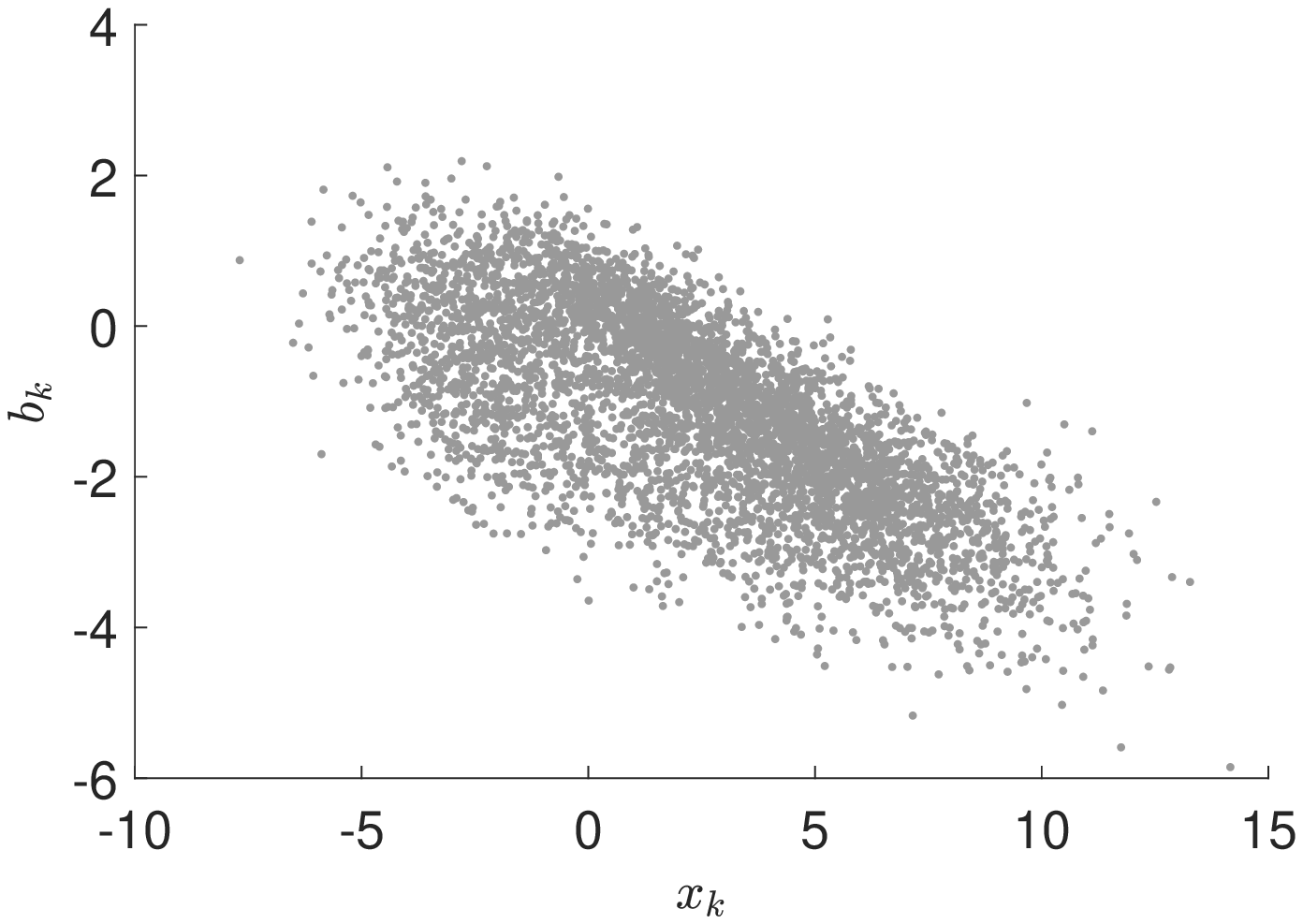}}
\hfill
\subfloat[\label{fig:scatterplot2}]{\includegraphics[width=0.49\linewidth]{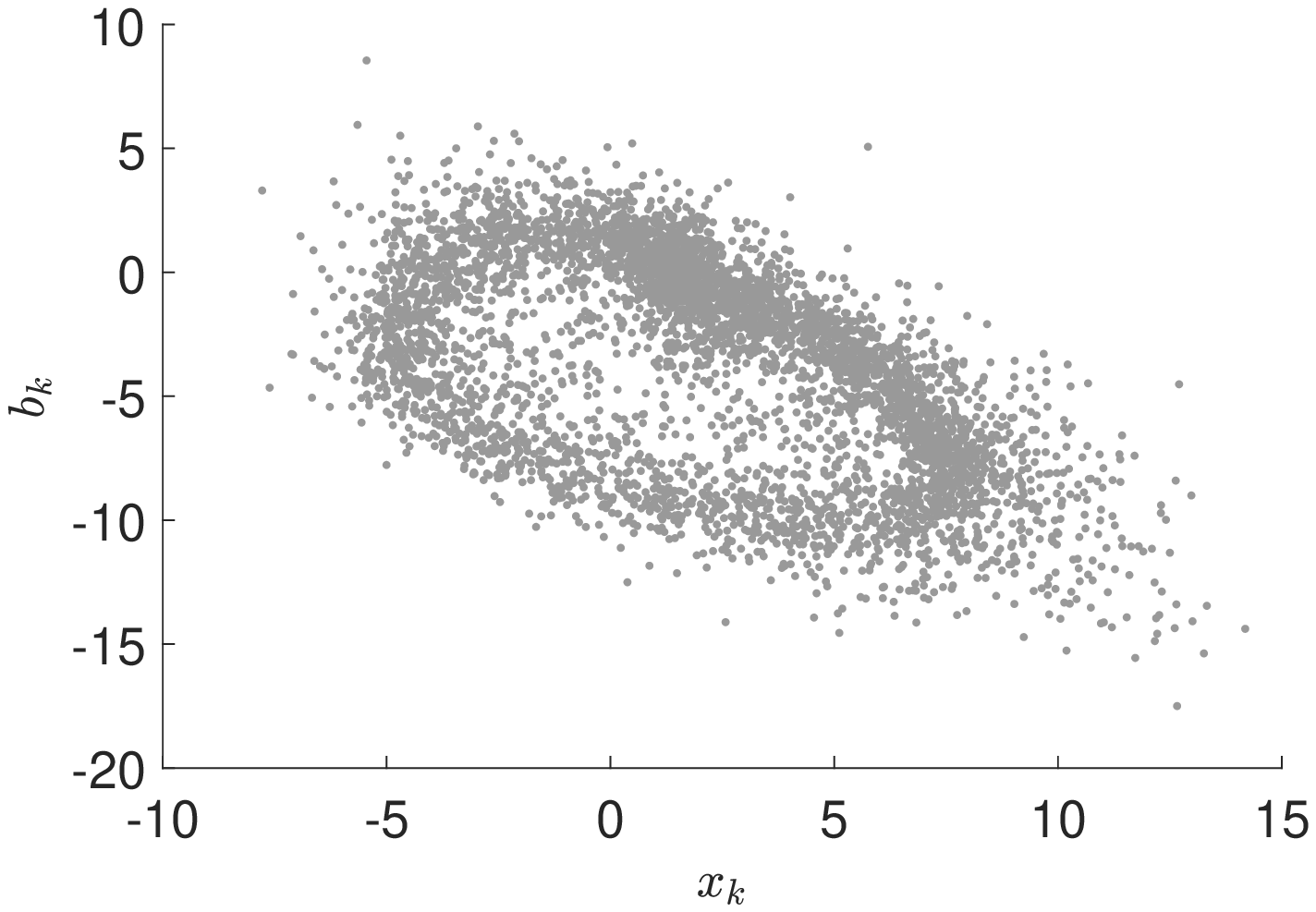}}
\caption{Strong correlation between $x_k$ and $b_k$ shown by scatter plots for the reference deterministic L96 \cref{eq:l96slow,eq:l96fast,eq:l96defresidual}: (a) unimodal, and (b) trimodal configurations.}
\label{fig:l961prediction}
\end{figure}

\begin{figure}[htb]
\centering
\captionsetup[subfloat]{justification=centering}
\subfloat[\label{fig:gaussdistr}]{\includegraphics[width=0.49\linewidth]{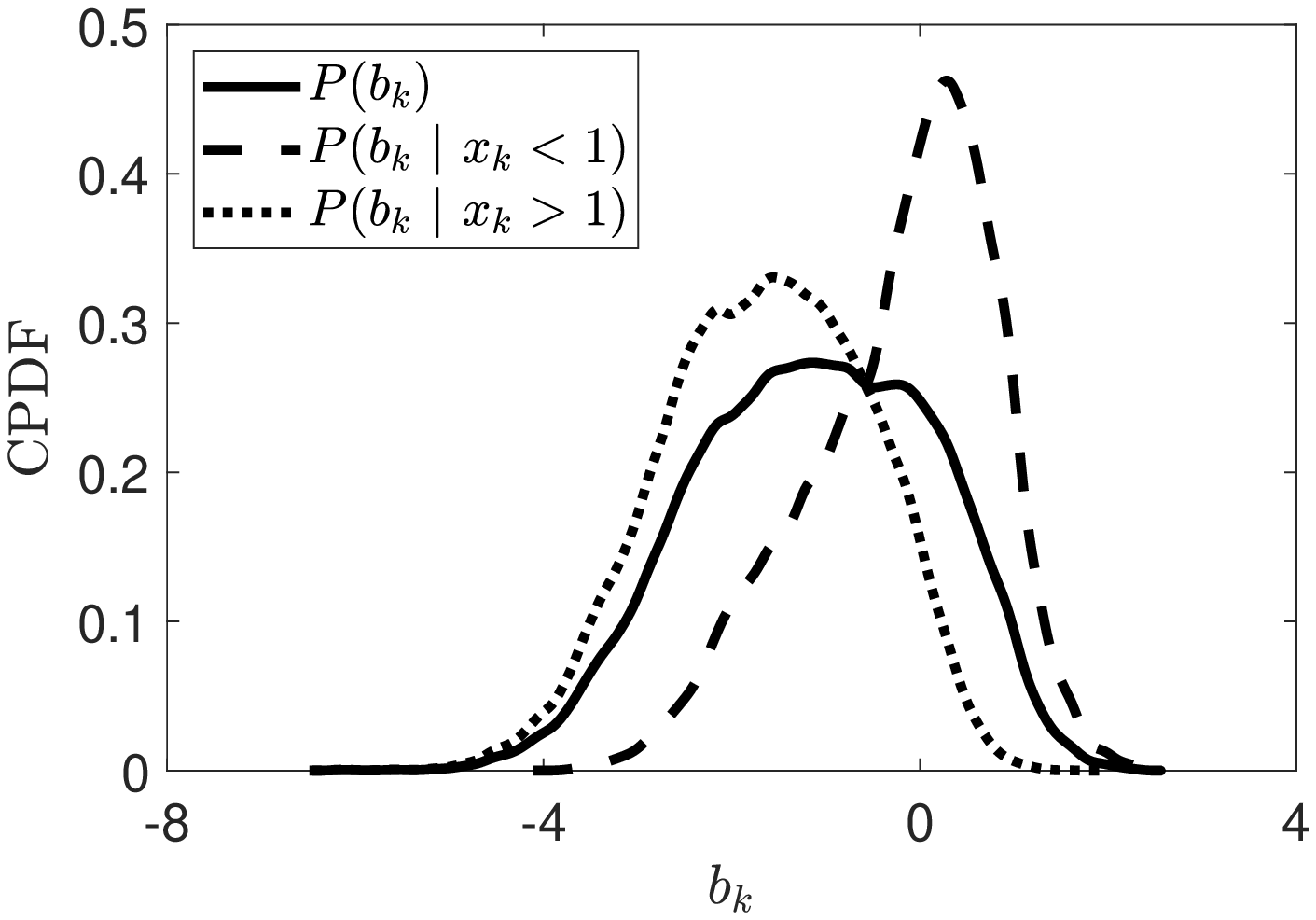}}
\hfill
\subfloat[\label{fig:gaussdistr2}]{\includegraphics[width=0.49\linewidth]{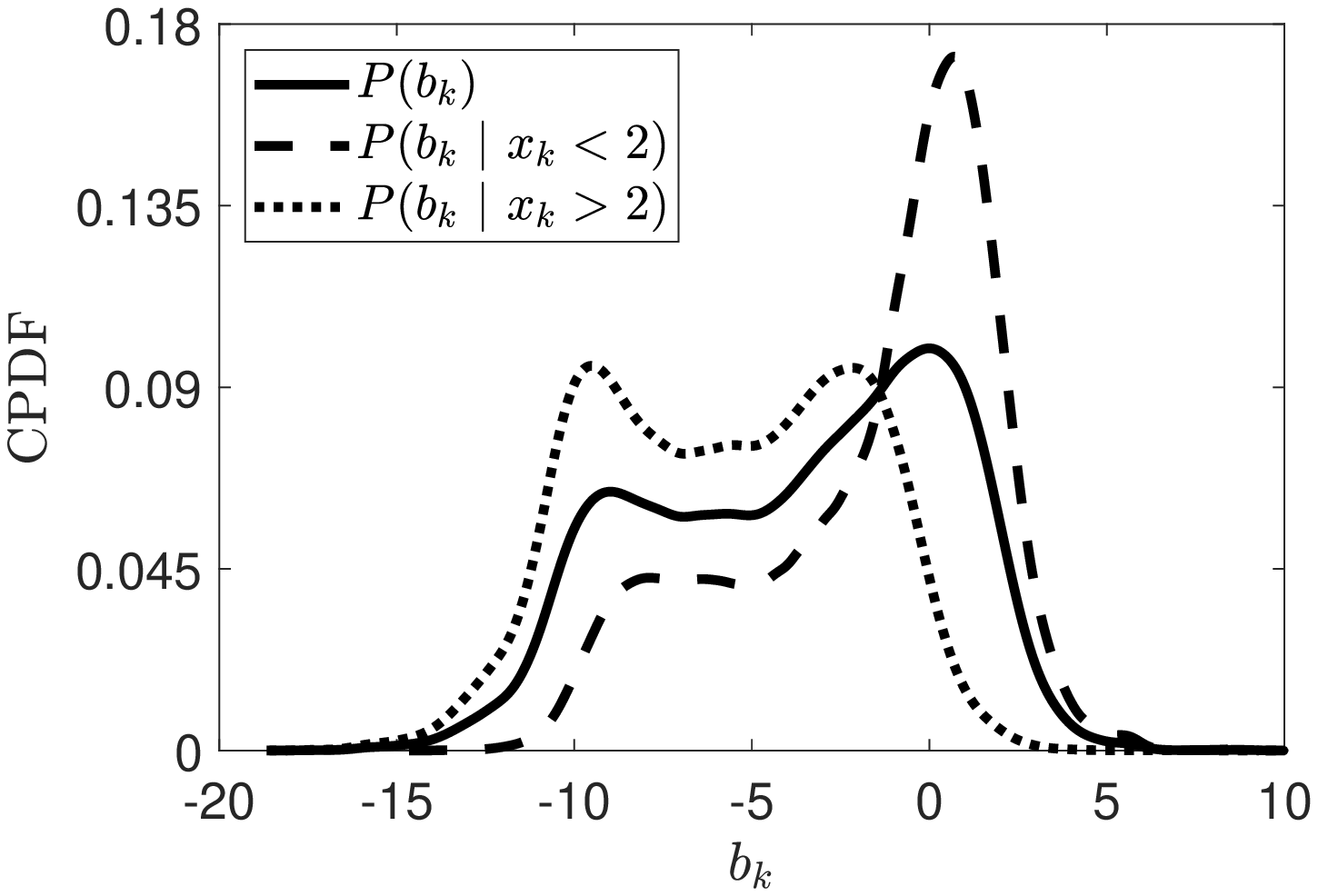}}
\caption{CPDFs for $b_k$ dependent on $x_k$ for the reference deterministic reference L96  \cref{eq:l96slow,eq:l96fast,eq:l96defresidual}: (a) unimodal, and (b) trimodal configurations}
\label{fig:l962prediction}
\end{figure}

The sample data $(\vecbigX, \vecbigB)$ of the two different deterministic L96 reference simulations reveal strong correlations between $x_k$ and $b_k$, as illustrated for one such $k$ in Figures \ref{fig:scatterplot} and \ref{fig:scatterplot2}. Because the statistical properties of $x_k$ are identical for all $k$, this figure is equivalent to that for any other $k$. Additionally, Figures \ref{fig:gaussdistr} and \ref{fig:gaussdistr2} show that the conditional probability density function (CPDF) $P(b_k | x_k)$ can change significantly for different ranges of $x_k$-values. Therefore, because $x_k$ is a resolved variable in both the deterministic and stochastic L96 models, $x_k$ is a valuable predictor variable for (the distribution of) $b_k$.
Clearly, the presence of a good predictor is not guaranteed, and identifying one may be nontrivial for some problems or application fields.  For ocean modeling, this was explored in \citep{berloff2005dynamically,porta2014toward,zanna2017scale}.

Because the (conditional) distributions of $b_k$ in Figure \ref{fig:gaussdistr} resemble normal distributions, we assume the underlying distribution of $b_k$ to be Gaussian. This starting point will test the robustness of our parameterization, because the trimodal configuration exhibits distinct multi-modality in $b_k$. While Figure \ref{fig:scatterplot2} does suggest a clear correlation between $x_k$ and $b_k$ for the trimodal configuration, there is a distinct circular pattern present in the scatter plot. The marginal distributions of $b_k$ and $x_k$ are also clearly trimodal, see Figures \ref{fig:gaussdistr2} and \ref{fig:distr_controlN}, respectively. 

\ifx\s\f \noindent {\color{darkgreen} Summary: covariate restriction, goal of each, spatio-temporal displacements} \fi

\ifx\s\f \noindent {\color{darkgreen} Summary: Definition stochastic model restricted to two possible covariates} \fi

The stochastic L96 model is obtained by forcing a reduced version (without $b_k$) of (\ref{eq:l96slow}) with the VARX $\widetilde{\vecB}$ (\ref{eq:completestochasticmodel}) that aims to approximate $b_k$ in each $k$. Following the model reduction approach as described in Section \ref{sec:varx}, the stochastic L96 model then becomes:
\begin{align}
\widetilde{x}^{n+1}_k &= T\left( \Delta t, \, \widetilde{x}^n_{k-1} (\widetilde{x}^n_{k+1} - \widetilde{x}^n_{k-2}) - \widetilde{x}^n_k + F + \widetilde{b}^n_k \right) \label{eq:stochl96x} \\
\widetilde{\vecB}^{n} &= \vecA_0 + A_1 \widetilde{\vecB}^{n-1} + \dots + A_p \widetilde{\vecB}^{n-p} + D \widetilde{\vecX}^n + \Sigma \vecXi^n, \label{eq:stochl96r}
\end{align}

where $T$ is the numerical integration scheme of choice (see Section \ref{sec:practical}).

\subsubsection{Order selection - lag time choice}\label{sec:results_lagchoice}

The order selection of the VARX$(p)$ $\widetilde{\vecB}$ in (\ref{eq:varx}) determines the temporal decorrelation of the VARX. By choosing the order $p$ appropriately, one can match the stochastic model with the decorrelation timescale of the reference model. In this section, let us consider the choice for nonzero lag times $p$, i.e. the order selection for the VARX$(p)$. The reference $x_k$ has strongly oscillating, slowly decaying correlations (see also Figures \ref{fig:ACF_controlG} and \ref{fig:ACF_controlN} later on). To model this behavior perfectly one would need a high order VARX process. However, estimating a $\text{VARX}(1)$ that is numerically stable is rather straightforward whereas estimating a stable $\text{VAR}(p)$ of arbitrary order $p$ is difficult due to the constraint (\ref{eq:varxstationarity}). This constraint can only be verified a posteriori; we are not aware of estimation methods that guarantee  (\ref{eq:varxstationarity}) is satisfied a priori.
Therefore, we opt for a single nonzero $A_p$ and, in doing so, interpret the process as $\text{VARX}(1)$ over an interval $p$ times larger than the sampling interval. This leaves us with the choice for the nonzero lag contribution $p$.

\begin{figure}[htb]
\centering
\captionsetup[subfloat]{justification=centering}
\subfloat[\label{fig:PACF_refG}]{\includegraphics[width=0.5\linewidth]{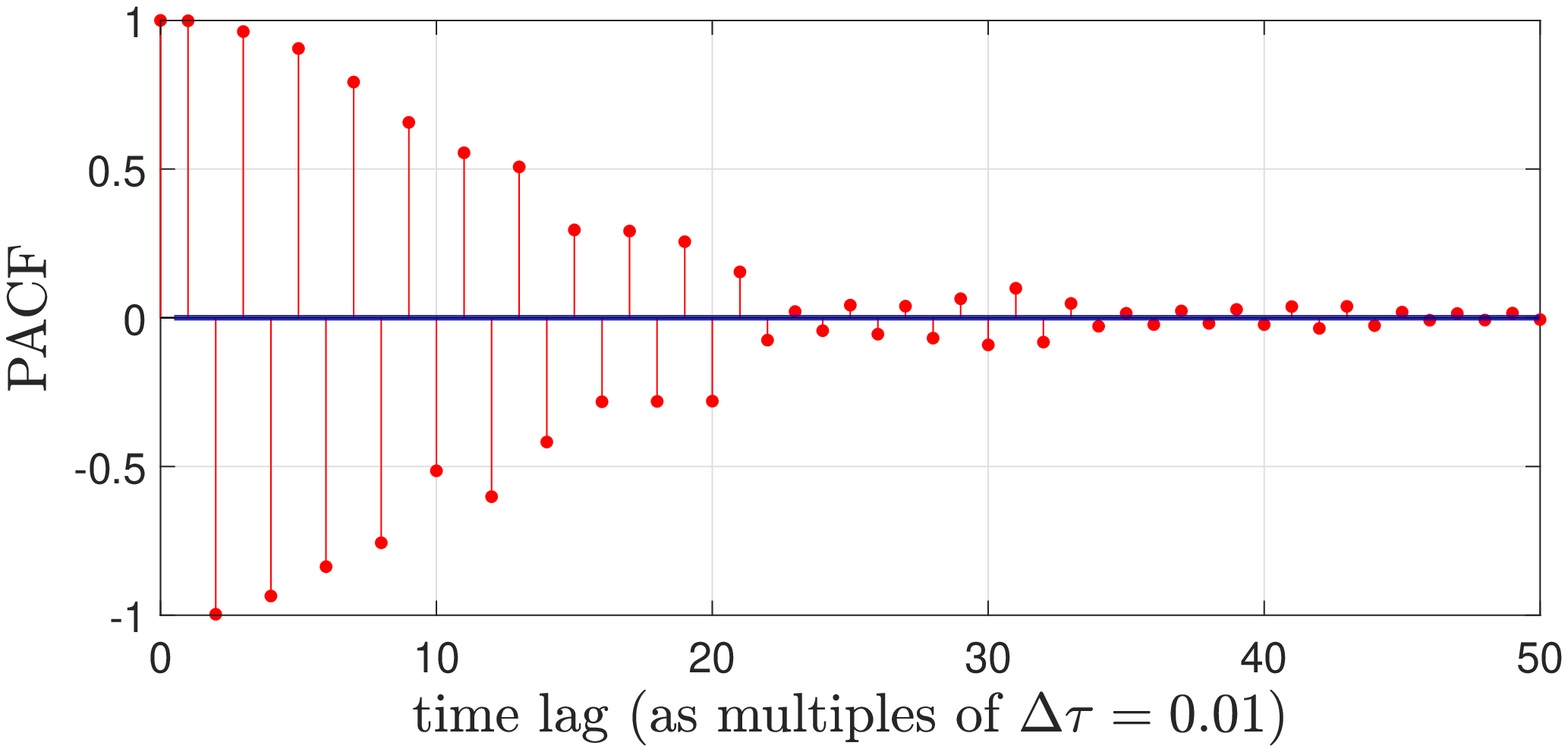}}
\hfill
\subfloat[\label{fig:PACF_refN}]{\includegraphics[width=0.5\linewidth]{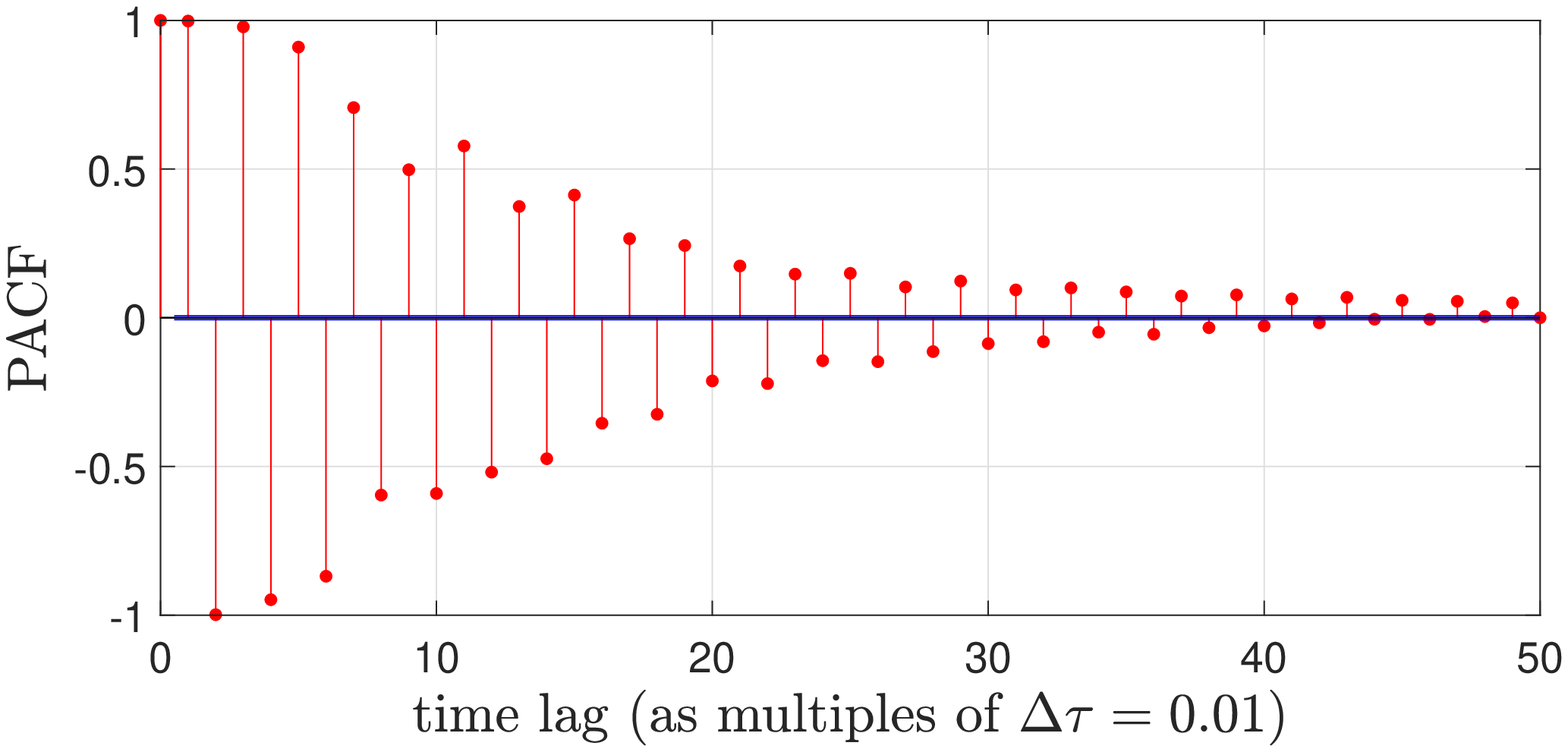}}
\caption{The partial autocorrelation functions (PACFs) for $b_k$ for \textbf{(a)} the unimodal and \textbf{(b)} the trimodal ``resolving'' deterministic reference L96 simulations (see Table \ref{tab:parameters}).}
\label{fig:pacfs}
\end{figure}

Because we choose the coefficient matrices $A_p$ and $D$ diagonal (see Section \ref{sec:modelingchoices}), connections to univariate autoregressive (AR) models are easily made, particularly in the case of $\Sigma_D$ (as it is also diagonal). In univariate time series analyses, the order $p$ of $\text{AR}(p)$ models is often determined with the Box--Jenkins method \citep{box2015time}. Both the ACF and partial autocorrelation function (PACFs) are used to determine the order of an AR model for approximating timeseries data. Here the partial autocorrelation of lag $n'$ is the autocorrelation between $b_k^n$ and $b_k^{n+n'}$ that is not accounted for by lags $1$ through $n'-1$, i.e. the partial autocorrelation is a conditional correlation that controls for all shorter lags:

\begin{equation}
\text{PACF}(b_k, l) = \frac{\Cov(b_k^n, b_k^{n-l} \, | \, b_k^{n-1}, \dots, b_k^{n-l+1})}{\sqrt{\Var(b_k^n \, | \, b_k^{n-1}, \dots, b_k^{n-l+1})\Var(b_k^{n-l} \, | \, b_k^{n-1}, \dots, b_k^{n-l+1})}}
\end{equation}

Although the \emph{sample} (P)ACFs do not necessarily describe the same autoregressive properties as the \emph{analytical} (P)ACFs, they typically are used in model selection. It is common practice that when the ACF shows sinusoidal behavior with no clear decay to 0 (as is the case for both the unimodal and trimodal deterministic reference simulations, see \Crefrange{fig:ACF_controlG}{fig:ACF_controlN}), the order $p$ of the modeling $\text{AR}(p)$ is chosen at the last spike in the PACF after which the PACF no longer returns to this same level \citep{hyndman2014forecasting}. The PACFs of the ``resolving'' unimodal and trimodal $b_k$ are plotted in Figure \ref{fig:pacfs}. The PACF in Figure \ref{fig:PACF_refG} shows such a spike around $\Delta \tau = 0.14$, hence we pick $p=14$ for the unimodal case. The PACF in the trimodal case, Figure \ref{fig:PACF_refN}, decays very gradually, showing no clear steep monotonous decline. This necessitates, according the Box--Jenkins method, the choice for a relatively long time scale of approximately $\Delta \tau = 0.3$, i.e. $p=30$.

\section{Practical implementation of stochastic parameterization}\label{sec:practical}
\ifx\s\f \noindent {\color{darkgreen} Summary: interpretation as CPDFs, graphic representation} \fi

The stochastic L96 model \Crefrange{eq:stochl96x}{eq:stochl96r} is forced by the VARX model $\widetilde{\vecB}$ in (\ref{eq:stochl96r}) dependent on exogenous ($\widetilde{\vecX}$) and endogenous (past $\widetilde{\vecB}$) states. The reference data $(\vecbigX, \vecbigB)$ is used to approximate appropriate CPDFs from which $\widetilde{\vecB}$ is sampled, determined by the selection of endogenous and exogenous variables. For example, if we select $D$ and $A_1$ to be scalar matrices and $A_{i} = 0$ for $i > 1$, then by (\ref{eq:stochl96r}) $\widetilde{\vecB}^n$ is sampled from the Gaussian approximation of the CPDF $P(\vecB^{n} \ | \ \vecX^n = \widetilde{\vecX}^n, \vecB^{n-1} = \widetilde{\vecB}^{n-1})$, or $P(\vecB^{n} \ | \ \widetilde{\vecX}^n, \widetilde{\vecB}^{n-1})$ for short. 

\ifx\s\f \noindent {\color{darkgreen} Summary: Algorithmic details, pseudo-code} \fi

We solve the L96 system directly using a classical second-order Runge--Kutta integration scheme \citep{fatkullin2004computational}. The regression coefficients and covariance matrix of the VARX (\ref{eq:stochl96r}) are precomputed with least squares. The VARX is integrated over time together with the L96 system and applied to the timestepping of $\widetilde{x}_k$ (\ref{eq:stochl96x}). The pseudo-code for our stochastic L96 model is shown in Figure \ref{fig:algorithm}. All parameters used in our deterministic and stochastic simulations are listed in Table \ref{tab:parameters}. 

\begin{figure}[ht]
\begin{algo}{
\textit{input}: \>\> $\vecbigX$ \>\> : concatenated vector of sample data for $x_k^n$, size $NK \times 1$.\\
\>\> $\vecbigB$ \>\> : concatenated vector of sample data for $b_k^n$, size $NK \times 1$. \\
\\
/* Precompute the VARX coefficients $\vecA_0, A_i$, and $D$ */ \\
$\vecbigZ = (\vecOnes, \vecbigX(\text{lag } 0, \vecbigB(\text{lag } 1), \dots, \vecbigB(\text{lag } p))$, the regressor variable matrix, size $K(N - p) \times (p+2)$ \\
$(\vecA_{0}, D, A_{1}, \dots, A_{p}) = (\vecbigZ^T \vecbigZ)^{-1}\vecbigZ^T \vecbigB$ \\
\\
/* Either compute $\Sigma_D$ */ \\
$\vecbigR = \text{VEC}(\vecB^{n} - \vecA_0 - A_1 \vecB^{n-1} - \dots - A_p \vecB^{n-p} - D \vecX^n)$, where $\vecbigR$ the concatenated residual vector ($K(N - p) \times 1$) \\
$\Sigma_D = \sqrt{\Var(\vecbigR)} \, I_{K \times K}$ \\
\\
/* Or compute $\Sigma_L$ */ \\
/* Let [$\vecbigR$] denote the reshaped $((N-p) \times K)$-matrix corresponding to $\vecbigR$ */ \\
$\Sigma_L = \text{Chol}(\Cov([\vecbigR]))$, where $\Cov([\vecbigR])$ a $(K \times K)$-matrix \\
\\
$(\widetilde{\vecX}^{-p+1, \dots, 0},\widetilde{\vecB}^{-p+1, \dots, 0}) = (\vecX^{-p+1, \dots, 0},\vecB^{-p+1, \dots, 0})$ \\
\textbf{for} $i:=0$ \textrm{to} $N-1$ \textbf{do} \\
\>/* Sample $\widetilde{\vecB}$ */ \\
\> $\widetilde{\vecB}^{n} = \vecA_0 + A_1 \widetilde{\vecB}^{n-1} + \dots + A_p \widetilde{\vecB}^{n-p} + D \widetilde{\vecX}^n + \Sigma \vecXi^n$, where $\vecXi^n \sim \mathcal{N}(0, \vecbigI)$  \\ 
\\
\>/* Update $\widetilde{x}$ with second order Runge--Kutta, \\
\>\> notation: let $\vecX_{+d}$ denote the module rotation of points $x_k$, e.g. $\vecX_{+1} := (x_{2}, \dots, x_{K},x_1)$) */ \\
\> $\widetilde{\vecX}' = \widetilde{\vecX}^n + \frac{\Delta \tau}{2} (\widetilde{\vecX}^n_{-1} (\widetilde{\vecX}^n_{+1} - \widetilde{\vecX}^n_{-2}) - \widetilde{\vecX}^n + F + \widetilde{\vecB}^{n}) $ 
\\
\> $\widetilde{\vecX}^{n+1} = \widetilde{\vecX}^n + \Delta \tau (\widetilde{\vecX}'_{-1} (\widetilde{\vecX}'_{+1} - \widetilde{\vecX}'_{-2}) - \widetilde{\vecX}' + F + \widetilde{\vecB}^{n}) $ \\
\textbf{endfor} \\
}\end{algo}
\caption{Algorithm for the time integration of the stochastic L96 model. Notation: $\vecbigB(\text{lag } i)$ is the vector of concatenated sample data $\vecB^j$ for all $p + 1 - i \le j \le N-i$, i.e. the sample data of $b_k$ at $i$ time steps in the past (where the first $p-i$ vectors $\vecB$ of $\vecbigB$ are skipped to make each $\vecbigB(\text{lag } i)$ equal in length). Similar notation is used for $\vecbigX(\text{lag } 0)$ to denote the vector of concatenated sample data $\vecX^j$ for all $p + 1 \le j \le N$.}
\label{fig:algorithm}
\end{figure}

\begin{table}[htb]
\caption{Parameter settings for all deterministic and stochastic L96 models}
\centering
\begin{tabularx}{0.95\textwidth}{l X r r}
\noalign{\vskip 1pt}
Parameter & Explanation & unimodal L96 & trimodal L96 \\
\hline
\noalign{\vskip 2pt}
$\epsilon$ & scale separation & $0.5$ & $0.5$ \\
$K$ & $\#$ discretized large-scale spatial points / $\#$ resolved $x$-variables & $18$ & $32$ \\
$J$ & $\#$ discretized small-scale spatial points / $\#$ unresolved $y$-variables per $x$-variable & $20$ & $16$ \\
$F$ & forcing on the $x$ variables & $10$ & $18$ \\
$h_x$ & scale coupling constant  & $-1$ & $-3.2$ \\
$h_y$ & scale coupling constant  & $1$ & $1$ \\
$\Delta t$ & integration time step full L96 model & $10^{-3}$ & $10^{-3}$ \\
$\Delta \tau$ & integration time step reduced L96 model & $10^{-2}$ & $10^{-2}$ \\
$\delta t$ & sampling interval & $10^{-2}$ & $10^{-2}$ \\
$N$ & Number of integration time steps in a simulation & $10^6 + p$ & $10^6 + p$ \\
\hline
\end{tabularx}
\label{tab:parameters}
\end{table}

We choose the sampling interval $\delta t$ of the reference data $(\vecbigX, \vecbigB)$ to be larger than the integration time step $\Delta t$ of the full L96 model. We pick $\delta t = 10 \Delta t$, same as in  \citet{crommelin2008subgrid}. This reduces the amount of data that must be handled, at the price of loosing some high-frequency (short timescale) information. However, as  we set the integration time step of the reduced model equal to the sampling interval, i.e. $\Delta \tau = \delta t$ (see Table \ref{tab:parameters}), these are very short timescales that are not resolved by the reduced model anyway.

\section{Numerical results}\label{sec:results}
In this section we compare the statistical behavior of the reduced model with VARX stochastic parameterization \Crefrange{eq:stochl96x}{eq:stochl96r} with the reference model \Crefrange{eq:l96slow}{eq:l96fast}. Recall from Section \ref{sec:l96} that the statistics of $x_k$ are identical for all $k$. Therefore, the statistical properties determined for $x_k$ describe the full statistics of $\vecX$, i.e. equal for all $k$. Let $\mu := \E(x_k)$ and $\sigma := \sqrt{\E((x_k)^2) - \E(x_k)^2}$ denote the mean and standard deviation of $x_k$, respectively, where $\E$ denotes the average over time. We assess the following statistical criteria of the variable of interest $x$ of the models:

\begin{itemize}
\item The probability density function (PDF) of $x_k$.
\item The autocorrelation coefficient (ACF) of $x_k$: $\ACF(\tau):= \sigma^{-2}\E\left[ (x_k^t - \mu)(x_k^{t+\tau} - \mu) \right]$.
\item The cross-correlation coefficient (CCF) between $x_k$ and $x_{k+1}$: $\CCF := \sigma^{-2}\E\left[ (x_k^t - \mu)(x_{k+1}^t - \mu) \right]$.
\item The mean wave amplitude $\E (\abs{u_m})$ for each wave number $ 0 \le m \le K/2$, where a time series for the wavenumber vector $\vecU := \widehat{\vecX}$ is obtained by calculating the Fourier transform of $\vecX$ at every time step. 
\item The wave variance $\E ( \abs{ u_m - \E(u_m)}^2 )$, 
\end{itemize}

For the VARX model in (\ref{eq:stochl96r}) we use several different settings, each described explicitly in the following subsections. We show a representative selection of the results. All VARX models have a single nonzero $A_p$ for chosen lag time $p$ to circumvent VARX estimation stability issues (as discussed earlier). Furthermore, all VARX models until Section \ref{sec:results_densecovariance} have a diagonal covariance structure, i.e. a diagonal matrix $\Sigma_D$. To reduce the number of parameters we choose the coefficient matrices $A_p$ and $D$ to be diagonal in all cases.

First, in Section \ref{sec:results_control}, we illustrate for completeness the contrast between the deterministic reference L96 simulations and simulations with the simplest possible stochastic model, denoted (\ref{eq:b_bench}), in which the $\tilde b_k$ are independent white noise terms. 
Then, in Section \ref{sec:results_onevariable} we discuss results for stochastic model simulations with a single regressor: either only endogenous (\ref{eq:b_singleb}) or exogenous (\ref{eq:b_singlex}), respectively. Next, we demonstrate that with both regressors (\ref{eq:b_double_Gauss_diag}) (called ``double regressor'') the unimodal L96 reference statistics are reproduced very accurately in Section \ref{sec:results_diagonalcovariance}. However, we also show that (\ref{eq:b_double_tri_diag}) does not perform well in case of  the trimodal L96 model configuration. In Section \ref{sec:results_densecovariance}, we therefore compare the (\ref{eq:b_double_tri_diag}) and (\ref{eq:b_double_tri_dense}) simulations, and show that by allowing for a non-diagonal structure of the covariance we also succeed at reproducing the statistics of the trimodal L96 model accurately. In section \ref{sec:results_narmax}, we compare the results for our VARX models to those for the NARMAX models proposed in \citet{chorin2015discrete}. While the NARMAX models perform very accurately for the unimodel L96 test case, we show that the NARMAX models do not perform well for the trimodal L96 configuration. Neither the trimodal distribution of $x_k$ nor the wave statistics were reproduced accurately. All our simulations here use the parameter configurations as listed in Table \ref{tab:parameters}.

\subsection{White noise parameterizations}\label{sec:results_control}

We start with an `unconditioned' stochastic parameterization, that is to say a parameterization in which $\tilde b_k^n$ is not conditioned on its own past state(s) nor on $\widetilde{\vecX}^n$: 
\begin{equation}\label{eq:b_bench}\tag{\emph{WN}}
\widetilde{\vecB}^{n} = \sigma I \, \vecXi^n,
\end{equation}

where $\vecXi^n$ is a vector of independent normally distributed random variables. Note that this model is equivalent to choosing $A_0, \dots, A_p, D = 0$ and $\Sigma = \sigma I$ in (\ref{eq:stochl96r}). In this simplest possible stochastic parameterization, the time evolution of each $\tilde b_k$ is a series of Brownian motion increments, i.e. a white noise process, therefore we denote it (\ref{eq:b_bench}). We include it here to verify the added value of conditioning in the more complicated parameterizations discussed later on.

\begin{figure}[htb]
\centering
\captionsetup[subfloat]{justification=centering}
\subfloat[\label{fig:distr_controlG}]{\includegraphics[width=0.4\linewidth]{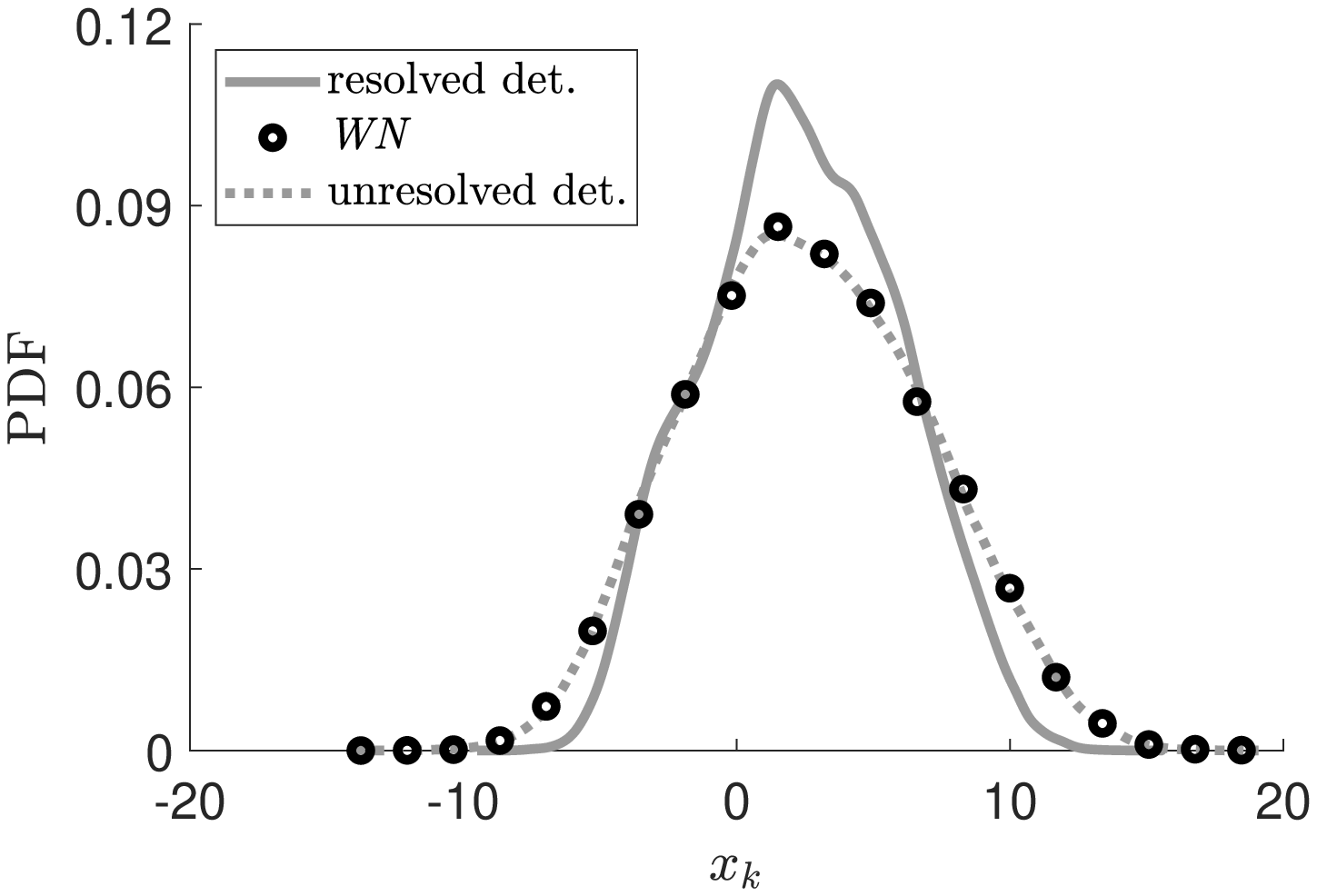}}
\hfill
\subfloat[\label{fig:distr_controlN}]{\includegraphics[width=0.4\linewidth]{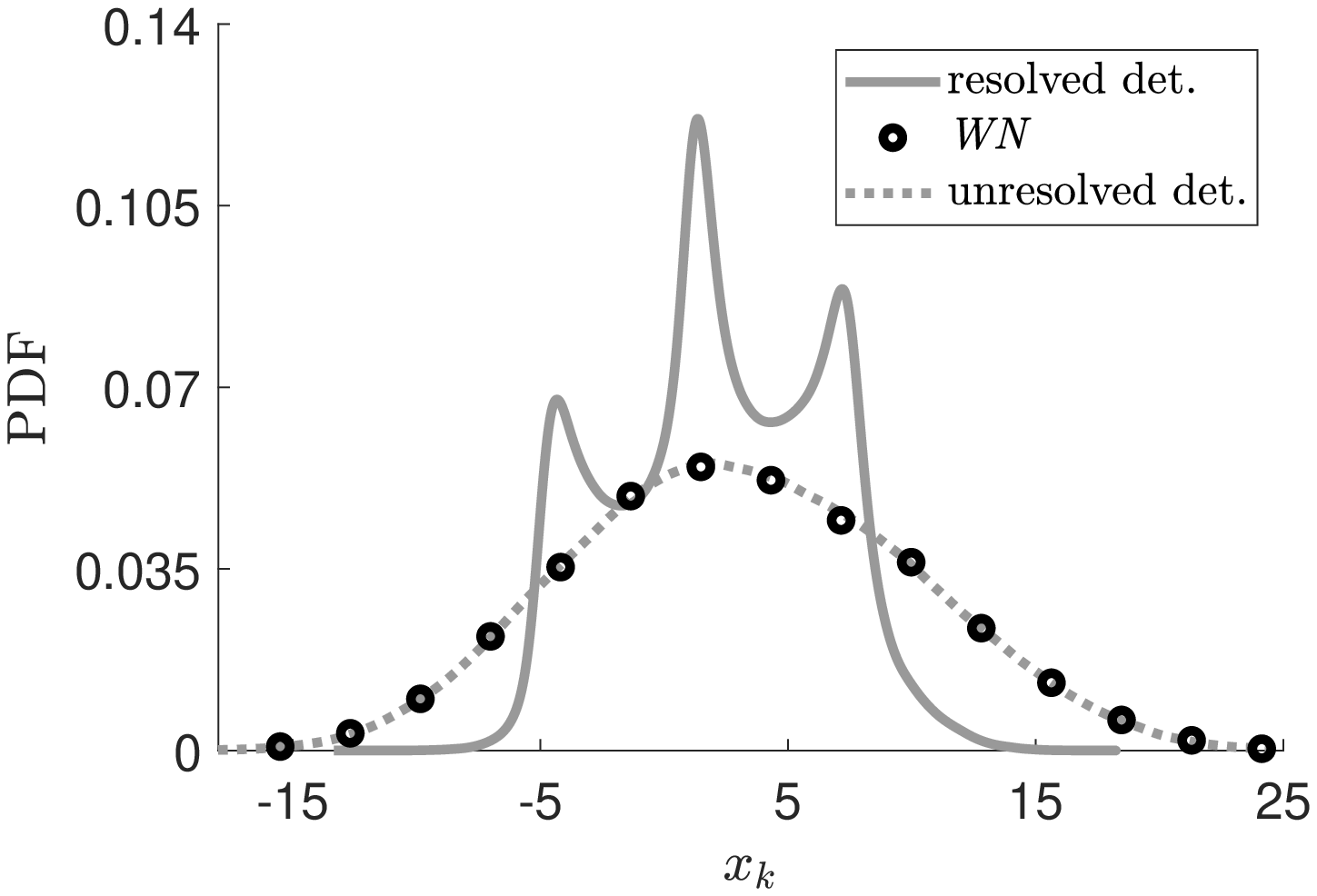}}
\caption{Comparison between PDFs of interest for \textbf{(a)} the unimodal L96 and \textbf{(b)} the trimodal L96 configuration (see Table \ref{tab:parameters}) resulting from unconditioned stochastic simulations using (\ref{eq:b_bench}), as well as ``resolving'' and ``unresolving'' ($b_k = 0$) deterministic reference simulations.}
\label{fig:distr_control}
\end{figure}

In Figures \ref{fig:distr_controlG} and \ref{fig:distr_controlN} we plot the distributions of $x_k$ for the two L96 model configurations. First, the ``resolved'' reference simulation obtained with the full L96 model \Crefrange{eq:l96slow}{eq:l96fast}, second, the ``unresolved'' reference simulation, i.e. (\ref{eq:l96slow}) with $b_k = 0$. The former is what we aim to reproduce with our reduced models. The latter of the two we include as a worst-case reference, the result of a reduced model with no parameterization at all to account for the missing unresolved scales.

The overall shape of the distribution of $x_k$ in the unimodal L96 model is reproduced (although the details are not well captured), both with the (\ref{eq:b_bench}) parameterization and without any parameterization (the ``unresolved'' case), see Figure \ref{fig:distr_controlG}. This result is in line with, e.g, \citet{crommelin2008subgrid} and \citet{chorin2015discrete}, where it was also found that the distribution of $x_k$ is not very difficult to reproduce with a reduced model, in case of the L96 unimodal configuration. However, Figure \ref{fig:distr_controlN} shows that the distribution of $x_k$ for the trimodal L96 configuration is not reproduced at all by (\ref{eq:b_bench}), nor by the ``unresolved'' case.  

\begin{figure}[p]
\centering
\captionsetup[subfloat]{justification=centering}
\subfloat[\label{fig:ACF_controlG}]{\includegraphics[width=0.49\linewidth]{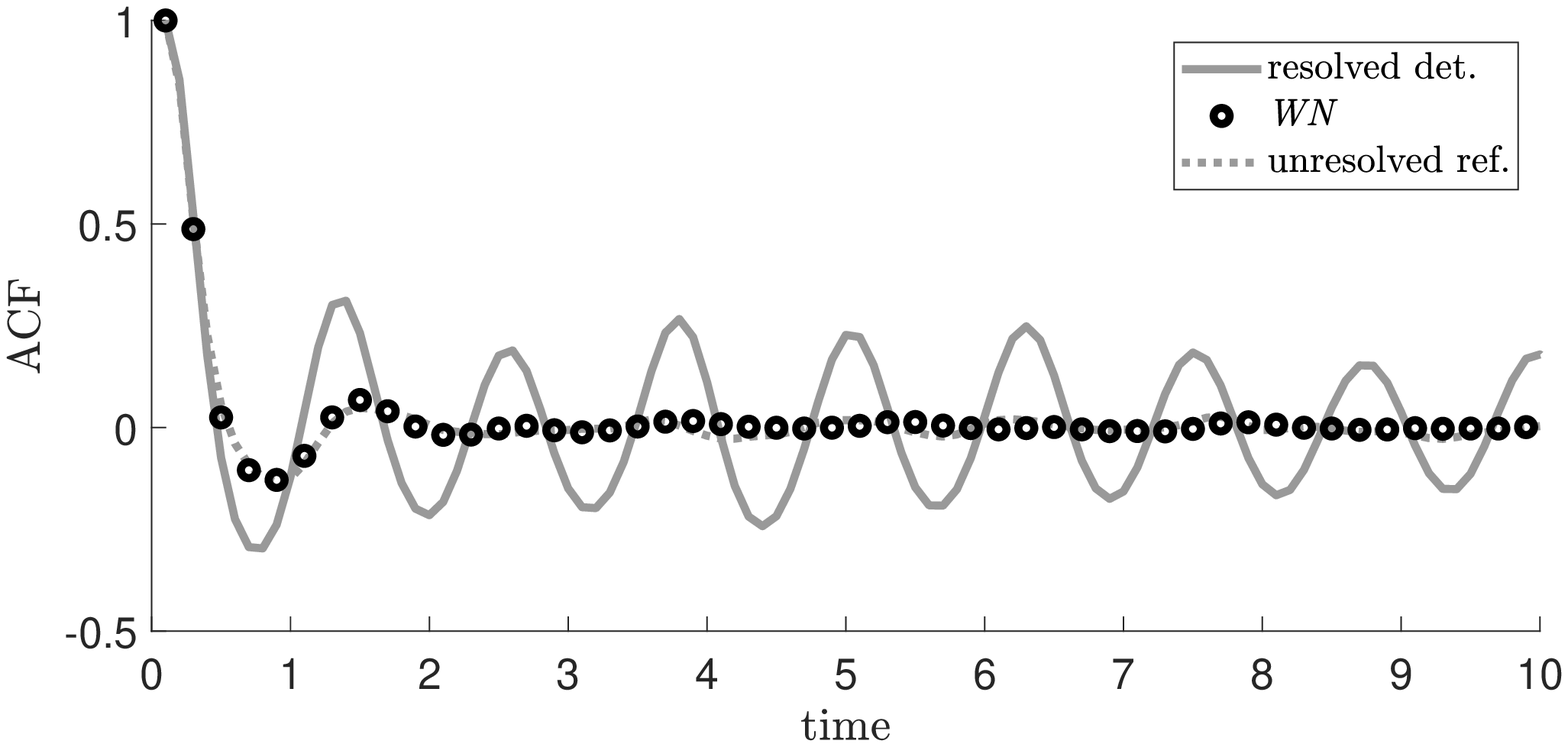}}
\hfill
\subfloat[\label{fig:ACF_controlN}]{\includegraphics[width=0.49\linewidth]{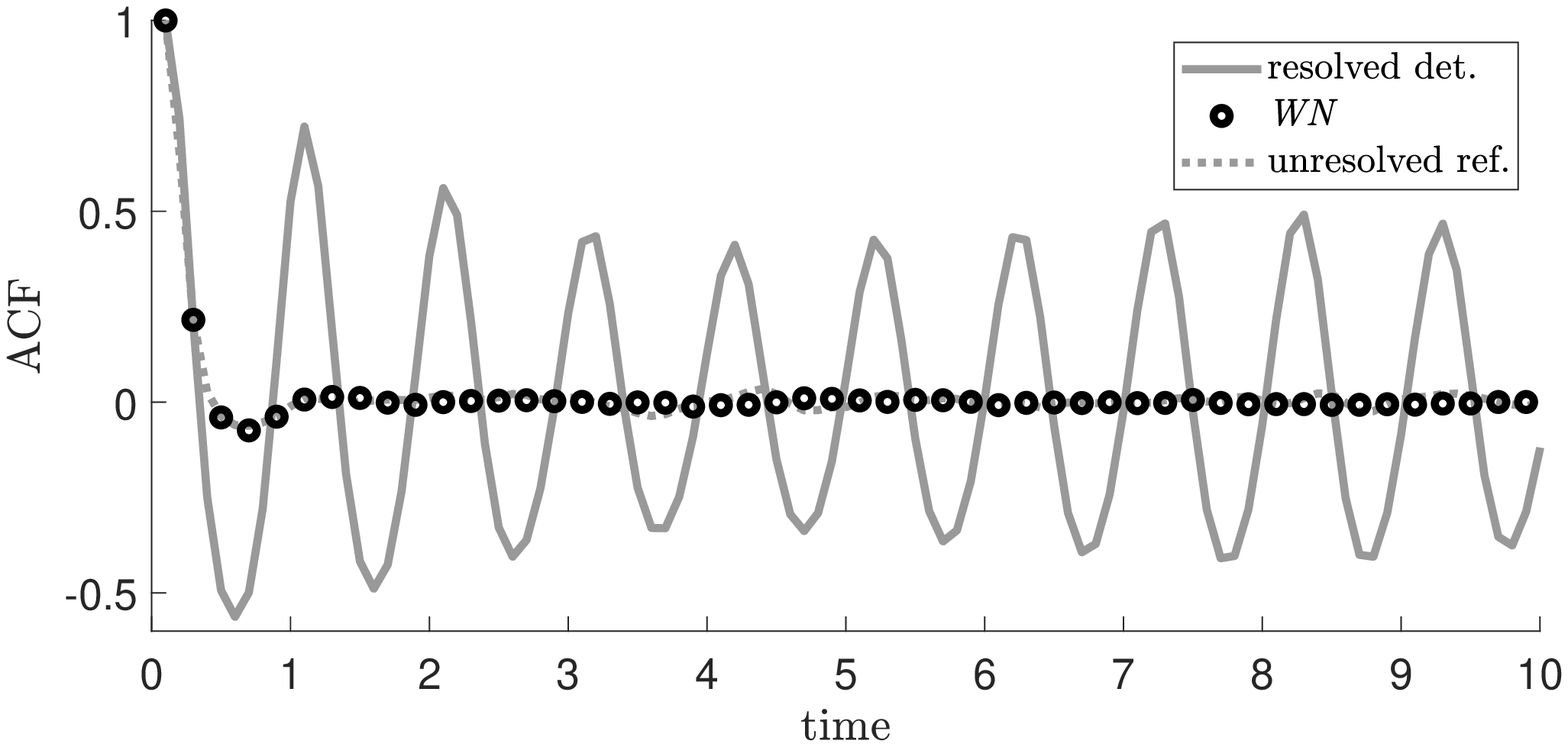}}
\hfill
\subfloat[\label{fig:CCF_controlG}]{\includegraphics[width=0.49\linewidth]{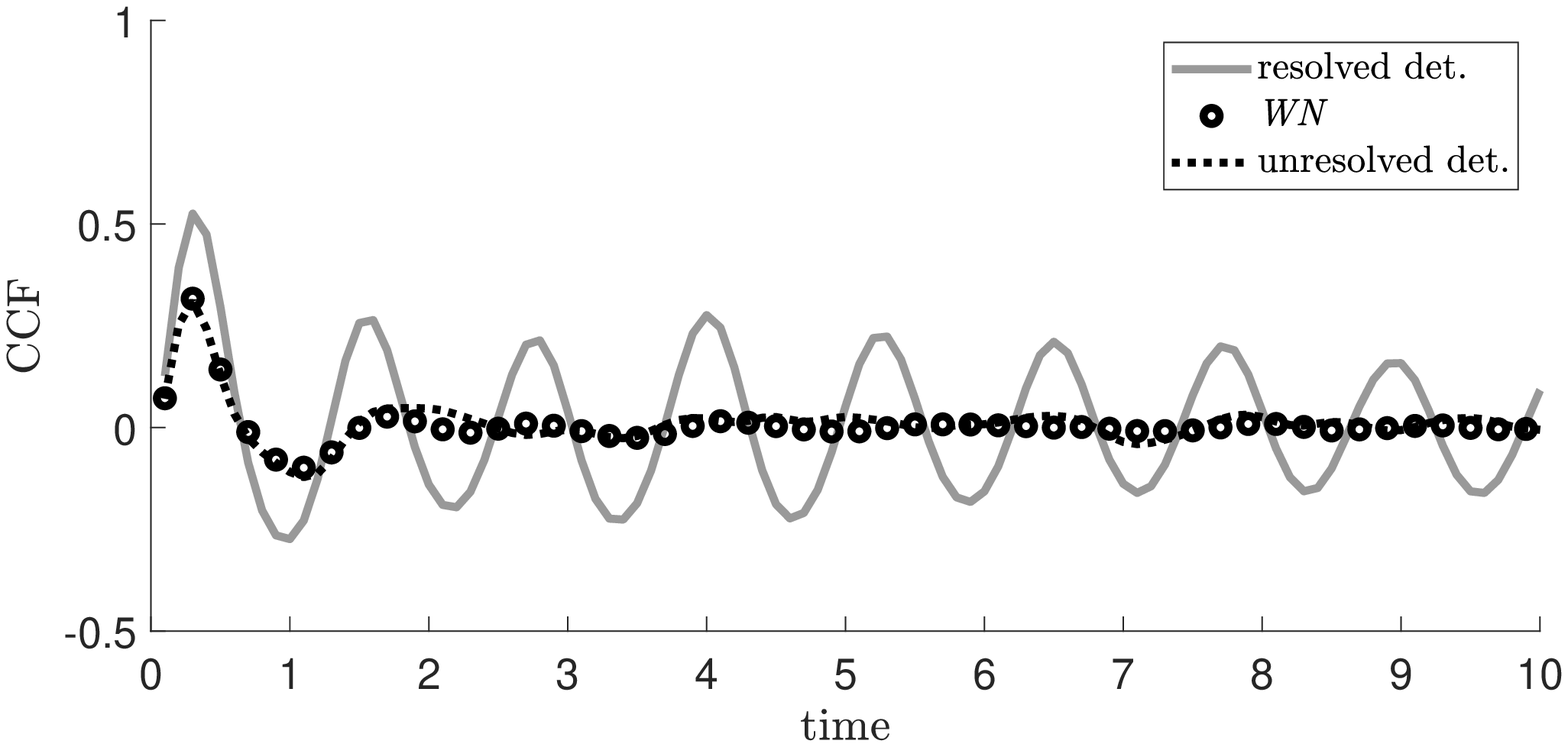}}
\hfill
\subfloat[\label{fig:CCF_controlN}]{\includegraphics[width=0.49\linewidth]{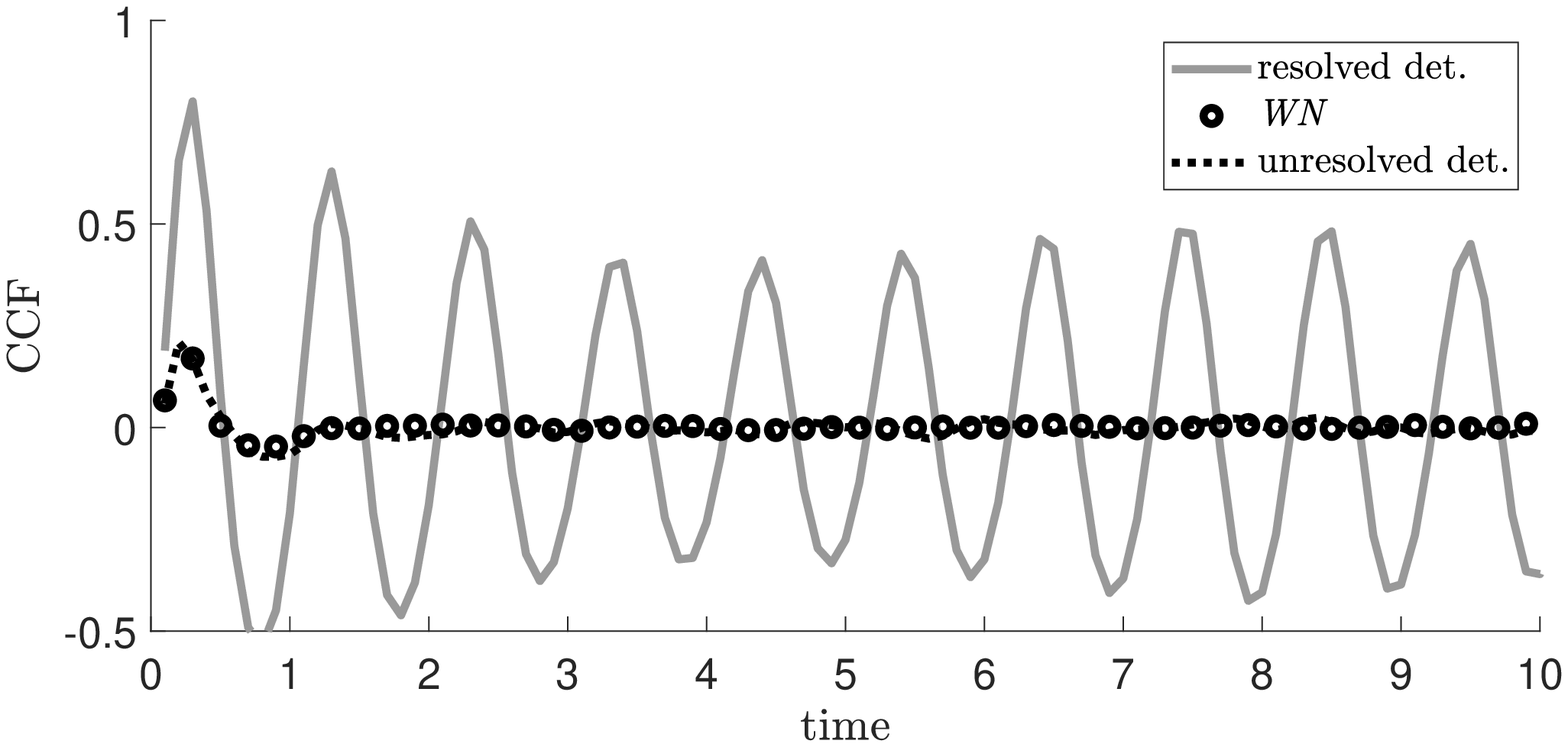}}
\hfill
\subfloat[\label{fig:wavemean_controlG}]{\includegraphics[width=0.4\linewidth]{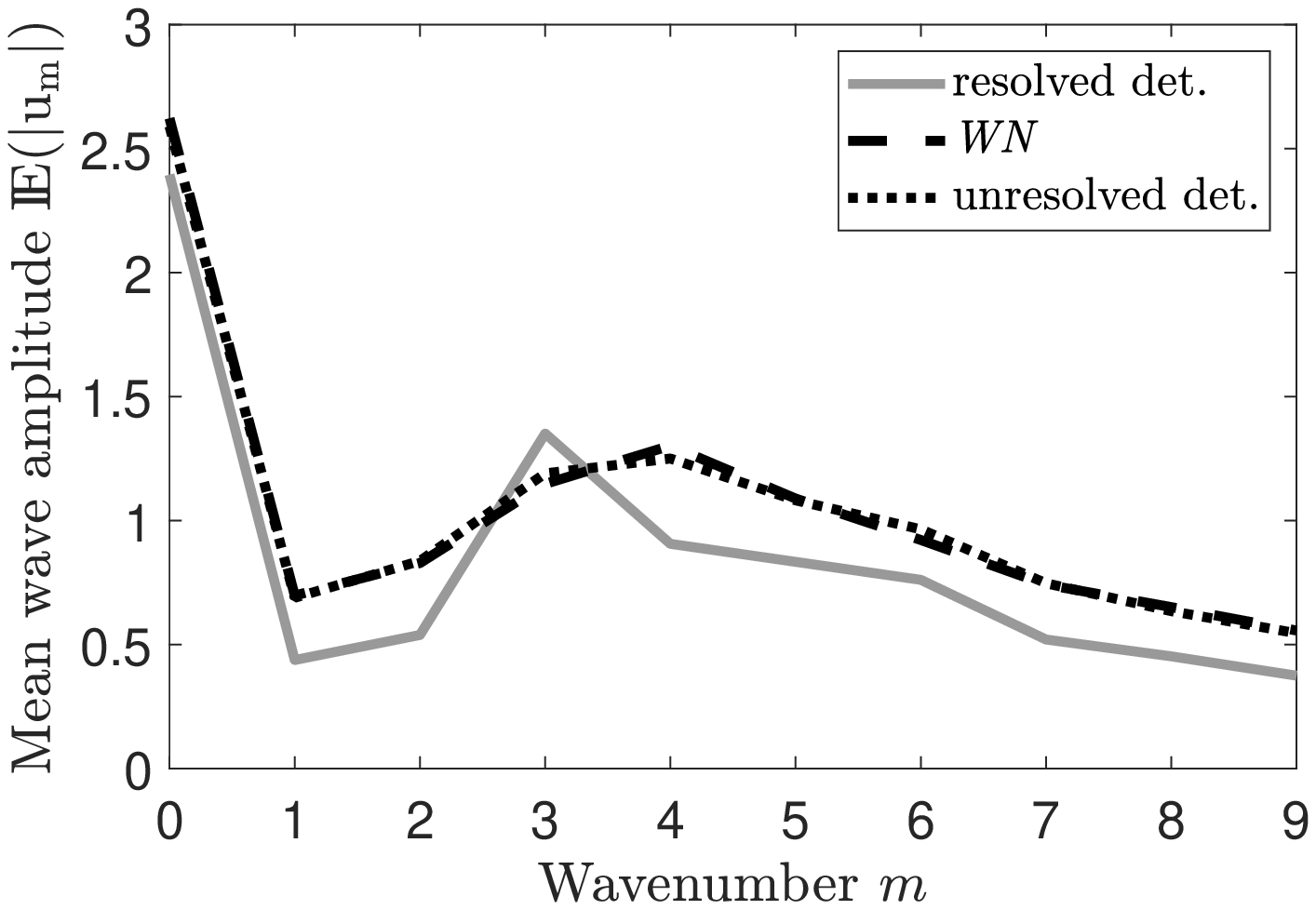}}
\hfill
\subfloat[\label{fig:wavemean_controlN}]{\includegraphics[width=0.4\linewidth]{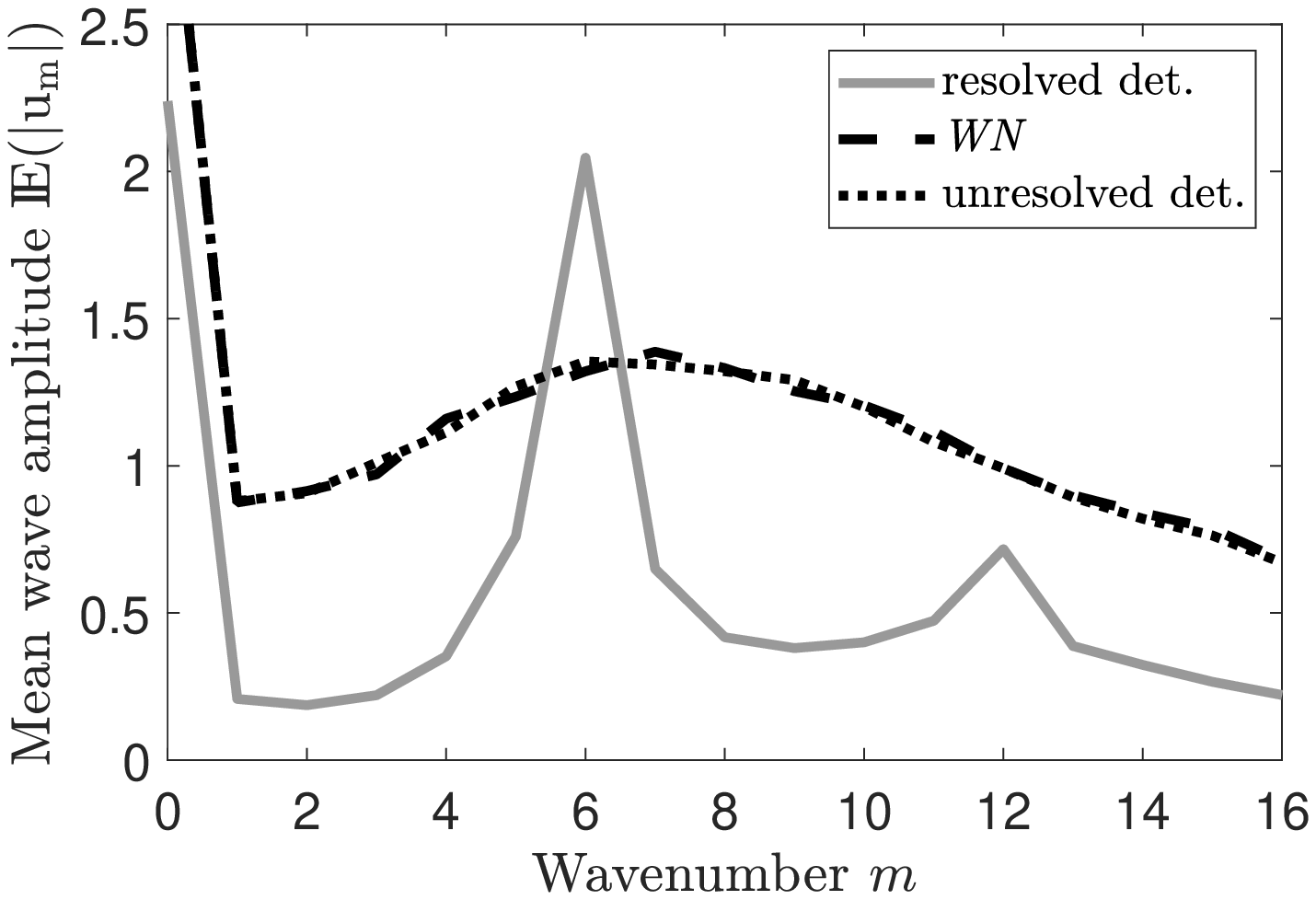}}
\hfill
\subfloat[\label{fig:wavevar_controlG}]{\includegraphics[width=0.4\linewidth]{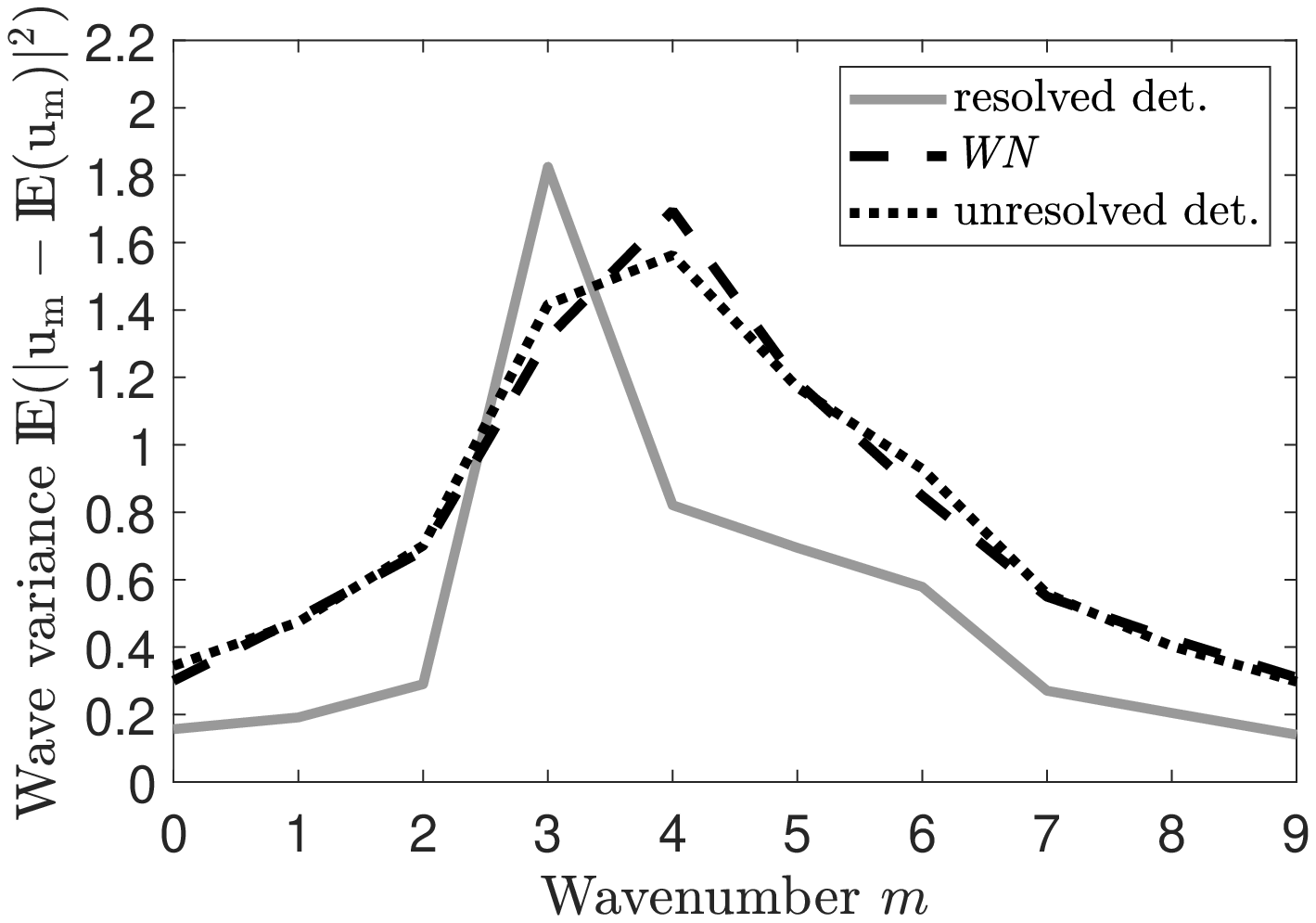}}
\hfill
\subfloat[\label{fig:wavevar_controlN}]{\includegraphics[width=0.4\linewidth]{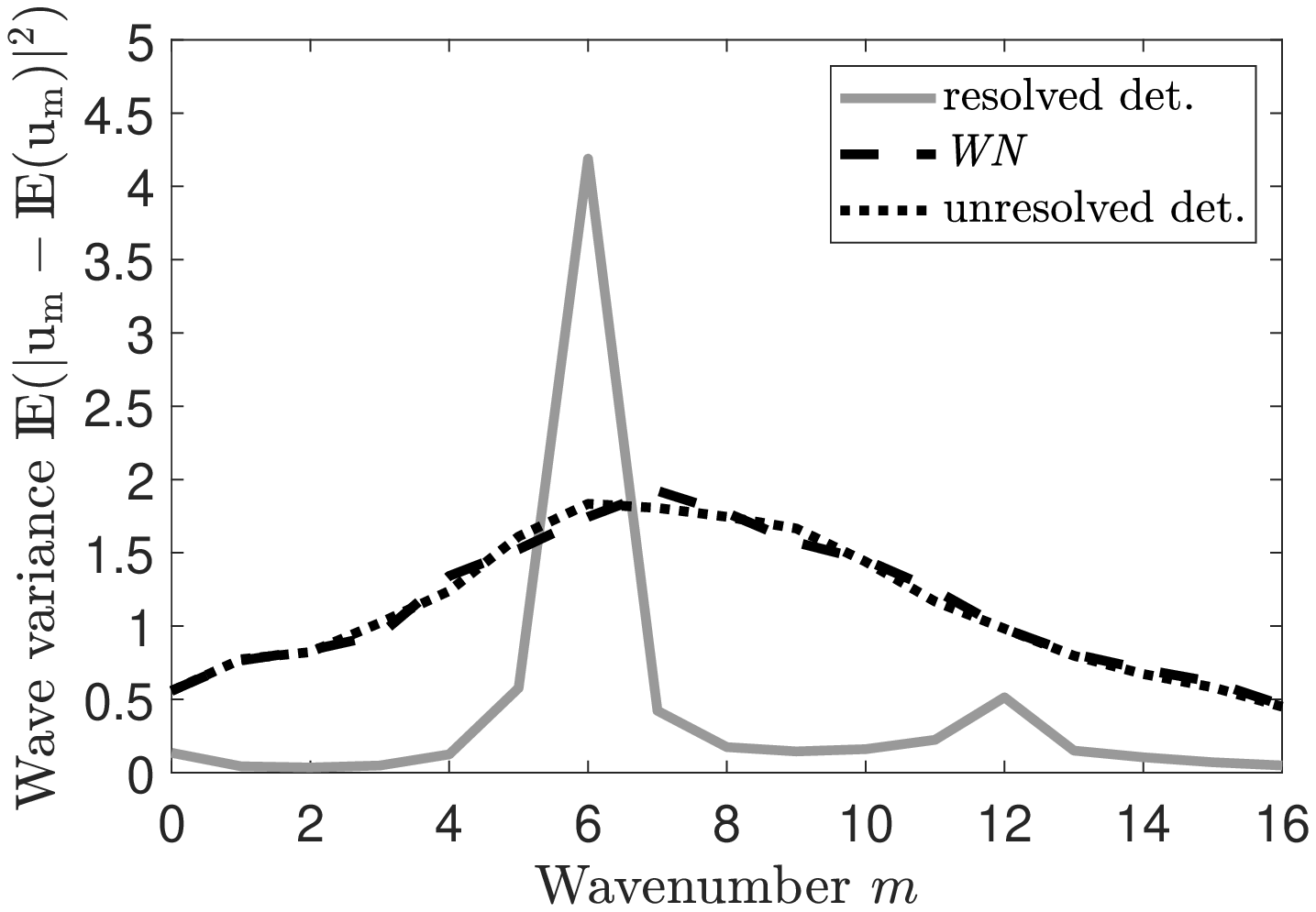}}
\caption{Comparison between ACFs \textbf{(a)}-\textbf{(b)}, CCFs \textbf{(c)}-\textbf{(d)}, wave mean amplitude \textbf{(e)}-\textbf{(f)}, and wave variance \textbf{(g)}-\textbf{(h)} for the unconditioned stochastic simulations (\ref{eq:b_bench}) as well as the ``resolving'' and ``unresolving'' deterministic reference simulations for unimodal \textbf{(a)}, \textbf{(c)}, \textbf{(e)}, and \textbf{(g)} and trimodal \textbf{(b)}, \textbf{(d)}, \textbf{(f)}, \textbf{(h)} configurations.}
\label{fig:rest_control}
\end{figure}

Interestingly, Figures \ref{fig:distr_control} and \ref{fig:rest_control} show that the (\ref{eq:b_bench}) parameterization  introduces no significant changes to the long-term statistics of the ``unresolved'' model without any parameterization. Thus, the perturbations of the white noise are not able to alter the dynamics of the ``unresolved'' model.

Similarly, the ACF, CCF, and wave criteria are not reproduced to any satisfactory degree with (\ref{eq:b_bench}), see Figures \ref{fig:ACF_controlG}-\ref{fig:wavevar_controlN}. In Figures \ref{fig:ACF_controlG} and \ref{fig:ACF_controlN} one sees that the reduced model with (\ref{eq:b_bench}) exhibits ACFs that are very similar to those of the unresolved deterministic model; neither the amplitudes nor the long decorrelation scales shown by the resolved deterministic simulation are reproduced. The CCFs in Figures \ref{fig:CCF_controlG} and \ref{fig:CCF_controlN} show the same problems. The mean wave amplitudes and wave variances of (\ref{eq:b_bench}) in Figures \ref{fig:wavemean_controlG}-\ref{fig:wavemean_controlN} and Figures \ref{fig:wavevar_controlG}-\ref{fig:wavevar_controlN}, respectively, show that the reduced models have more uniform spread over the larger wave numbers and do not peak at the correct wavenumbers, compared to the resolved deterministic model.

\subsection{Single regressor parameterizations}\label{sec:results_onevariable}
Next, we consider reduced model simulations with single regressors for the VARX models. We expect that the state-dependency, temporally correlated mean, and exogenous predictor variable of the VARX model will improve the performance, capturing more of the features from the resolving L96 reference simulation.

First, let us consider the autoregressive model (\ref{eq:b_singleb}) consisting of multiple independent AR$(1)$ processes due to its diagonal drift coefficient matrix $(A_1)$: 
\begin{equation}\label{eq:b_singleb}\tag{\emph{Multi} AR$(1)$}
\widetilde{\vecB}^{n} = \vecA_0 + A_1 \widetilde{\vecB}^{n-1} + \Sigma_D \vecXi^n. 
\end{equation}

Second, we consider a vector of independent white noise processes with drift:
\begin{equation}\label{eq:b_singlex}\tag{\emph{WND}}
\widetilde{\vecB}^{n} = \vecA_0 + D \widetilde{\vecX}^n + \Sigma_D \vecXi^n,
\end{equation}
 
\begin{figure}[htb]
\centering
\captionsetup[subfloat]{justification=centering}
\subfloat[\label{fig:distr_singleG}]{\includegraphics[width=0.40\linewidth]{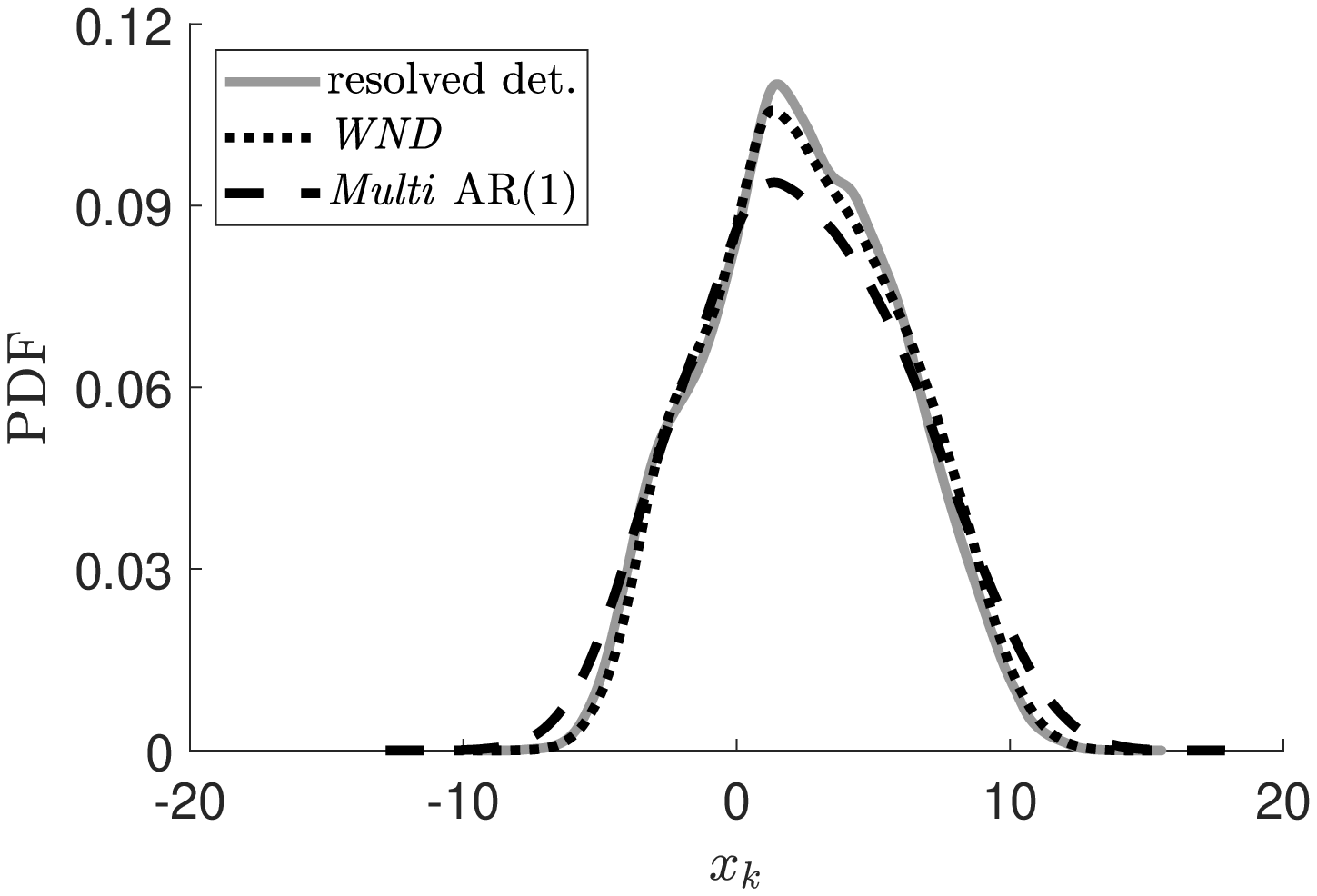}}
\hfill
\subfloat[\label{fig:distr_singleN}]{\includegraphics[width=0.40\linewidth]{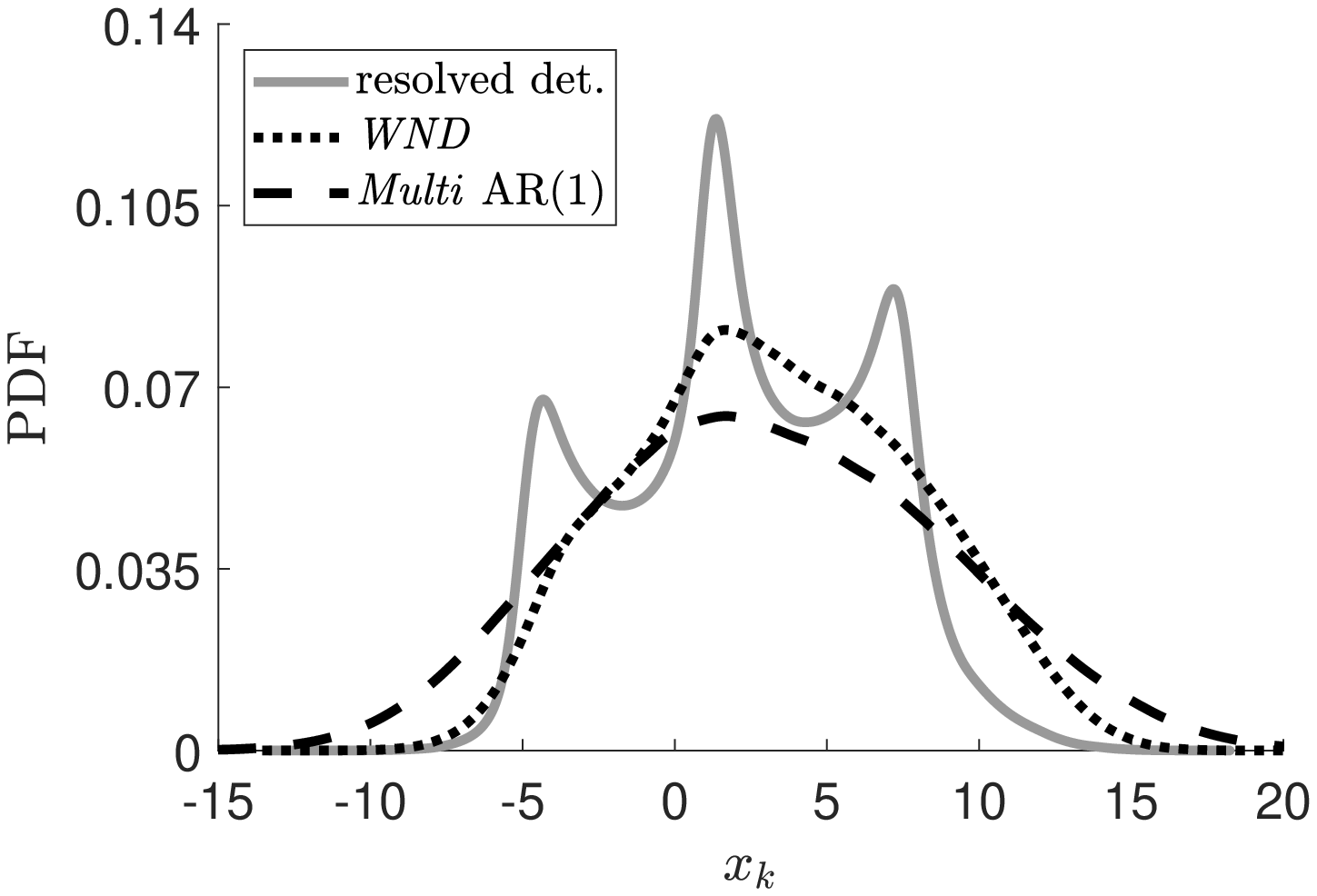}}
\hfill
\subfloat[\label{fig:ACF_singleG}]{\includegraphics[width=0.49\linewidth]{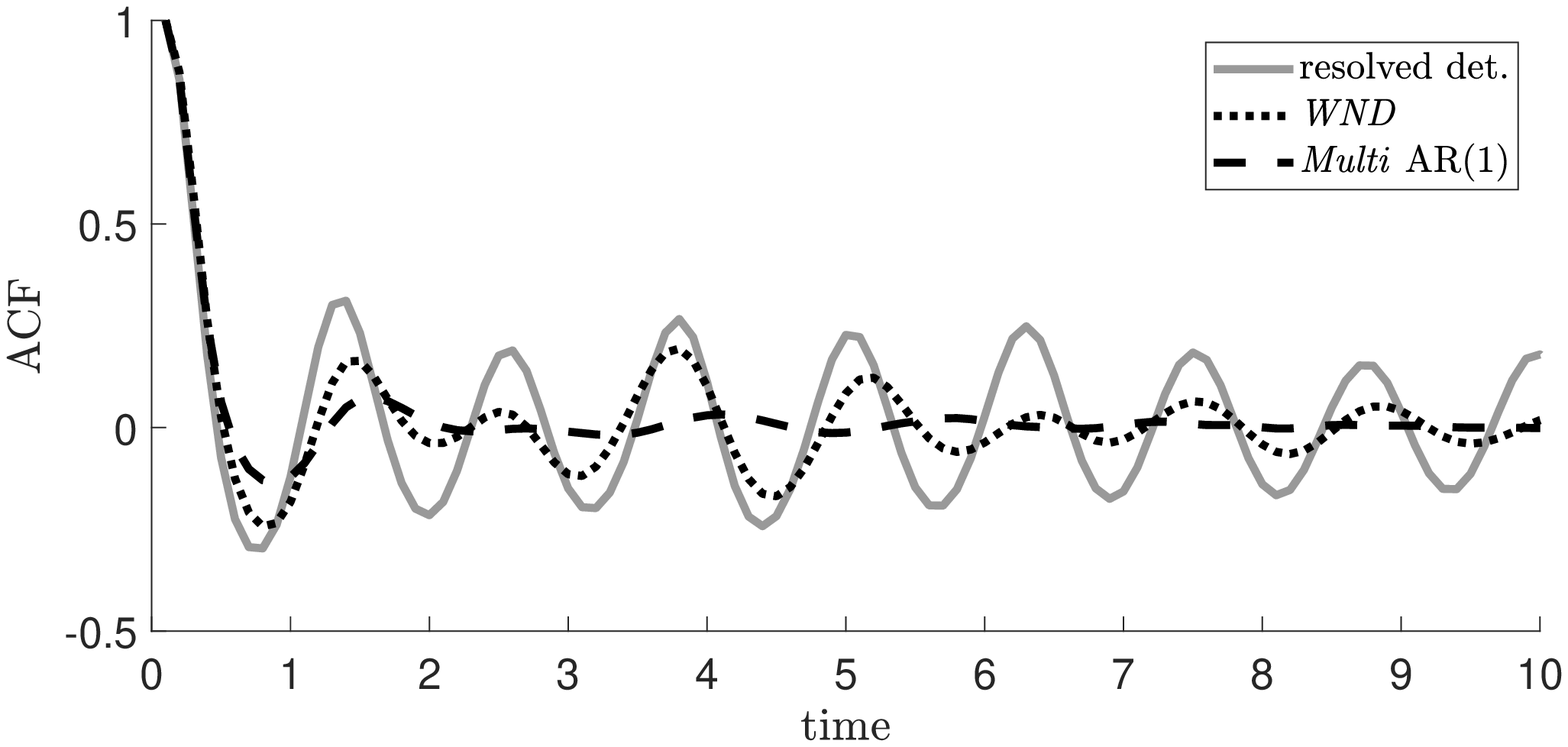}}
\hfill
\subfloat[\label{fig:ACF_singleN}]{\includegraphics[width=0.49\linewidth]{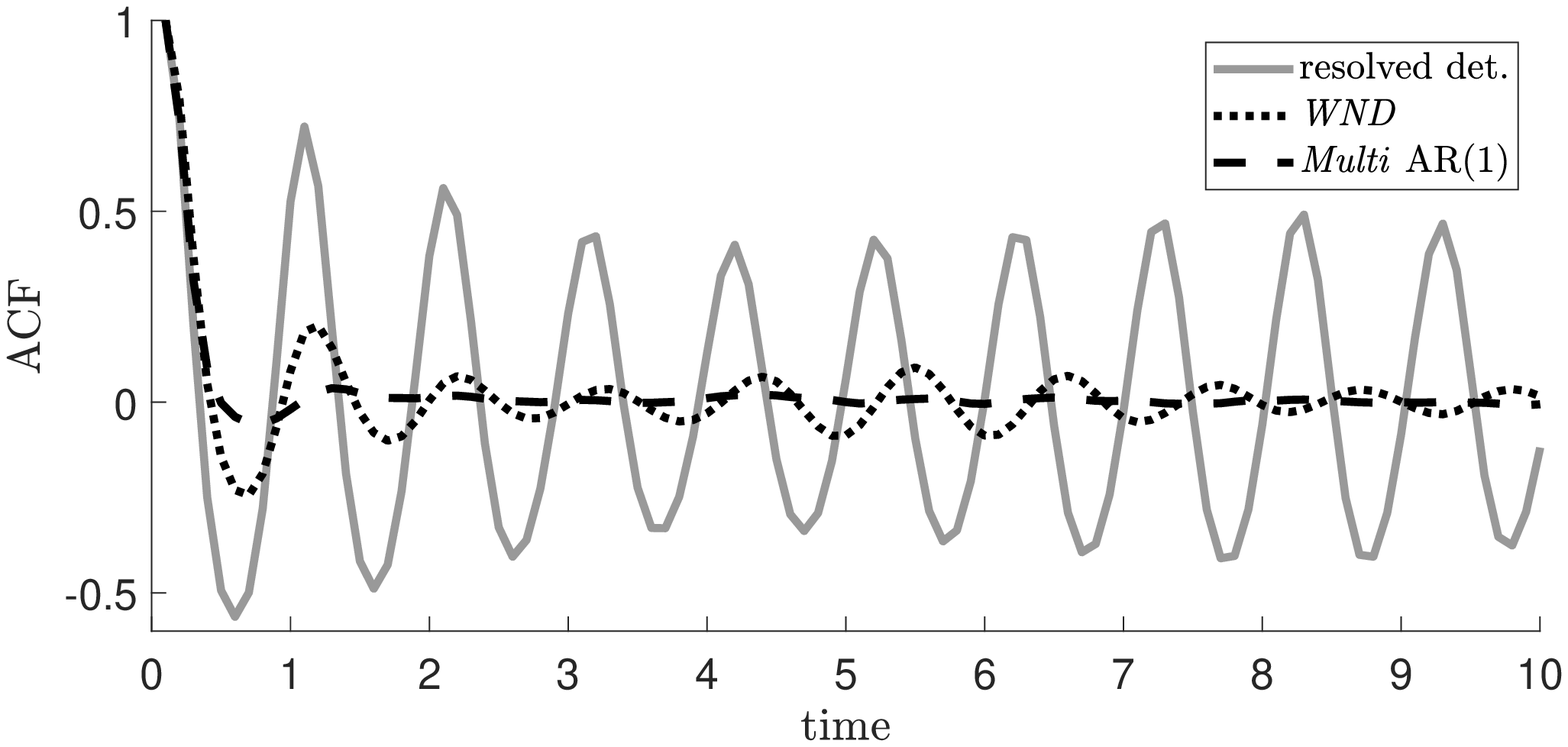}}
\caption{Comparison between \textbf{(a)}-\textbf{(b)} the PDFs and \textbf{(c)}-\textbf{(d)} the ACFs of interest for both the unimodal and trimodal L96 configurations (see Table \ref{tab:parameters}), respectively, resulting from stochastic simulations with single regressor variables.}
\label{fig:singleconditioning}
\end{figure}

As can be seen from the criteria plotted in Figure \ref{fig:singleconditioning}, the (\ref{eq:b_singleb}) model does not significantly improve over (\ref{eq:b_bench}) (cf. \Crefrange{fig:distr_controlG}{fig:ACF_controlN}). By contrast, Figure \ref{fig:distr_singleG} shows that the (\ref{eq:b_singlex}) model reproduces the unimodal distribution of $x_k$ significantly better than the (\ref{eq:b_bench}) model. This is due to the $x_k$ dependence of (\ref{eq:b_singlex}). It suggests that the exogenous variable $\vecX$ indeed holds predictive value for $\mathcal{B}_{xy}$ (as suggested in Section \ref{sec:introduction}). Also, the (\ref{eq:b_singleb}) model is independent of $\widetilde{\vecX}$, unlike (\ref{eq:b_singlex}). However, while (\ref{eq:b_singlex}) reproduces the distribution of $x_k$ accurately in the unimodal case (Figure \ref{fig:distr_singleG}), it fails to do so in the trimodal case (Figure \ref{fig:distr_singleN}). Furthermore, (\ref{eq:b_singlex}) improves only slightly on the ACFs of $x_k$ when compared to (\ref{eq:b_bench}) (see Figure \ref{fig:ACF_singleG}). These same conclusions are reached for the CCFs and wave criteria (not shown). To introduce more spatio-temporal consistency in the VARX we test combinations of endogenous and exogenous regressors in the next section.

\subsection{Double regressor parameterizations}\label{sec:results_twovariables}

\subsubsection{Diagonal covariance}\label{sec:results_diagonalcovariance}

As motivated in Section \ref{sec:results_lagchoice}, we suggest the (\ref{eq:b_double_Gauss_diag}) model here for parameterization in the case of the unimodal L96 configuration:
\begin{equation}\label{eq:b_double_Gauss_diag}\tag{VARX$(14)$ $\Sigma_D$}
\widetilde{\vecB}^{n} = \vecA_0 + A_{14} \widetilde{\vecB}^{n-14} +  D \widetilde{\vecX}^n + \Sigma_D \vecXi^n. 
\end{equation}

\begin{figure}[htb]
\centering
\captionsetup[subfloat]{justification=centering}
\subfloat[\label{fig:distr_doubleG}]{\includegraphics[width=0.33\linewidth]{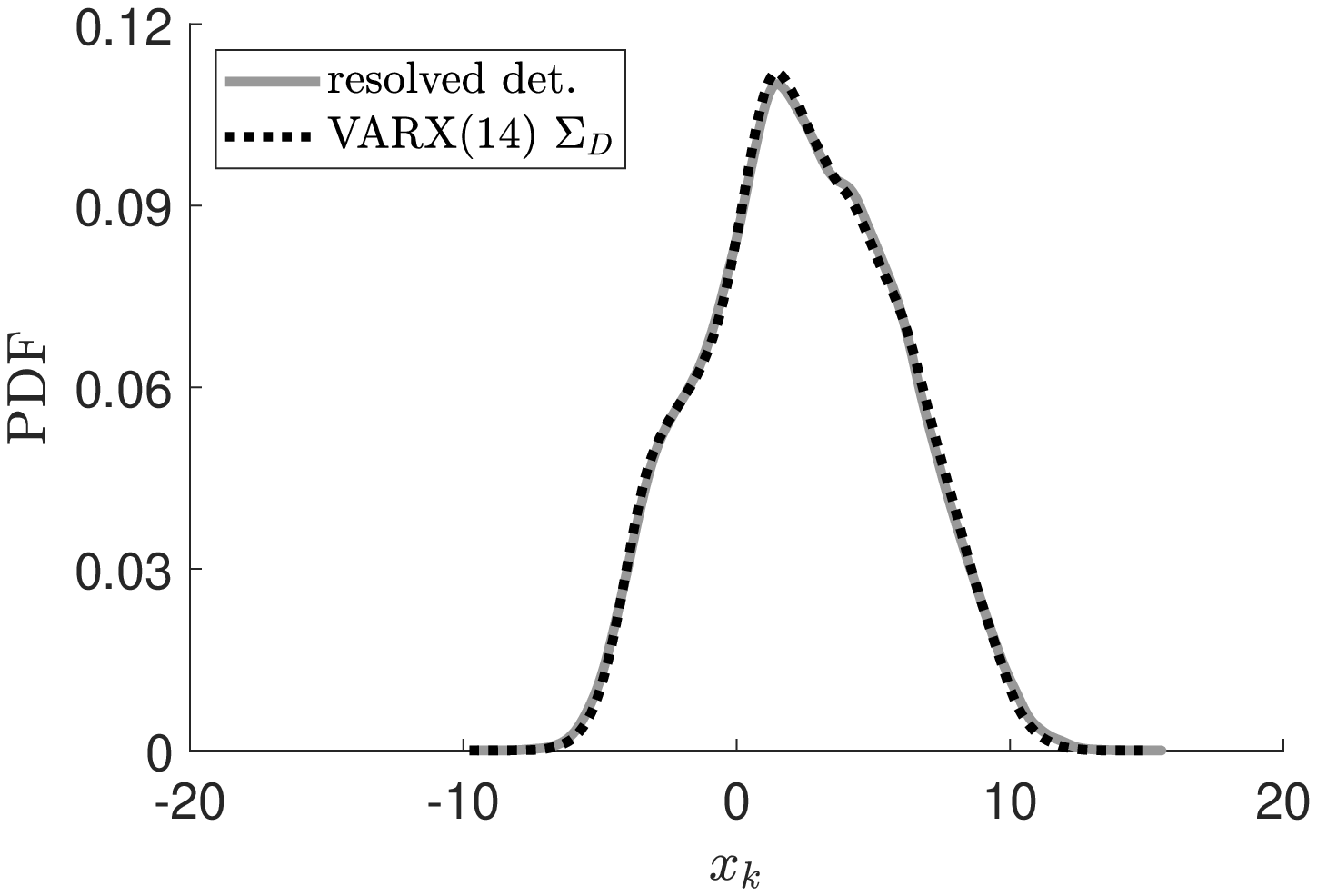}}
\hfill
\subfloat[\label{fig:wavemean_doubleG}]{\includegraphics[width=0.33\linewidth]{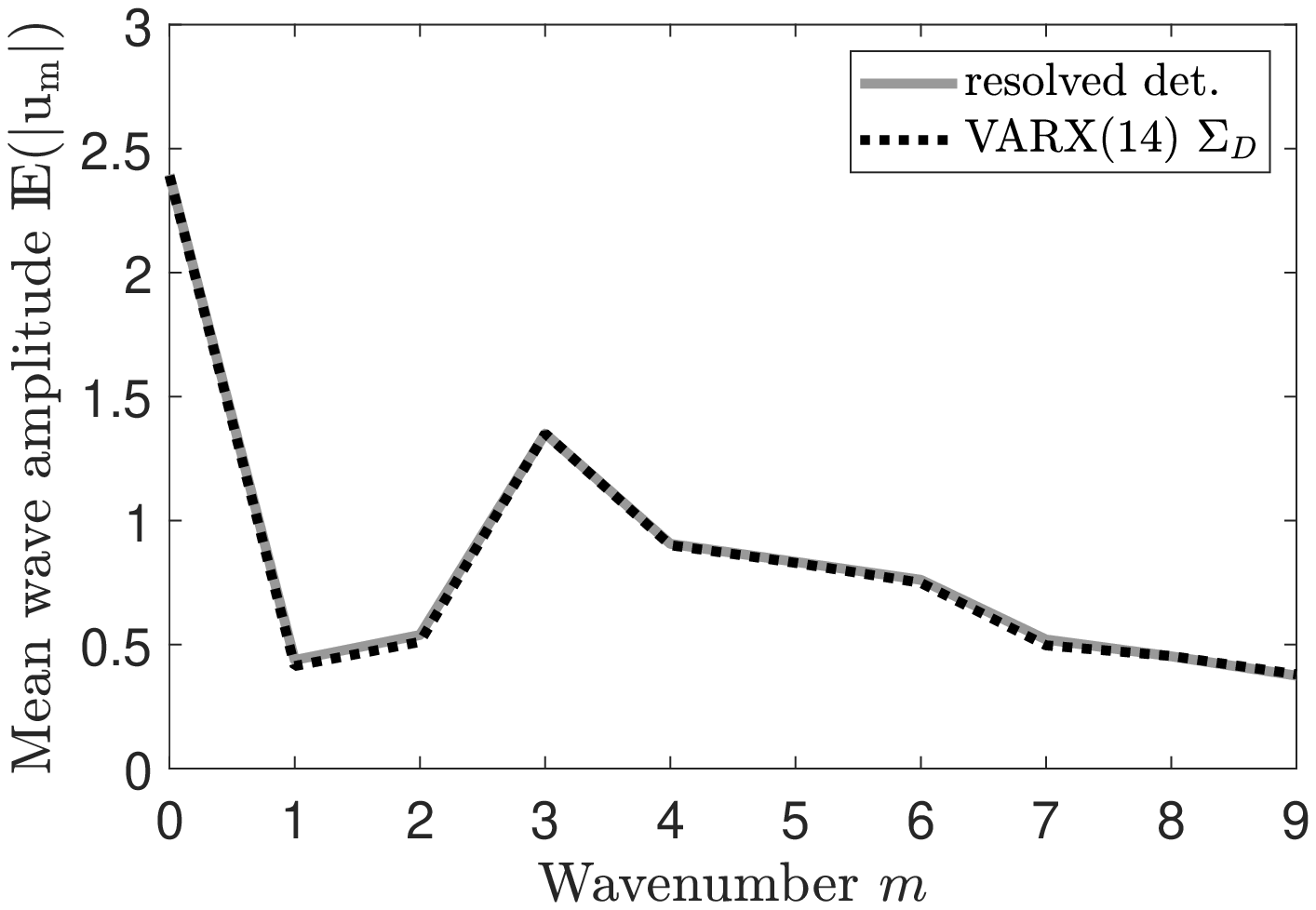}}
\hfill
\subfloat[\label{fig:wavevar_doubleG}]{\includegraphics[width=0.33\linewidth]{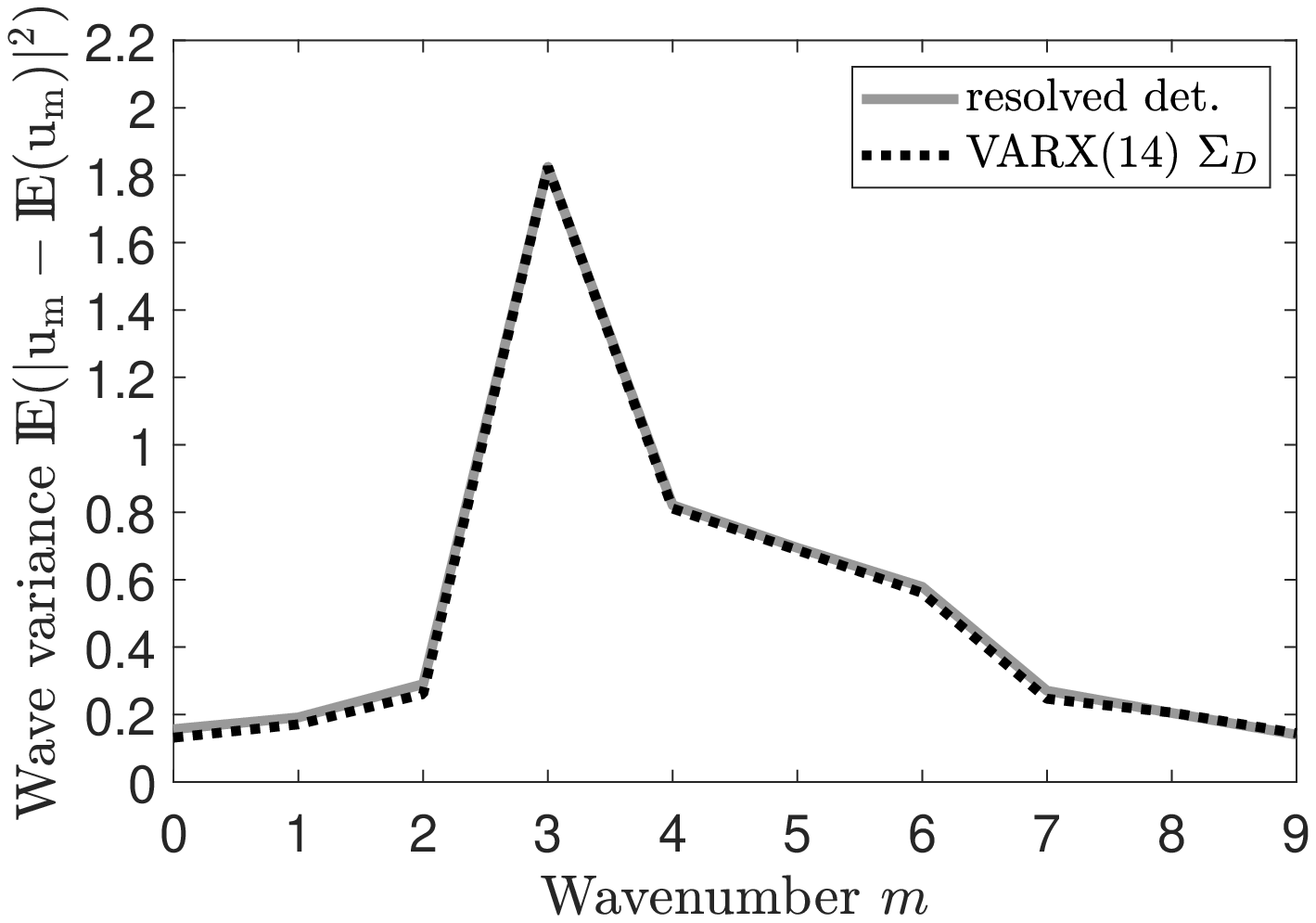}}
\hfill
\subfloat[\label{fig:ACF_doubleG}]{\includegraphics[width=0.5\linewidth]{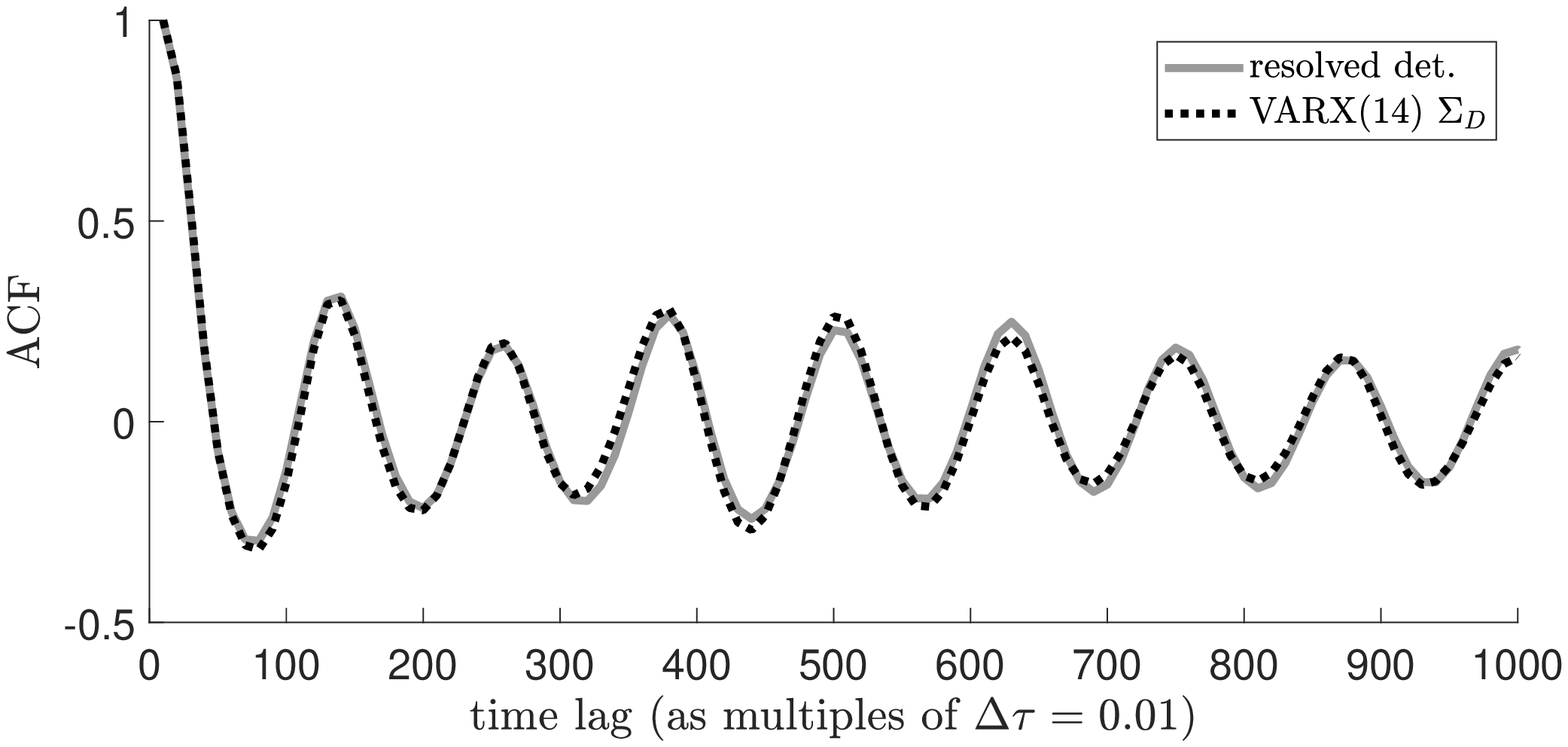}}
\hfill
\subfloat[\label{fig:CCF_doubleG}]{\includegraphics[width=0.5\linewidth]{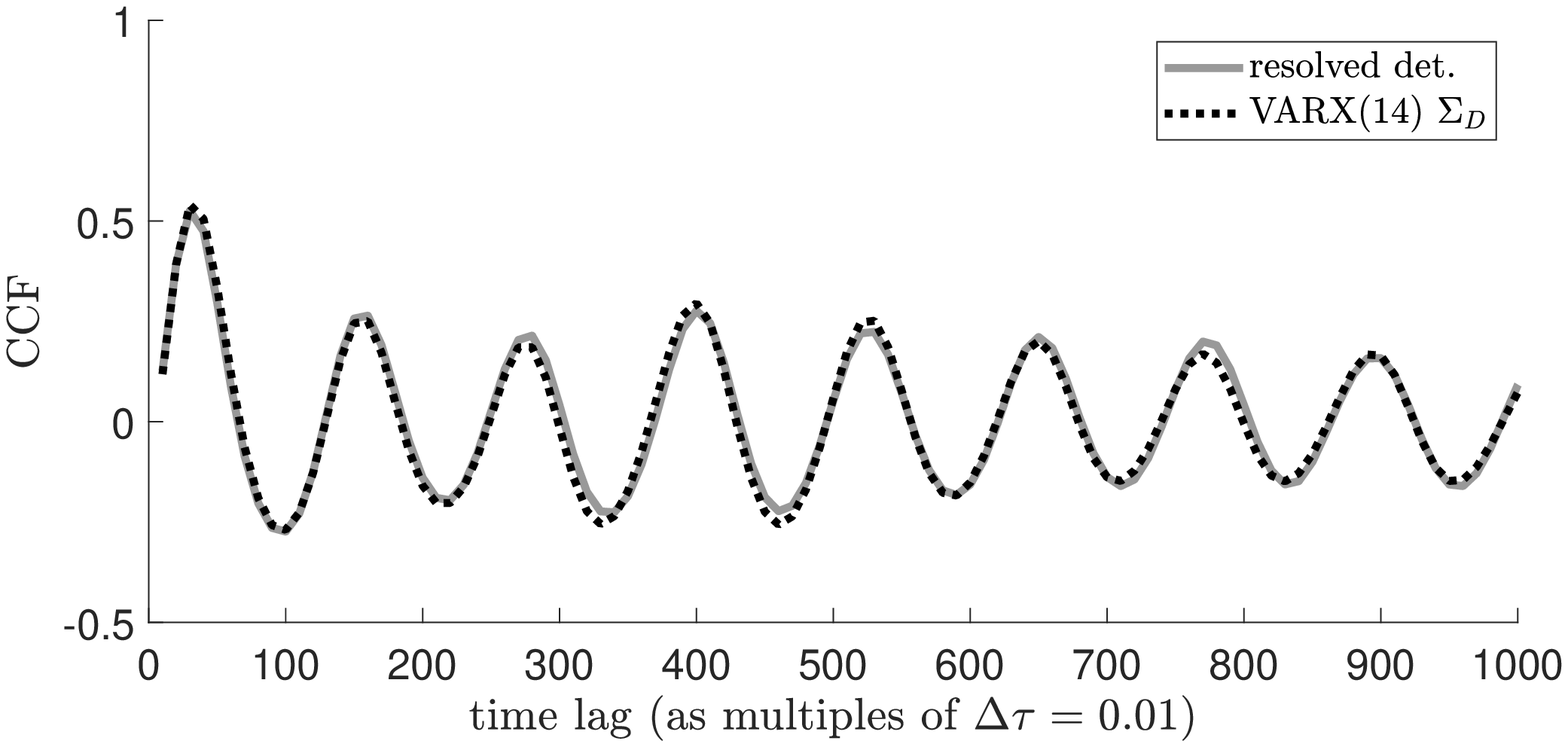}}
\caption{Comparison of \textbf{(a)} PDFs, \textbf{(b)} mean wave amplitude, \textbf{(c)} wave variance, \textbf{(d)} ACFs, and \textbf{(e)} CCFs of the reduced model using (\ref{eq:b_double_Gauss_diag}) and of the unimodal deterministic reference.}
\label{fig:doubleconditioningG_diagonal}
\end{figure}

Figure \ref{fig:doubleconditioningG_diagonal} shows that the state-dependence and temporal correlation introduced by $D$ and $A_{14}$ in (\ref{eq:b_double_Gauss_diag}) result in near-perfect approximations of the reference statistics. Not only does the distribution of $\tilde{x}_k$ match perfectly to the reference (Figure \ref{fig:distr_doubleG}), but also the wave criteria (Figures \ref{fig:wavemean_doubleG} and \ref{fig:wavevar_doubleG}) and correlations (Figures \ref{fig:ACF_doubleG} and \ref{fig:CCF_doubleG}) match almost exactly. We emphasize the accuracy of the reproduced long sinusoidal decorrelation structure visible in \ref{fig:ACF_doubleG} and \ref{fig:CCF_doubleG}, a particularly challenging feature of the reference L96 simulations. 

However, this strong performance does not extend fully to the trimodal L96 configuration.  For this configuration we suggested $p=30$ in Section \ref{sec:results_lagchoice}, i.e. the following (\ref{eq:b_double_tri_diag}) model:
\begin{equation}\label{eq:b_double_tri_diag}\tag{VARX$(30)$ $\Sigma_D$}
\widetilde{\vecB}^{n} = \vecA_0 + A_{30} \widetilde{\vecB}^{n-30} + D \widetilde{\vecX}^n + \Sigma_D \vecXi^n. 
\end{equation} 

\begin{figure}[htb]
\centering
\captionsetup[subfloat]{justification=centering}
\subfloat[\label{fig:distr_doubleN}]{\includegraphics[width=0.33\linewidth]{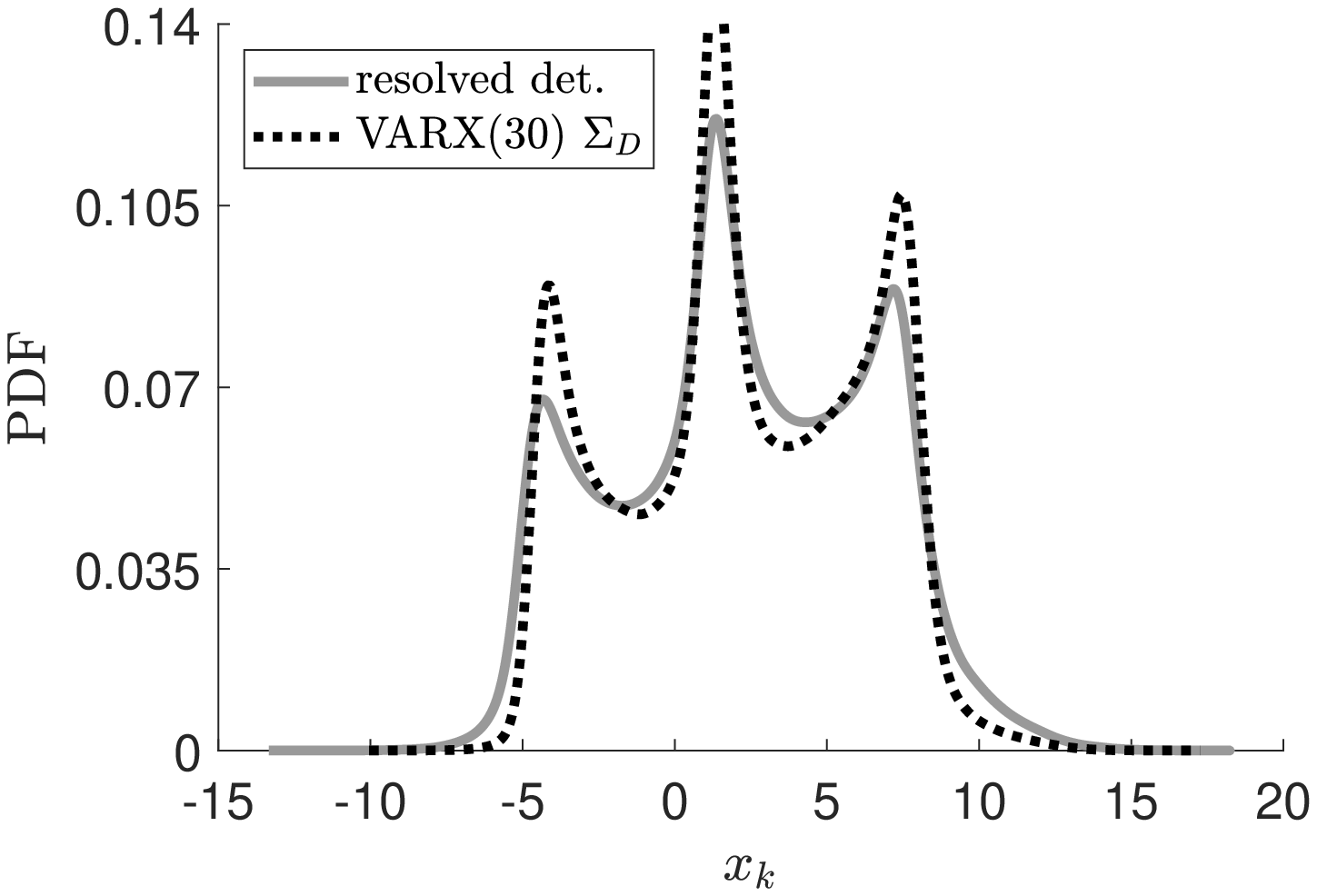}}
\hfill
\subfloat[\label{fig:wavemean_doubleN}]{\includegraphics[width=0.33\linewidth]{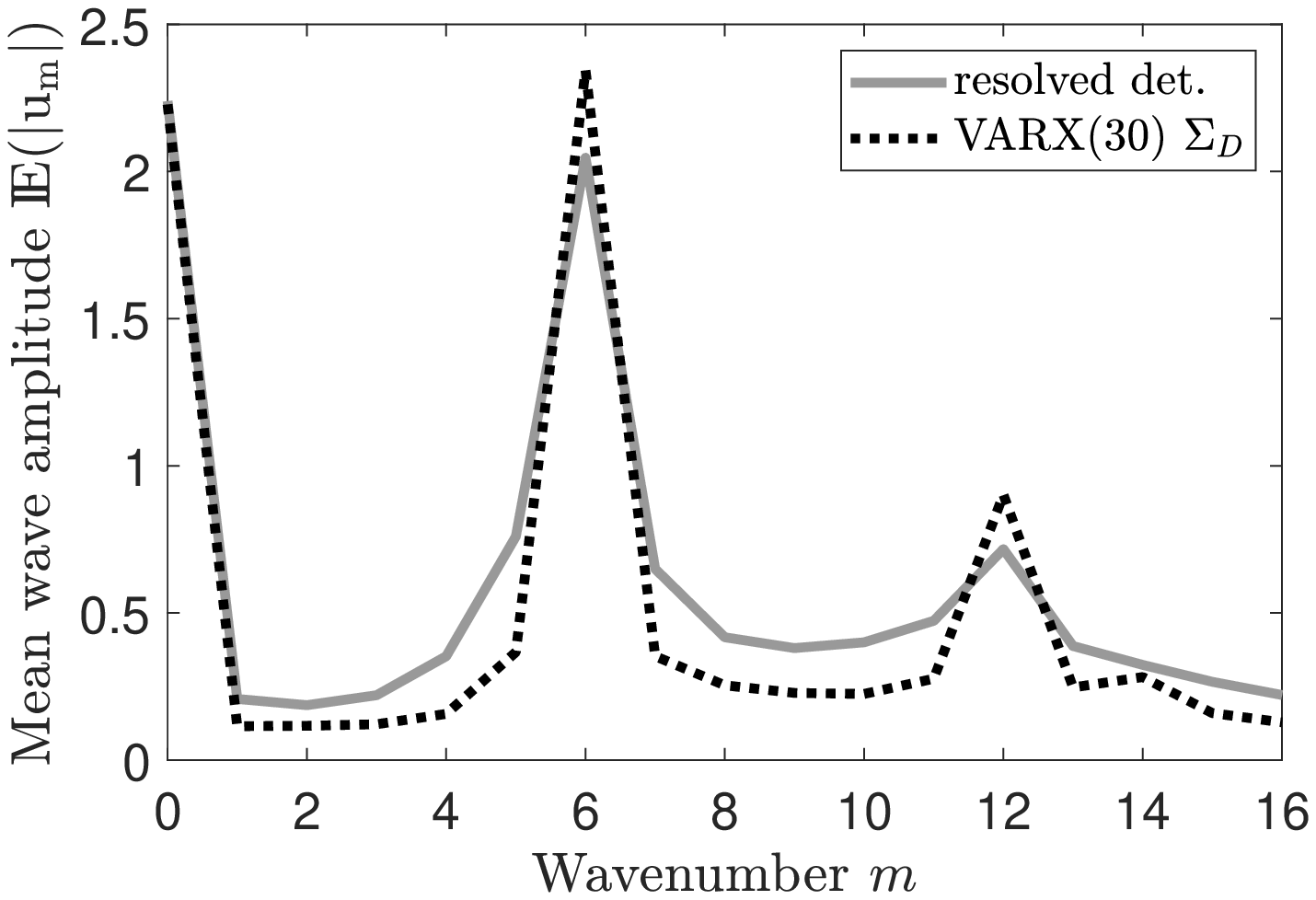}}
\hfill
\subfloat[\label{fig:wavevar_doubleN}]{\includegraphics[width=0.33\linewidth]{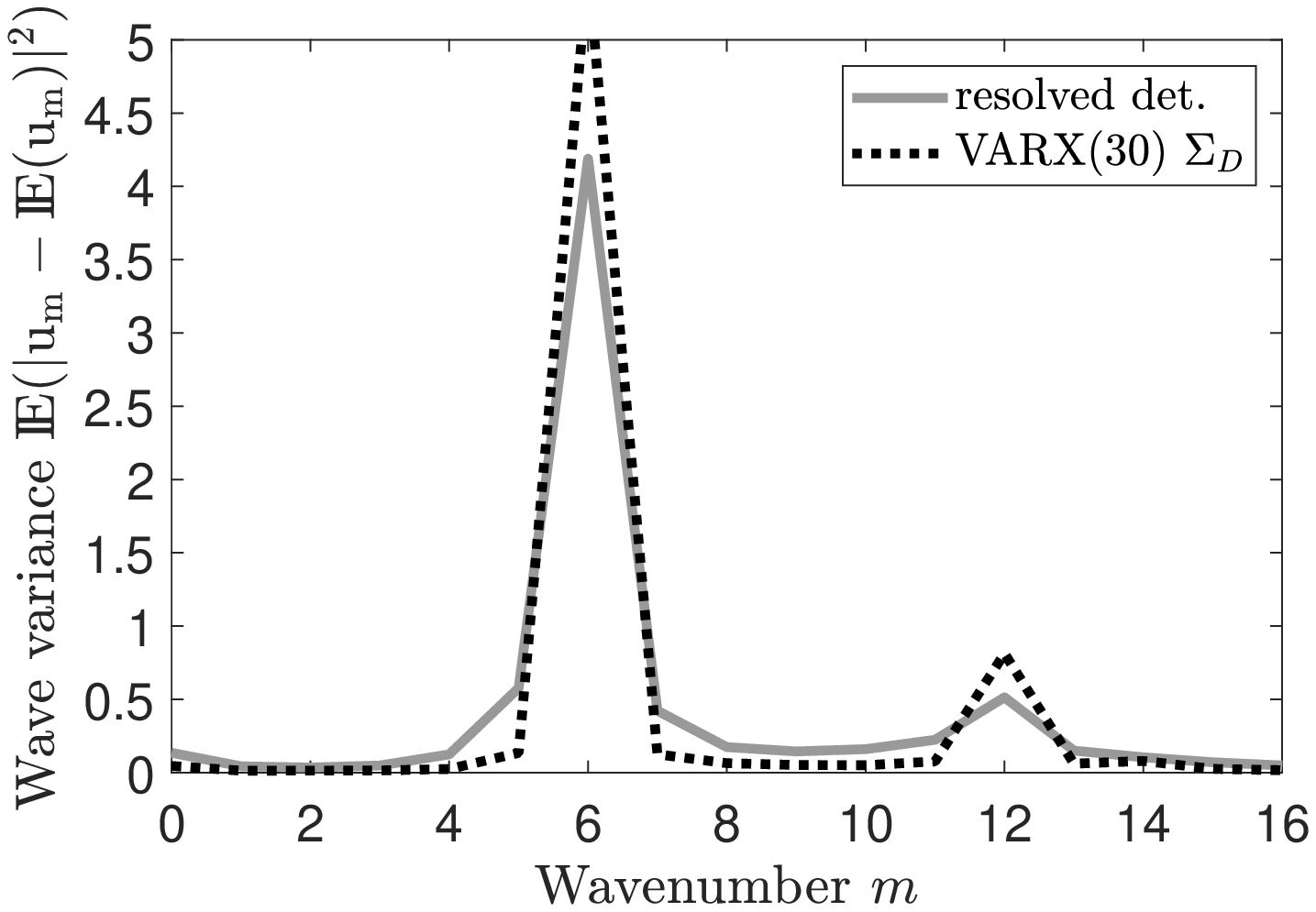}}
\hfill
\subfloat[\label{fig:ACF_doubleN}]{\includegraphics[width=0.5\linewidth]{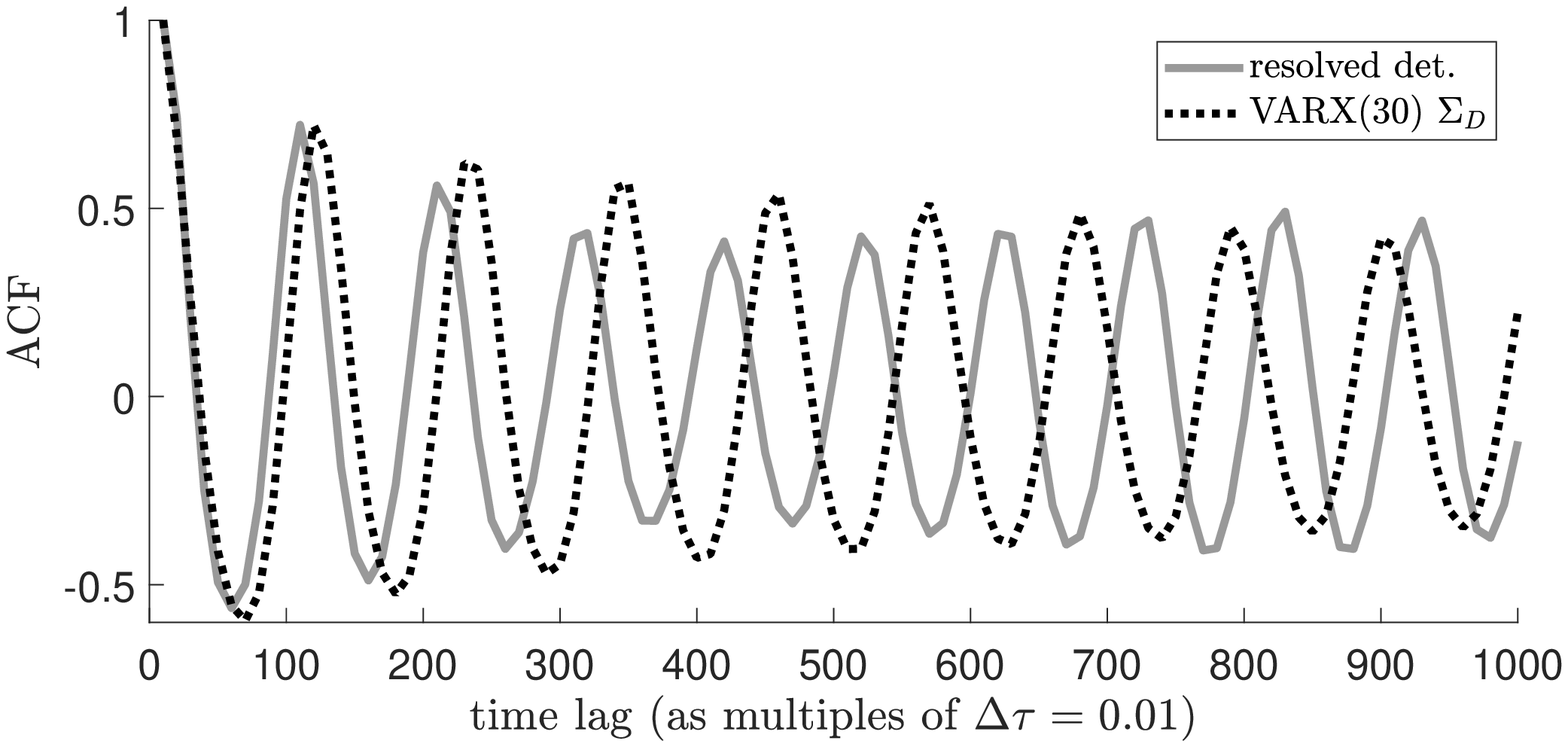}}
\hfill
\subfloat[\label{fig:CCF_doubleN}]{\includegraphics[width=0.5\linewidth]{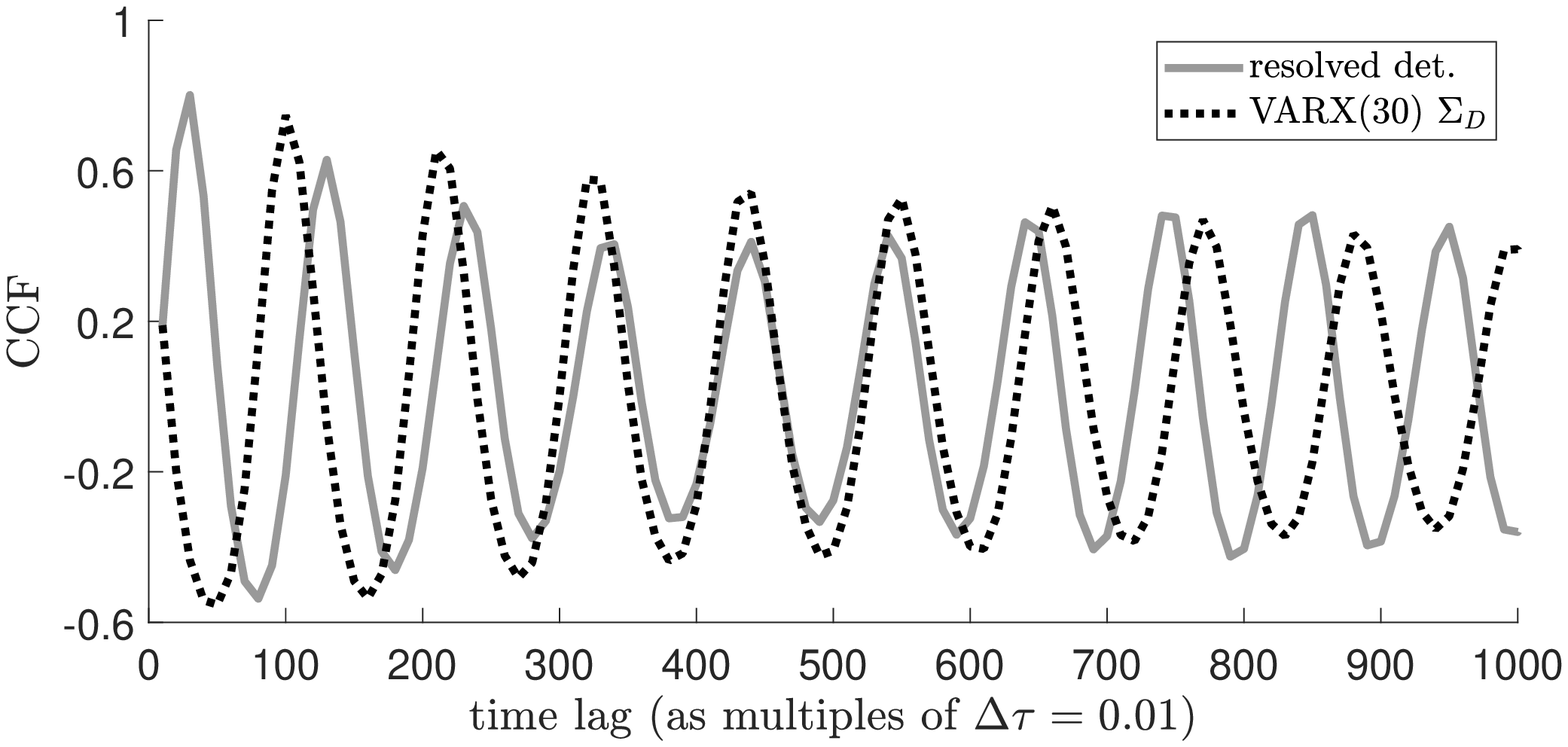}}
\caption{Comparison of \textbf{(a)} PDFs, \textbf{(b)} mean wave amplitude, \textbf{(c)} wave variance, \textbf{(d)} ACFs, and \textbf{(e)} CCFs of the reduced model using (\ref{eq:b_double_tri_diag}) and of the trimodal deterministic reference.}
\label{fig:doubleconditioningN_diagonal}
\end{figure}

The results with this model for parameterization are shown in Figure \ref{fig:doubleconditioningN_diagonal}. The PDF of $x_k$ (Figure \ref{fig:distr_doubleN}), the wave mean (\ref{fig:wavemean_doubleN}), the wave variance (\ref{fig:wavevar_doubleN}) and ACF (\ref{fig:ACF_doubleN}) are qualitatively correct, but not fully accurate. 
For example, the wave variance (Figure \ref{fig:wavevar_doubleN}) has peaks at wavenumbers 6 and 12 that are too high. Also, the oscillation periods of the ACF and CCF are too long (by circa 10$\%$) with the reduced model.

\subsubsection{Fully dense covariance}\label{sec:results_densecovariance}

The trimodal L96 configuration has strongly non-Gaussian features, making this a particularly challenging test case for our approach to  use VARX (i.e. Gaussian) processes for parameterization.
As displayed in Figures \ref{fig:distr_controlN} and \ref{fig:distr_singleN}, the trimodal nature of the PDF for $x_k$ is not captured at all
with the  (\ref{eq:b_bench}), (\ref{eq:b_singlex}) and (\ref{eq:b_singleb}) parameterizations. The results with (\ref{eq:b_double_tri_diag}) in the previous section are a major improvement. In this section we aim to improve further by 
using a fully-dense covariance matrix $\Sigma_L \Sigma_L^T$ instead of a diagonal one, as described in Section \ref{sec:covariance}:
\begin{equation}\label{eq:b_double_tri_dense}\tag{VARX$(30)$ $\Sigma_L$}
\widetilde{\vecB}^{n} = \vecA_0 + A_{30} \widetilde{\vecB}^{n-30} + D \widetilde{\vecX}^n + \Sigma_L \vecXi^n. 
\end{equation} 

\begin{figure}[htb]
\centering
\captionsetup[subfloat]{justification=centering}
\subfloat[\label{fig:distr_doubleN_2}]{\includegraphics[width=0.33\linewidth]{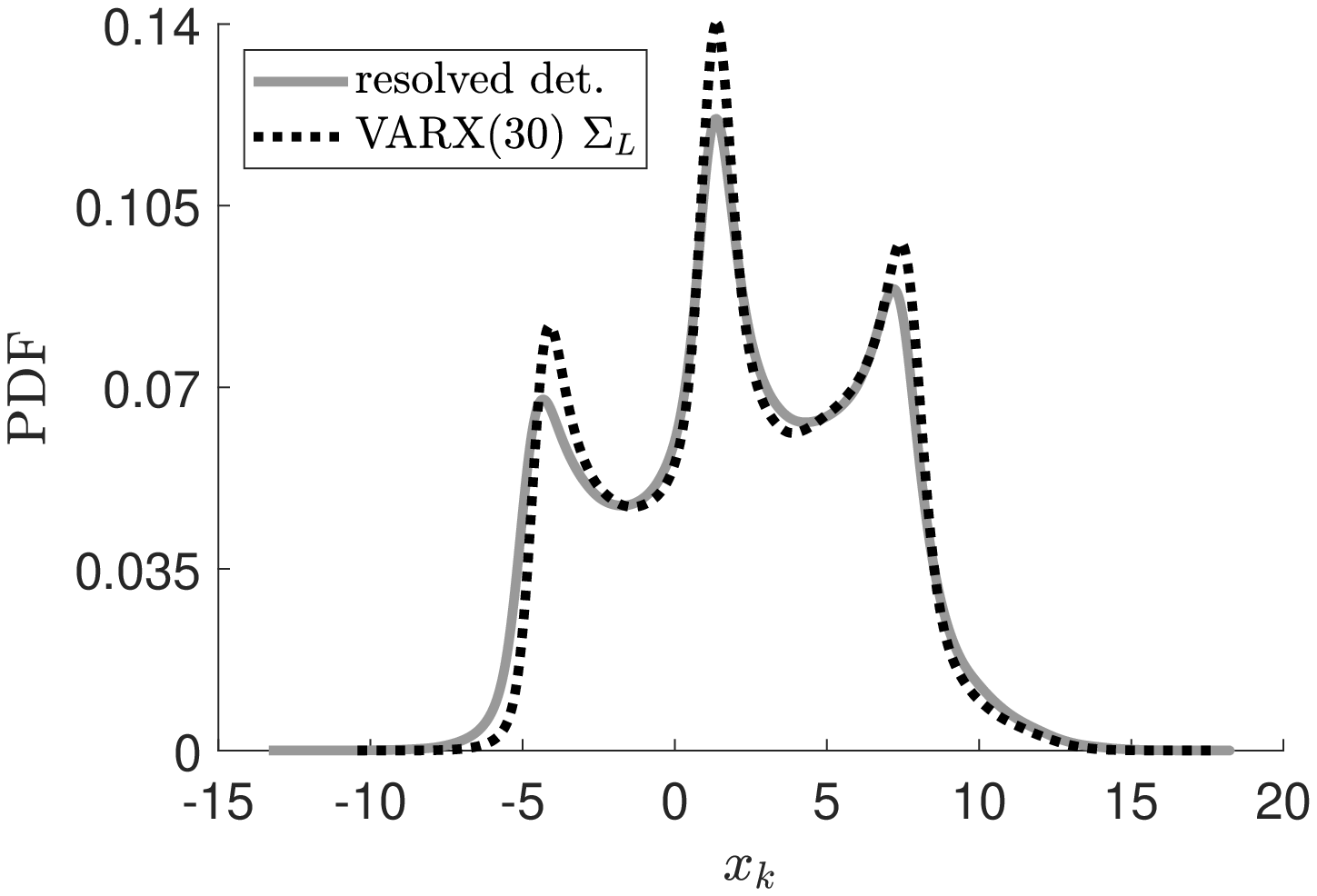}}
\hfill
\subfloat[\label{fig:wavemean_doubleN_2}]{\includegraphics[width=0.33\linewidth]{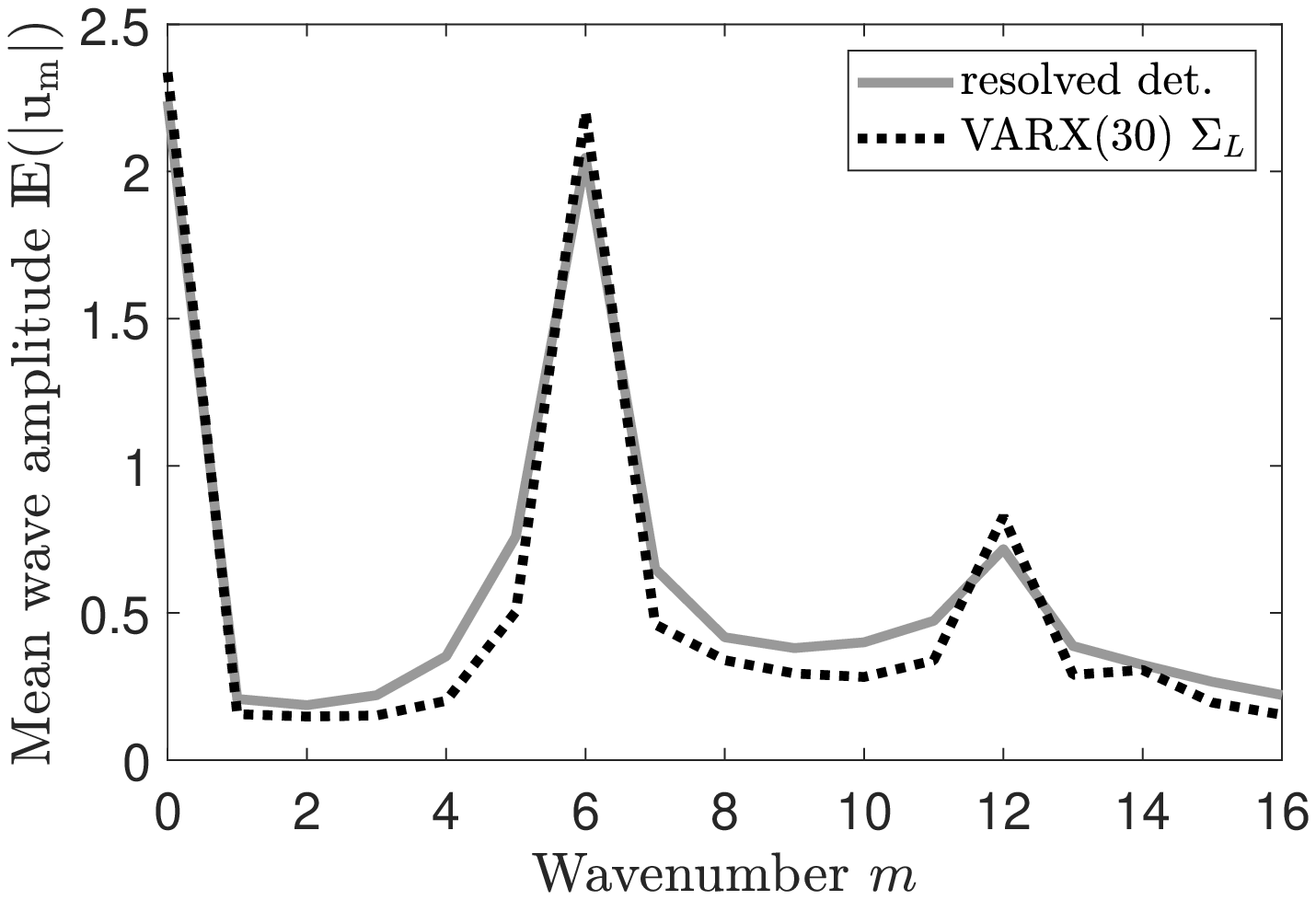}}
\hfill
\subfloat[\label{fig:wavevar_doubleN_2}]{\includegraphics[width=0.33\linewidth]{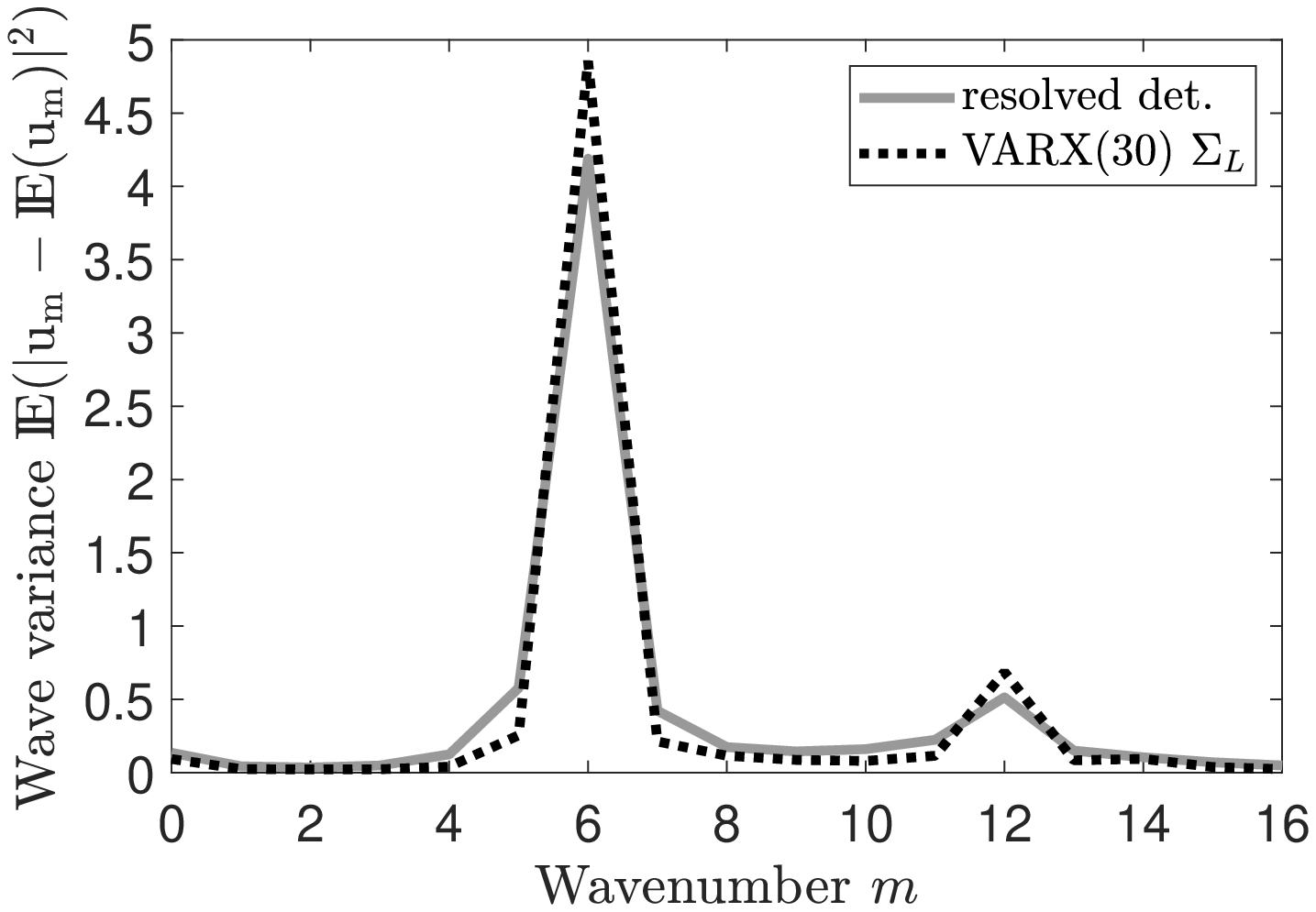}}
\hfill
\subfloat[\label{fig:ACF_doubleN_2}]{\includegraphics[width=0.5\linewidth]{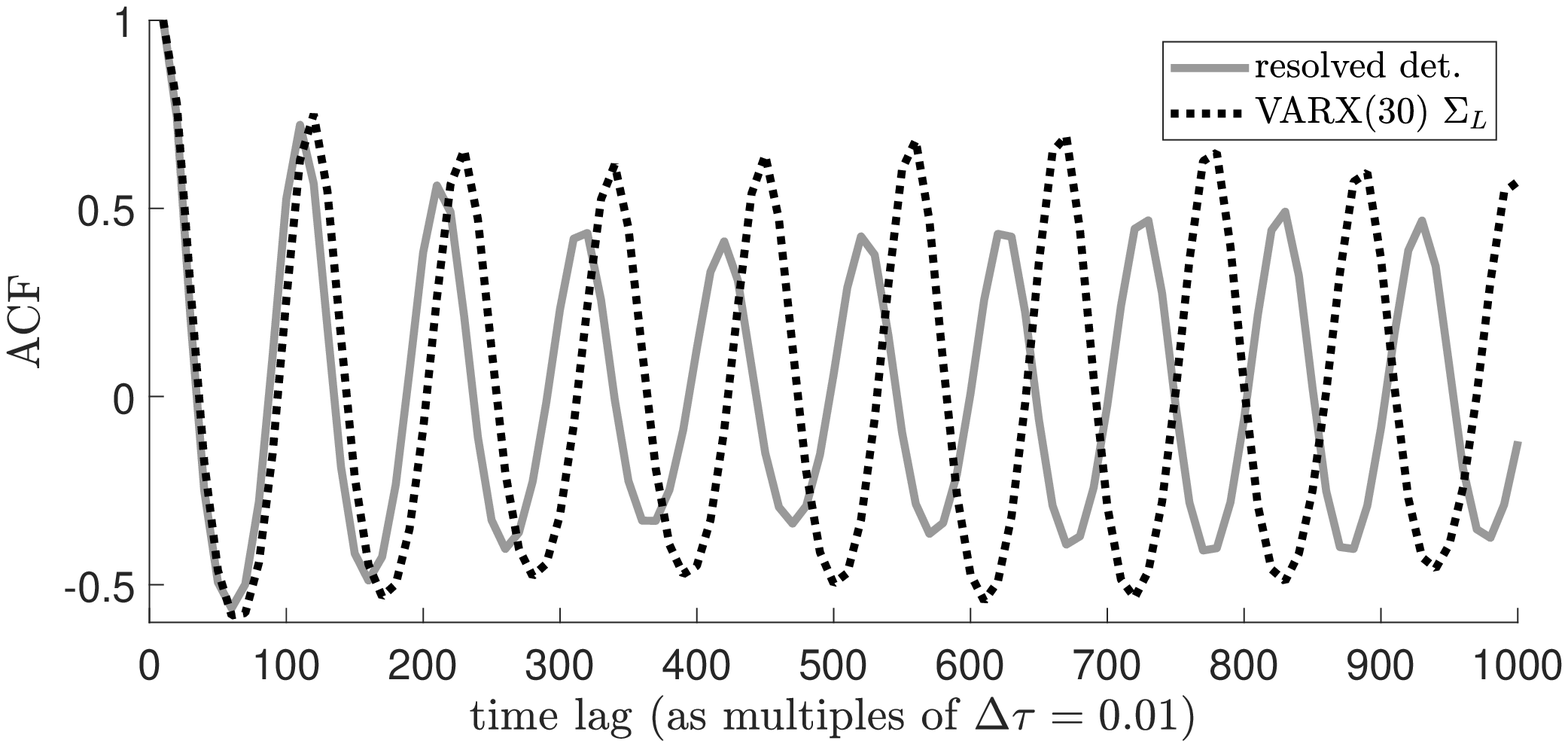}}
\hfill
\subfloat[\label{fig:CCF_doubleN_2}]{\includegraphics[width=0.5\linewidth]{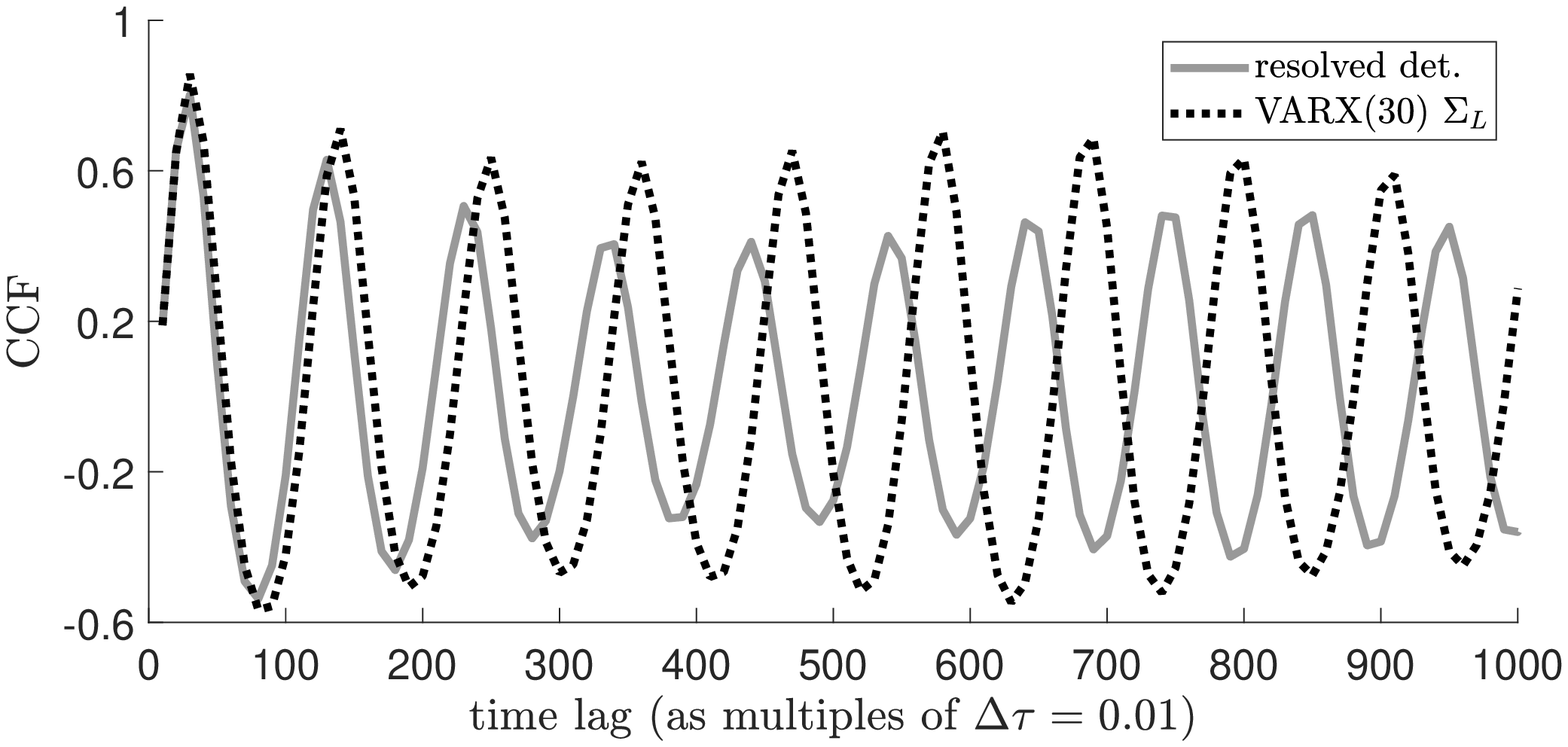}}
\caption{Comparison of \textbf{(a)} PDFs, \textbf{(b)} mean wave amplitude, \textbf{(c)} wave variance, \textbf{(d)} ACFs, and \textbf{(e)} CCFs of the reduced model using (\ref{eq:b_double_tri_dense}) and of the trimodal deterministic reference.}
\label{fig:doubleconditioningN_full}
\end{figure}

Figure \ref{fig:doubleconditioningN_full} shows the results using (\ref{eq:b_double_tri_dense}). The trimodal structure in the PDF of $x_k$ is reproduced accurately, as shown in Figure \ref{fig:distr_doubleN_2}. The main deviation from the trimodal L96 reference is a slightly higher kurtosis in the PDF for $\widetilde{x}_k$. Furthermore, the oscillations in the ACF and CCF have somewhat shorter period compared to those resulting from  (\ref{eq:b_double_tri_diag}), and align better (albeit not perfectly) with the reference trimodal L96 model, compare in particular Figures \ref{fig:CCF_doubleN} and \ref{fig:CCF_doubleN_2} to see an improved CCF reproduction. Finally, the mean amplitude and variance of most wave numbers differ only slightly from the reference values in Figures \ref{fig:wavemean_doubleN_2} and \ref{fig:wavevar_doubleN_2}. Altogether, the results, while not perfect, are very satisfactory for this highly challenging test case.

\subsection{NARMAX parameterizations}\label{sec:results_narmax}
As motivated in Section \ref{sec:varx_comparisons}, we compare the VARX parameterizations from Sections \ref{sec:results_onevariable} and \ref{sec:results_twovariables} to the NARMAX parameterization proposed in \citet{chorin2015discrete}. Specifically, we compare to the performance of the two configurations of NARMAX used in \citet{chorin2015discrete} defined by the function $\Phi$ (for further details on the NARMAX description see \citet{chorin2015discrete}):

\begin{align}
\Phi^n &= \mu + a_1 z^{n-1} + b_{1,1} x^{n-1} + b_{2,1} x^{n-2} + d_1 \xi^{n-1}, \label{eq:narmax_1201}\tag{$\text{NARMAX}_{1,2,0,1}$} \\
\Phi^n &= \mu + a_1 z^{n-1} + b_{1,1}x^{n-1} + b_{1,2} (x^{n-1})^2 + b_{1,3} (x^{n-1})^3 + c_{1,1} (R_{\delta} (x^{n-1})), \label{eq:narmax_1110}\tag{$\text{NARMAX}_{1,1,1,0}$}
\end{align}

\noindent where $\xi^{n}$ are independent Gaussian random variables with zero mean and variance $\sigma^2$, $R_{\delta}(x)$ represents the resolved features of the L96 model that are only dependent on $x$, and $\mu, \sigma^2, a_i, b_i, c_i, d_i$ are the parameters to be estimated. The NARMAX parameterization is applied independently to each grid point $k$. Because the L96 model is spatially homogeneous, the estimated NARMAX parameters are equal for all grid points $k$. 

\citet{chorin2015discrete} show that the NARMAX models above perform very well for the unimodal L96 configuration (see Table \ref{tab:parameters}), using different sampling intervals. The (\ref{eq:narmax_1201})  model gives good results with $\delta t = 10^{-2}$, whereas  (\ref{eq:narmax_1110})  performs well  with $\delta t = 5 \cdot 10^{-2}$. It is not dicussed in \citet{chorin2015discrete} how these specific configurations of NARMAX were selected. The choice of configuration is important though: we applied (\ref{eq:narmax_1110}) to the case with $\delta t = 10^{-2}$ (including re-estimation of parameters) and found it to be less accurate than (\ref{eq:narmax_1201}) (results not shown).



Analogous to the tests in Section \ref{sec:results_twovariables} we test the performance of the NARMAX models also with the trimodal L96 configuration (see Table \ref{tab:parameters}). The estimated model parameters resulting from the maximum likelihood estimation (see \citet{chorin2015discrete}) are shown in Table \ref{tab:narmax_regression}. 

\begin{figure}[htb]
\centering
\captionsetup[subfloat]{justification=centering}
\subfloat[\label{fig:distr_NARMAX}]{\includegraphics[width=0.33\linewidth]{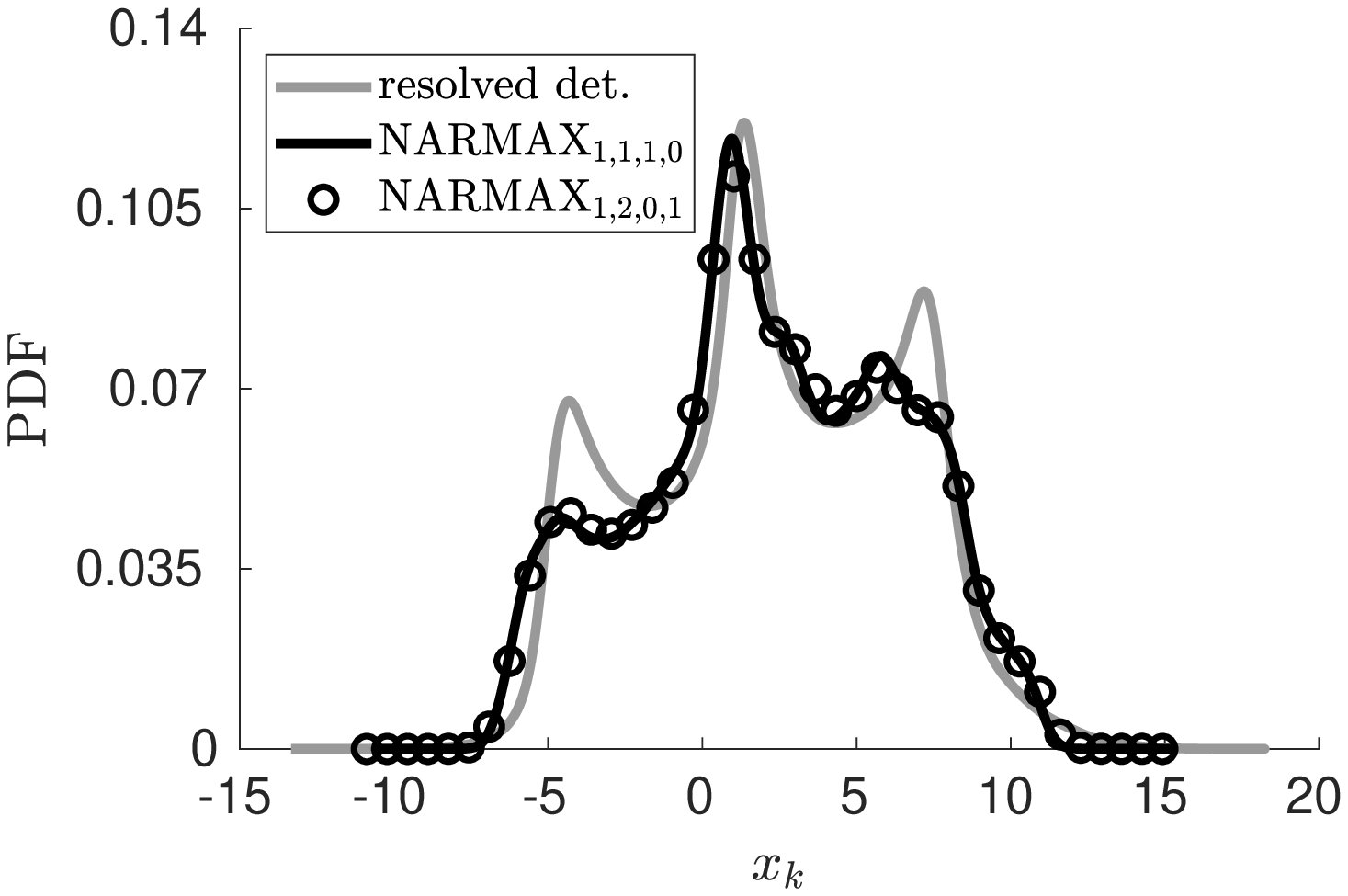}}
\hfill
\subfloat[\label{fig:wavemean_NARMAX}]{\includegraphics[width=0.33\linewidth]{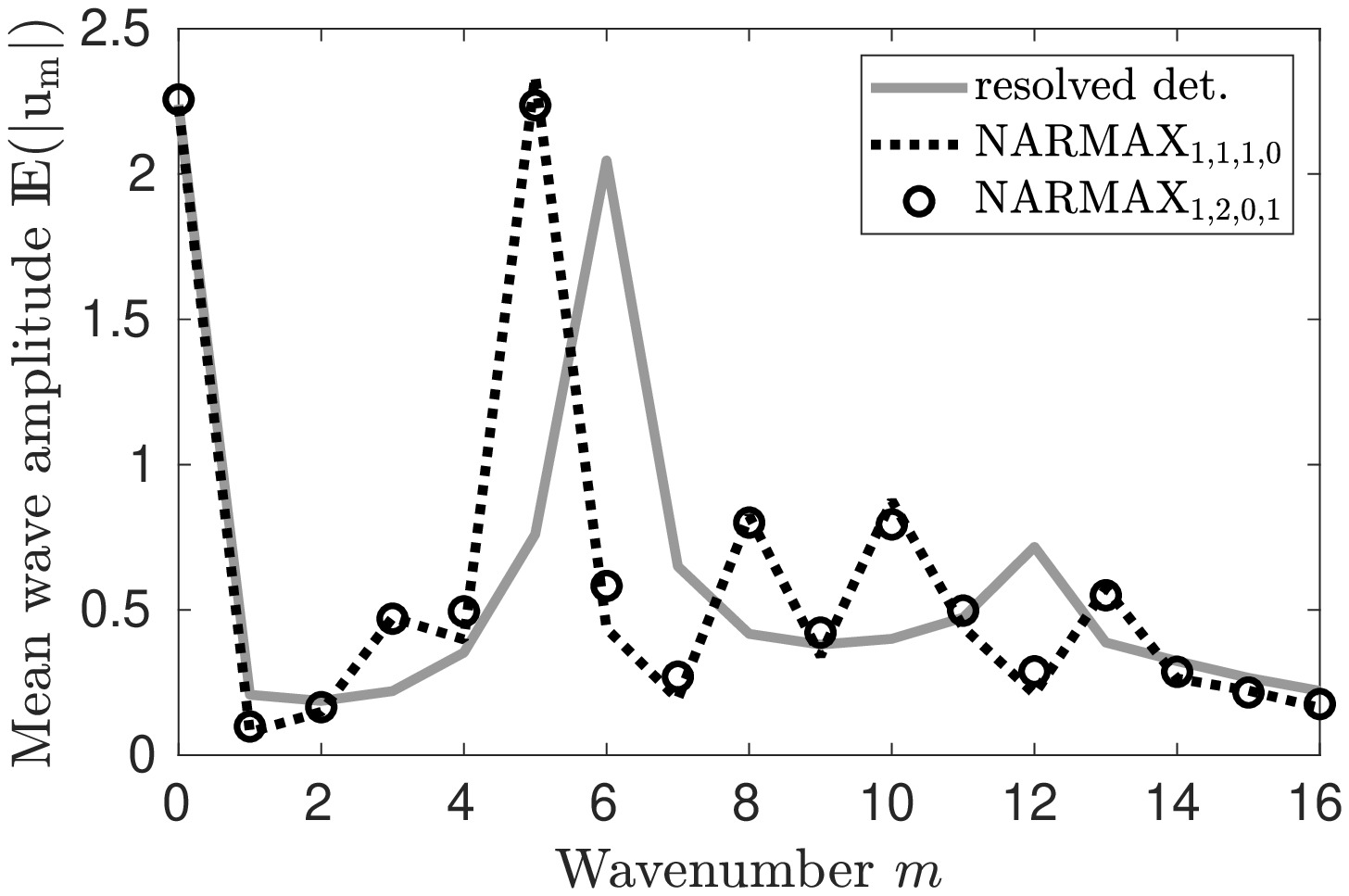}}
\hfill
\subfloat[\label{fig:wavevar_NARMAX}]{\includegraphics[width=0.33\linewidth]{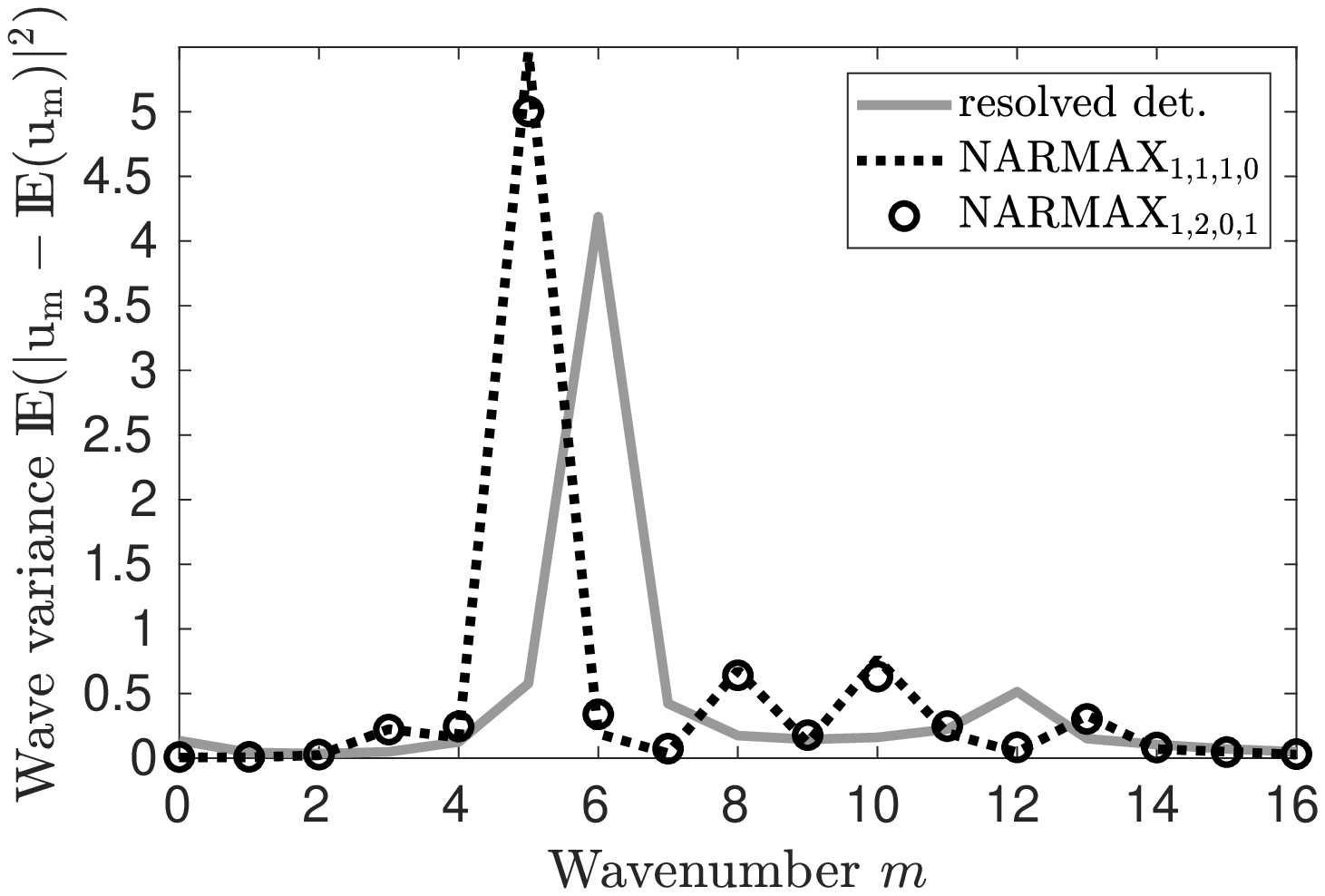}}
\hfill
\subfloat[\label{fig:ACF_NARMAX}]{\includegraphics[width=0.5\linewidth]{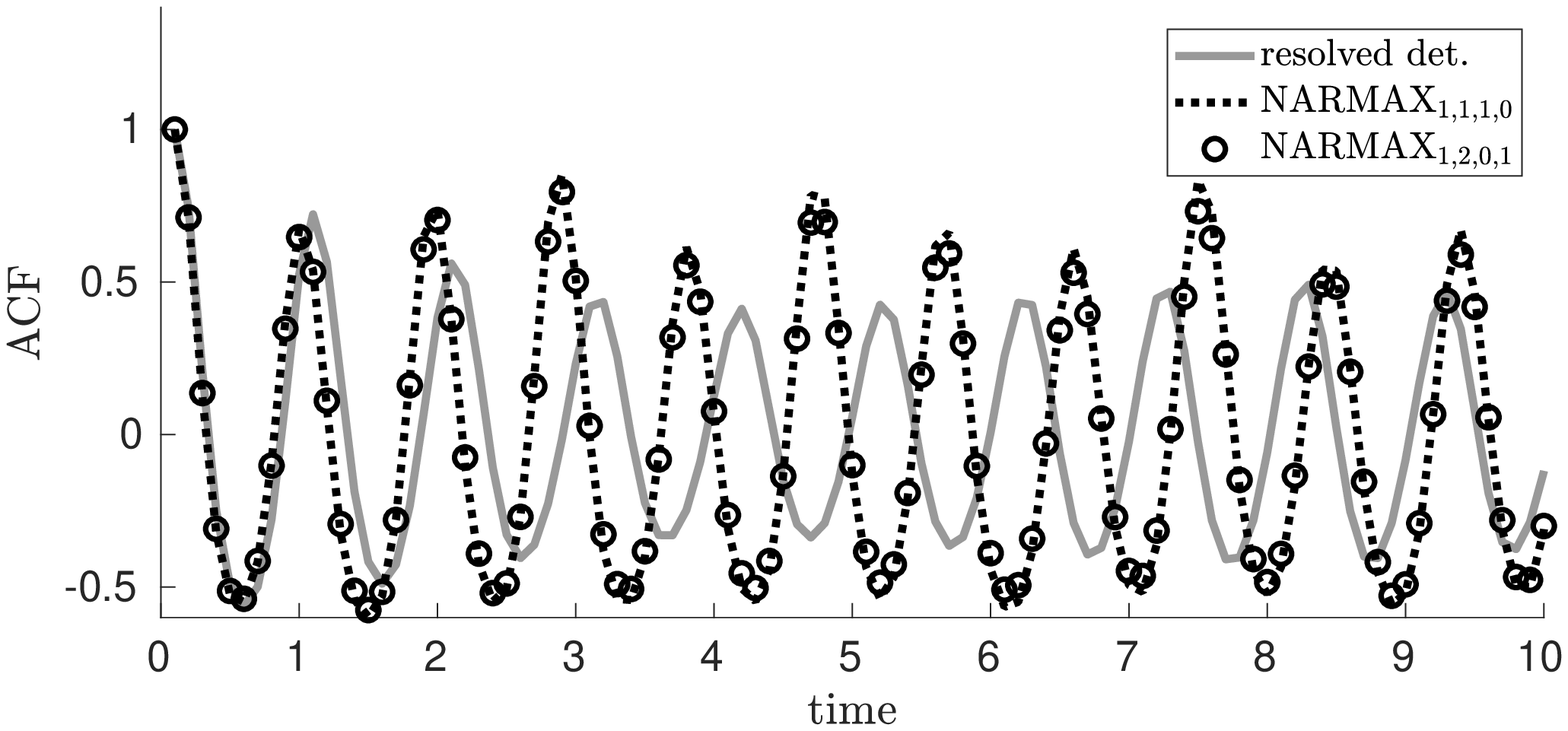}}
\hfill
\subfloat[\label{fig:CCF_NARMAX}]{\includegraphics[width=0.5\linewidth]{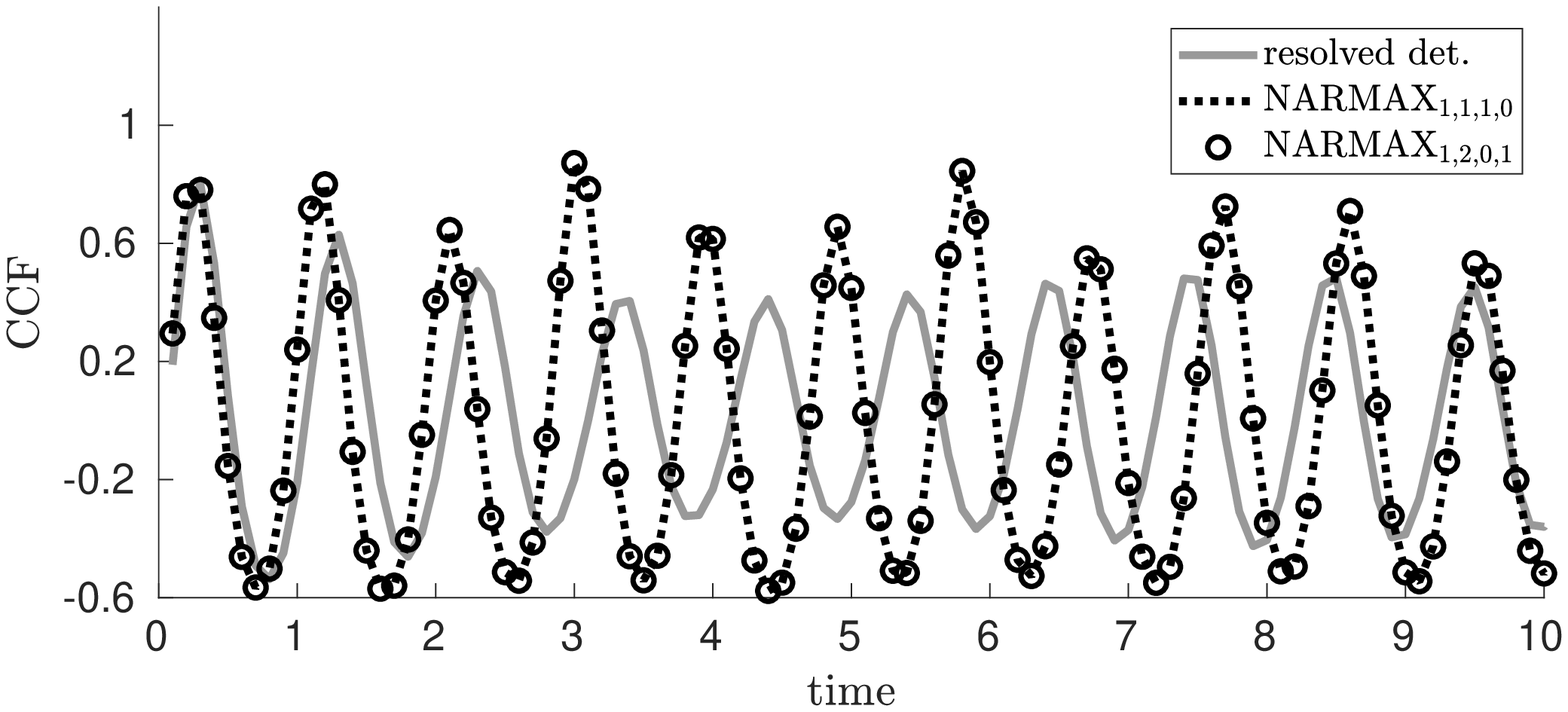}}
\caption{Comparison of \textbf{(a)} PDFs, \textbf{(b)} mean wave amplitude, \textbf{(c)} wave variance, \textbf{(d)} ACFs, and \textbf{(e)} CCFs of the NARMAX models proposed in \citet{chorin2015discrete} and of the trimodal deterministic reference.}
\label{fig:narmax_full}
\end{figure}

Figure \ref{fig:narmax_full} shows that the NARMAX models have comparable performance for the trimodal L96 configuration. Neither of the NARMAX models reproduces the trimodal distribution of $x_k$ accurately, as shown in Figure \ref{fig:distr_NARMAX}. However, they do reproduce accurately the mean and variance of the distribution.

\begin{table}[htb]
\caption{Estimated parameters in the NARMAX models for $\delta t = 0.01$}
\centering
\begin{tabularx}{0.95\textwidth}{l X X X X X X X}
\noalign{\vskip 1pt}
(\ref{eq:narmax_1201}) & $a_1$  & $b_{1,1}$ & $b_{2,1}$ & $d_1$ & $ $ & $\mu$ & $\sigma^2$ \\
\hline
\noalign{\vskip 2pt}
 & 0.9780 & -0.1276 & 0.1134 & 0.9998 & - & 0.0096 & 0.0028  \\
\hline
\noalign{\vskip 2pt}
(\ref{eq:narmax_1110}) & $a_1$  & $b_{1,1}$ & $b_{1,2}$ & $b_{1,3}$ & $ c_{1,1} $ & $\mu$ & $\sigma^2$ \\
\hline
\noalign{\vskip 2pt}
 & 0.9729 & -0.0669 & -0.0001 & 0.0001 & -0.0028 & 0.0467 & 0.0106 \\
\hline
\end{tabularx}
\label{tab:narmax_regression}
\end{table}

Figures \ref{fig:wavemean_NARMAX} and \ref{fig:wavevar_NARMAX} show that the wave statistics are also not reproduced accurately. The most prominent peak at wavenumber 5 is shifted, and some of the higher wavenumbers have overestimated mean and variance. For the correlation functions (ACF and CCF), both (\ref{eq:narmax_1201}) and (\ref{eq:narmax_1110}) result in oscillations with periods  that are somewhat too short (Figures \ref{fig:ACF_NARMAX} and \ref{fig:CCF_NARMAX}), whereas the VARX models in section \ref{sec:results_twovariables} gave periods that are a bit too long in the trimodal case  (e.g. Figure \ref{fig:doubleconditioningN_full}).

Overall, the VARX models (in particular (\ref{eq:b_double_tri_dense})) show better performance on the trimodal test case than the NARMAX models, with more accurate reproduction of the PDF and wave statistics. It must be noted that although we estimated the parameters of the NARMAX models specifically for the trimodal test case (see Table \ref{tab:narmax_regression}), we did not alter their configurations (i.e., the parameters $p,r,s,q$ that determine the structure of the NARMAX model). A different NARMAX configuration may be more optimal for the trimodal test case, however we have no guidance on how to select such a configuration. 


\section{Discussion}\label{sec:discussion}
In this study we proposed a method for data-driven stochastic parameterization using vector autoregressive processes with exogenous variable (VARX). This method is used to parameterize the feedback from unresolved processes in reduced models of multiscale dynamical systems. The choice for VARX is aimed specifically at spatially extended dynamical systems, for which it is important to capture spatial correlations, while keeping the number of parameters that must be estimated from data as low as possible.

We tested the proposed VARX parameterization method on the 2-layer L96 model (\ref{eq:l96slow}) - (\ref{eq:l96defresidual}), replacing the feedback vector $\vecB$ by a VARX $\widetilde{\vecB}$ so that the ``small-scale'' variables $y_{j,k}$ no longer had to be resolved. 
The process  $\widetilde{\vecB}$ was trained to emulate the dynamical effects of $\vecB$. With a proper formulation of $\widetilde{\vecB}$ the simulations of the reduced model were able to reproduce the statistical criteria of the reference simulation accurately. We note that these criteria  focus on long-term statistical properties rather than on the accuracy of short-term predictions.

The stochastic approach formulated in this study was developed with the aim to limit the amount of required computer memory and  number of parameters, as these can become computational bottlenecks in large, spatially extended systems (see e.g. \citet{verheul2017covariatebased}).
To this end, we modeled the VARX models with diagonal coefficient matrices $A_i$ and $D$. The covariance was estimated in a straightforward manner from the regression residuals. We considered both a diagonal and a fully dense covariance matrix. Our VARX model set-up is a particular case of a Gaussian process that uses generalized linear models (GLIMs) to describe its mean matrix, where the covariates of the GLIM represent spatio-temporal process variables. In this study we chose to formulate our approach in the more  specific terms of VARX processes.

In order to test the performance of the proposed stochastic parameterizations, we compared the reduced stochastic model simulations with two different configurations of the deterministic L96 reference model. First, the unimodal configuration, where ``unimodal'' refers to the overall shape of the probability distribution of $\vecX$, the variable of interest. This is a ``standard'' configuration of the L96 model that has also been used in previous studies. Second, to provide a very challenging test case and push our methodology to its limits, we also considered a \emph{trimodal} configuration of the L96 model. This is a non-standard configuration for the L96 model that exhibits three clear distinct peaks in the distribution of $\vecX$. The trimodal configuration tests the robustness of the proposed VARX process. As mentioned, the performance was assessed using a number of statistical criteria of the resolved model variable $\vecX$: the probability density function (PDF), the autocorrelations (ACFs), cross-correlations (CCFs), and the mean and variance of the wavenumber vector of $\vecX$.

In our results we compared different stochastic parameterizations for the reduced model \Crefrange{eq:stochl96x}{eq:stochl96r}. First, we tested both conditioning on the state vector $\vecX$ and self-conditioning on the stochastic process $\vecB$ in parameterizations (\ref{eq:b_singlex}) and (\ref{eq:b_singleb}), respectively. Here, self-conditioning refers to conditioning on the process itself at previous times. The results show that these regressors serve different roles in the conditioning. The state-dependent regressor $\vecX$ served effectively as predictor variable for the unresolved process, whereas the self-conditioning on $\vecB$ was instrumental in preserving temporal (de)correlations in the VARX. Each of these regressors by themselves was unsuccessful in giving satisfying results. However, combining the state-dependent and self-conditioning regressors proved very successful. The statistical criteria of the reference unimodal L96 model were reproduced very accurately using just a diagonal covariance matrix. For the trimodal test case, VARX with a diagonal covariance matrix gave qualitatively correct but not very accurate results. We showed quantitative improvement of results using a fully-dense covariance structure.


Finally, we also compared the performance of the VARX models to the NARMAX models proposed in \citet{chorin2015discrete}. As shown in \citet{chorin2015discrete}, the NARMAX models perform very accurately for the unimodal L96 test case. However, we showed that for the trimodal test case, the NARMAX models were not able to reproduce the trimodal distribution of the resolved variable $x_k$  accurately, nor its wave statistics.

The NARMAX models provide a parameterization for a single grid point, so they are applied independently to all grid points. By contrast, the VARX model can give a parameterization for the entire grid at once (as the VARX process is vector-valued), making it easier to include spatial correlations and spatial inhomogeneity. These spatial characteristics can be important for applications such as ocean modeling.




In future work we plan to develop the VARX stochastic parameterization methodology further. An important issue to consider is how to compute efficiently with a covariance structure that allows for spatial correlations without having to construct a fully dense matrix. This should involve a number of parameters that is at most linear in the number of spatial degrees of freedom, e.g. grid points. We intend to apply these methods in tests with a complex ocean model.

\medskip

{\bf Acknowledgements.} This research is funded by the Netherlands Organization for Scientific Research (NWO) through the Vidi project ``Stochastic models for unresolved scales in geophysical flows''. We thank Dr. Fei Lu for sharing his code for the NARMAX parameterization with us.
\medskip

\medskip

\bibliographystyle{apa}
\bibliography{bibliography}

\end{document}